\documentclass[12pt,letterpaper]{article}
\usepackage{amsmath,amssymb,array,calc,rotating,epsfig,psfrag, amscd, datetime,graphicx, caption,subcaption,comment,nicefrac}
\usepackage{color}
\usepackage[
      colorlinks=true,
      urlcolor=blue,    
      filecolor=blue,     
      citecolor=red,
      pdfstartview=FitV,
       bookmarksopen=true    
      ]{hyperref}
\usepackage[left=2cm,top=2cm,right=2cm,nohead]{geometry}
\def\di{{\rm d}}
\newcommand{\nc}{\newcommand}

\definecolor{cardinal}{rgb}{0.6,0,0}
\definecolor{darkgreen}{rgb}{0,0.5,0}
\definecolor{golden}{rgb}{0.92, 0.7, 0}
\definecolor{midnight}{rgb}{0, 0, 0.5}
\definecolor{darkblue}{rgb}{0.2, 0, 0.8}

\nc{\ra}{\rightarrow} 
\nc{\lra}{\leftrightarrow} 
\nc{\Ra}{\Rightarrow} 
\nc{\LRa}{\Leftightarrow} 
\nc{\blp}{{\big (}}
\nc{\brp}{{\big )}}
\nc{\Blp}{{\Big (}}
\nc{\Brp}{{\Big )}}
\nc{\bglp}{{\bigg (}}
\nc{\bgrp}{{\bigg )}}
\nc{\Bglp}{{\Bigg (}}
\nc{\Bgrp}{{\Bigg )}}
\nc{\slb}{{\rm [}}
\nc{\srb}{{\rm ]}}
\nc{\bslb}{{\rm \big [}}
\nc{\bsrb}{{\rm \big ]}}
\nc{\Bslb}{{\rm \Big [}}
\nc{\Bsrb}{{\rm \Big ]}}

\def\al{\alpha}

\def\eps{\epsilon}
\nc{\veps}{\varepsilon}
\def\gam{\gamma}

\def\lam{\lambda}
\def\om{\omega}

\nc{\vphi}{\varphi}
\def\tha{\theta}

\def\sig{\sigma}

\def\Gam{\Gamma}

\def\Lam{\Lambda}
\def\Om{\Omega}
\def\Sig{\Sigma}

\def\coeff#1#2{\relax{\textstyle {#1 \over #2}}\displaystyle}

\nc{\myvspace}{\rule[-1em]{0pt}{2.5em}}
\nc{\bea}{\begin{eqnarray}}
\nc{\eea}{\end{eqnarray}}
\nc{\be}{\begin{equation}}
\nc{\ee}{\end{equation}}
\nc{\barr}{\begin{array}}
\nc{\earr}{\end{array}}
\nc{\co}{{\cal o}}

\nc{\cA}{{\cal A}}
\nc{\cB}{{ \cal B}}

\def\cD{{\cal D}}

\nc{\cF}{{\cal F}}
\nc{\cG}{{\cal G}}
\def\cH{{\cal H}}
\def\cI{{\cal I}}

\def\cK{{\cal K}}
\nc{\cL}{{\cal L}}
\nc{\cM}{{\cal M}}

\def\cN{{\cal N}}

\def\cO{{\cal O}}

\nc{\cQ}{{\cal Q}}
\nc{\cR}{{\cal R}}
\def\cS{{\cal S}}

\def\cV{{\cal V}}
\def\cV{{\cal V}}
\def\cW{{\cal W}}

\def\cZ{{\cal Z}}
\nc{\cQd}{\cQ^{\dagger}}
\nc{\cRd}{\cR^{\dagger}}
\nc{\BB}{{\mathbb B}}
\nc{\CC}{{\mathbb C}}
\nc{\DD}{{\mathbb D}}
\nc{\EE}{{\mathbb E}}
\nc{\FF}{{\mathbb F}}
\nc{\GG}{{\mathbb G}}
\nc{\HH}{{\mathbb H}}
\nc{\JJ}{{\mathbb J}}
\nc{\MM}{{\mathbb M}}
\nc{\RR}{{\mathbb R}}
\nc{\PP}{{\mathbb P}}
\nc{\QQ}{{\mathbb Q}}
\nc{\UU}{{\mathbb U}}
\nc{\ZZ}{{\mathbb Z}}
\nc{\calone}{{\mathbb 1}}
\nc{\half}{\frac{1}{2}}
\nc{\quarter}{\coeff{1}{4}}
\nc{\del}{\partial}

\nc{\delbar}{\bar\partial}
\nc{\thalf}{\frac{t}{2}}
\nc{\Spin}{\operatorname{Spin}}
\nc{\SO}{\operatorname{SO}}

\nc{\Sp}{{\rm Sp}}
\nc{\com}[2]{{ \left[ #1, #2 \right] }}
\nc{\acom}[2]{{ \left\{ #1, #2 \right\} }}
\nc{\rr}{\rightarrow}
\nc{\p}{\partial}
\nc{\LT}{{\LL_\T}}
\nc{\Tr}{{\rm Tr}}
\nc{\tr}{{\rm tr}}
\nc{\Adag}{A^{\dagger}}
\nc{\AdagI}{A^{\dagger I}}
\nc{\AdagJ}{A^{\dagger J}}
\nc{\AdagK}{A^{\dagger K}}
\nc{\AdagL}{A^{\dagger L}}
\nc{\AdagM}{A^{\dagger M}}
\nc{\Bdag}{B^{\dagger}}
\nc{\BdagI}{B^{\dagger}_I}
\nc{\BdagJ}{B^{\dagger}_J}
\nc{\BdagK}{B^{\dagger}_K}
\nc{\BdagL}{B^{\dagger}_L}
\nc{\BdagM}{B^{\dagger}_M}
\nc{\Cdag}{C^{\dagger}}
\nc{\CdagI}{C^{\dagger I}}
\nc{\CdagJ}{C^{\dagger J}}
\nc{\CdagK}{C^{\dagger K}}
\nc{\Ddag}{D^{\dagger}}
\nc{\DdagI}{D^{\dagger I}}
\nc{\DdagJ}{D^{\dagger J}}
\nc{\DdagK}{D^{\dagger K}}
\nc{\bva}{\breve{a}}
\nc{\bvb}{\breve{b}}
\nc{\bvc}{\breve{c}}
\nc{\bvd}{\breve{d}}
\nc{\bve}{\breve{e}}
\nc{\bvf}{\breve{f}}
\nc{\bvg}{\breve{g}}
\nc{\bvh}{\breve{h}}
\nc{\bvi}{\breve{i}}
\nc{\bvj}{\breve{j}}
\nc{\bvk}{\breve{k}}
\nc{\bvl}{\breve{l}}
\nc{\bvm}{\breve{m}}
\nc{\bvn}{\breve{n}}
\nc{\bvo}{\breve{o}}
\nc{\bvp}{\breve{p}}
\nc{\brvq}{\breve{q}}
\nc{\bvr}{\breve{r}}
\nc{\bvs}{\breve{s}}
\nc{\bvt}{\breve{t}}
\nc{\bvu}{\breve{u}}
\nc{\bvv}{\breve{v}}
\nc{\bvw}{\breve{w}}
\nc{\bvx}{\breve{x}}
\nc{\bvy}{\breve{y}}
\nc{\bvz}{\breve{z}}

\nc{\bvA}{\breve{A}}
\nc{\bvB}{\breve{B}}
\nc{\bvC}{\breve{C}}
\nc{\bvD}{\breve{D}}
\nc{\bvE}{\breve{E}}
\nc{\bvF}{\breve{F}}
\nc{\bvG}{\breve{G}}
\nc{\bvH}{\breve{H}}
\nc{\bvI}{\breve{I}}
\nc{\bvJ}{\breve{J}}
\nc{\bvK}{\breve{K}}
\nc{\bvL}{\breve{L}}
\nc{\bvM}{\breve{M}}
\nc{\bvN}{\breve{N}}
\nc{\bvO}{\breve{O}}
\nc{\bvP}{\breve{P}}
\nc{\bvQ}{\breve{Q}}
\nc{\bvR}{\breve{R}}
\nc{\bvS}{\breve{S}}
\nc{\bvT}{\breve{T}}
\nc{\bvU}{\breve{U}}
\nc{\bvV}{\breve{V}}
\nc{\bvcV}{\breve{\cV}}
\nc{\bvW}{\breve{W}}
\nc{\bvX}{\breve{X}}
\nc{\bvY}{\breve{Y}}
\nc{\bvZ}{\breve{Z}}

\nc{\ul}[1]{{\underline{#1}}}
\nc{\tal}{\widetilde{\alpha}}
\nc{\tbeta}{\widetilde{\beta}}
\nc{\ttha}{\tilde{\theta}}
\nc{\ttau}{\tilde{\tau}}
\nc{\tTha}{\tilde{\Theta}}
\nc{\tphi}{\tilde{\phi}}
\nc{\tsig}{\tilde{\sig}}
\nc{\tom}{\widetilde{\om}}
\nc{\tOm}{\widetilde{\Om}}
\nc{\tlam}{\widetilde{\lam}}
\nc{\tLam}{\tilde{\Lam}}
\nc{\tSig}{\widetilde{\Sig}}
\nc{\tPhi}{\tilde{\Phi}}
\nc{\tPhibar}{\ol{\tPhi}}
\nc{\tPi}{\widetilde{\Pi}}
\nc{\tpsi}{\widetilde{\psi}}
\nc{\tPsi}{\tilde{\Psi}}
\nc{\tgam}{\widetilde{\gam}}
\nc{\tGam}{\widetilde{\Gam}}
\nc{\tzeta}{\tilde{\zeta}}
\nc{\tZeta}{\tilde{\Zeta}}
\nc{\teta}{\widetilde{\eta}}
\nc{\teps}{\tilde{\eps}}
\nc{\tveps}{\tilde{\veps}}
\nc{\tEta}{\tilde{\Eta}}
\nc{\tchi}{\tilde{\chi}}
\nc{\tChi}{\tilde{\Chi}}
\nc{\txi}{\tilde{\xi}}
\nc{\tXi}{\widetilde{\Xi}}
\nc{\tnu}{\tilde{\nu}}
\nc{\tmu}{\tilde{\mu}}

\nc{\ta}{\tilde a}
\nc{\tb}{\tilde b}
\nc{\tc}{\tilde c}
\nc{\te}{\tilde e}
\nc{\tf}{\widetilde f}
\nc{\tg}{\widetilde g}
\nc{\ti}{\tilde i}
\nc{\tj}{\tilde j}
\nc{\tk}{\widetilde k}
\nc{\tl}{\tilde l}
\nc{\tm}{\widetilde m}
\nc{\tn}{\tilde n}
\nc{\tp}{\tilde{p}}
\nc{\tq}{\widetilde{q}}
\nc{\trr}{{\tilde r}}
\nc{\ts}{{\tilde s}}
\nc{\tu}{{\tilde u}}
\nc{\tv}{{\tilde v}}
\nc{\tw}{{\tilde w}}
\nc{\tx}{{\tilde x}}
\nc{\ty}{{\tilde y}}
\nc{\tz}{\tilde z}
\nc{\tA}{{\widetilde A}}
\nc{\tAbar}{{\ol \tA}}
\nc{\tB}{{\widetilde B}}
\nc{\tC}{{\widetilde C}}
\nc{\tD}{{\widetilde D}}
\nc{\tE}{{\widetilde E}}
\nc{\tF}{{\widetilde F}}
\nc{\tG}{{\widetilde G}}
\nc{\tcG}{{\widetilde \cG}}
\nc{\tH}{{\widetilde H}}
\nc{\tcH}{{\widetilde \cH}}
\nc{\tI}{{\widetilde I}}
\nc{\tcI}{{\widetilde \cI}}
\nc{\tJ}{{\widetilde J}}
\nc{\tJbar}{{\ol {\tilde J}}}
\nc{\tK}{{\widetilde K}}
\nc{\tL}{{\widetilde L}}
\nc{\tcL}{{\widetilde \cL}}
\nc{\tcLbar}{{\ol \tcL}}
\nc{\tM}{{\widetilde M}}
\nc{\tN}{{\widetilde N}}
\nc{\tcN}{{\widetilde \cN}}
\nc{\tP}{{\widetilde P}}
\nc{\tQ}{{\widetilde Q}}
\nc{\tR}{{\widetilde R}}
\nc{\tS}{\widetilde{S}}
\nc{\tT}{\widetilde{T}}
\nc{\tU}{\widetilde{U}}
\nc{\tUU}{\widetilde{\UU}}
\nc{\tV}{\widetilde{V}}
\nc{\tcVbar}{\ol{\widetilde{\cV}}}
\nc{\tW}{\widetilde{W}}
\nc{\tcF}{\widetilde{{\cal F}}}
\nc{\tX}{\widetilde{X}}
\nc{\tY}{\widetilde{Y}}
\nc{\tcZ}{\tilde{\cZ}}
\nc{\tcZbar}{\ol{\tcZ}}

\nc{\ha}{\hat a}
\nc{\hb}{\hat b}
\nc{\hc}{\widehat c}
\nc{\hd}{\widehat d}
\nc{\he}{\widehat e}
\nc{\hf}{\widehat f}
\nc{\hg}{\widehat g}
\nc{\hh}{\widehat h}
\nc{\hm}{\widehat m}
\nc{\hn}{\widehat n}
\nc{\hp}{\widehat p}
\nc{\hq}{\widehat q}
\nc{\hr}{\widehat r}
\nc{\hs}{\widehat s}
\nc{\hu}{\widehat u}
\nc{\hv}{\widehat v}
\nc{\hw}{\widehat w}
\nc{\hx}{\widehat x}
\nc{\hy}{\widehat y}
\nc{\hz}{\widehat z}
\nc{\zhat}{\hat z}
\nc{\hA}{\widehat{A}}
\nc{\hB}{\widehat{B}}
\nc{\hC}{\widehat{C}}
\nc{\hD}{\widehat{D}}
\nc{\hcD}{\widehat{\cD}}
\nc{\hE}{\widehat{E}}
\nc{\hF}{\widehat{F}}
\nc{\hcF}{\widehat{\cF}}
\nc{\hG}{\widehat{G}}
\nc{\hcG}{\widehat{\cG}}
\nc{\hH}{\widehat{H}}
\nc{\hI}{\widehat{I}}
\nc{\hcI}{\widehat{\cI}}
\nc{\hJ}{\widehat{J}}
\nc{\hK}{\widehat{K}}
\nc{\hL}{\widehat{L}}
\nc{\hcL}{\widehat{\cL}}
\nc{\hM}{\widehat M}
\nc{\hcM}{\widehat{\cM}}
\nc{\hN}{\widehat{N}}
\nc{\hO}{\widehat{O}}
\nc{\hcO}{\widehat{\cO}}
\nc{\hP}{\widehat{P}}
\nc{\hQ}{\widehat{Q}}
\nc{\hcQ}{\widehat{\cQ}}
\nc{\hcR}{\widehat{\cR}}
\nc{\hR}{\widehat{R}}
\nc{\hS}{\widehat{S}}
\nc{\hcS}{\widehat{\cS}}
\nc{\hT}{\widehat{T}}
\nc{\hU}{\widehat{U}}
\nc{\hV}{\widehat V}
\nc{\hcV}{\widehat \cV}
\nc{\tcV}{\widetilde{\cV}}
\nc{\hX}{\widehat X}
\nc{\hcZ}{\widehat \cZ}
\nc{\hcZbar}{\ol{\widehat \cZ}}

\nc{\heta}{\widehat{\eta}}
\nc{\hal}{\widehat \alpha}
\nc{\hbeta}{\widehat \beta}
\nc{\heps}{\widehat \eps}
\nc{\hphi}{\widehat{\phi}}
\nc{\hkap}{\hat{\kappa}}
\nc{\hchi}{\widehat{\chi}}
\nc{\hpsi}{\widehat{\psi}}
\nc{\hgam}{\widehat{\gam}}
\nc{\hPhi}{\hat{\Phi}}
\nc{\hPsi}{\hat{\Psi}}
\nc{\hGam}{\hat{\Gam}}
\nc{\omhat}{\widehat{\om}}
\nc{\Omhat}{\widehat{\Om}}
\nc{\hsig}{\widehat{\sig}}
\nc{\hSig}{\widehat{\Sig}}
\nc{\htha}{\hat{\tha}}
\nc{\hrho}{\widehat{\rho}}
\nc{\hdel}{\widehat{\del}}
\nc{\hnabla}{\widehat{\nabla}}

\nc{\w}{\wedge}

\nc{\vb}{\vec b}
\nc{\vc}{\vec c}
\nc{\vd}{\vec d}
\nc{\ve}{\vec e}
\nc{\vf}{\vec f}
\nc{\vg}{\vec g}
\nc{\vh}{\vec h}
\nc{\vp}{\vec p}
\nc{\vq}{\vec q}
\nc{\vr}{\vec r}
\nc{\vs}{\vec s}
\nc{\vv}{\vec v}
\nc{\vw}{\vec w}
\nc{\vx}{\vec x}
\nc{\vy}{\vec y}
\nc{\vz}{\vec z}

\nc{\vB}{\vec B}
\nc{\vC}{\vec C}
\nc{\vD}{\vec D}
\nc{\vE}{\vec E}
\nc{\vF}{\vec F}
\nc{\vG}{\vec G}
\nc{\vH}{\vec H}
\nc{\vP}{\vec P}
\nc{\vQ}{\vec Q}
\nc{\vR}{\vec R}
\nc{\vS}{\vec S}
\nc{\vV}{\vec V}
\nc{\vW}{\vec W}
\nc{\vX}{\vec X}
\nc{\vY}{\vec Y}
\nc{\vZ}{\vec Z}

\nc{\ol}{\overline}
\nc{\abar}{\ol{a}}
\nc{\bbar}{\ol{b}}
\nc{\cbar}{\ol{c}}
\nc{\dbar}{\ol{d}}
\nc{\ebar}{\ol{e}}
\nc{\fbar}{\ol{f}}
\nc{\gbar}{\ol{g}}
\nc{\ibar}{\ol{\imath}}
\nc{\jbar}{\ol{\jmath}}
\nc{\kbar}{\ol{k}}
\nc{\lbar}{\ol{l}}
\nc{\mbar}{\ol{m}}
\nc{\nbar}{\ol{n}}
\nc{\pbar}{\ol{p}}
\nc{\qbar}{\ol{q}}
\nc{\rbar}{\ol{r}}
\nc{\sbar}{\ol{s}}
\nc{\ubar}{\ol{u}}
\nc{\vbar}{\ol{v}}
\nc{\wbar}{\ol{w}}
\nc{\xbar}{\ol{x}}
\nc{\ybar}{\ol{y}}
\nc{\zbar}{\ol{z}}

\nc{\Abar}{\ol{A}}
\nc{\Bbar}{\ol{B}}
\nc{\cBbar}{\ol{\cB}}
\nc{\Cbar}{\ol{C}}
\nc{\Dbar}{\ol{D}}
\nc{\Ebar}{\ol{E}}
\nc{\Fbar}{\ol{F}}
\nc{\Gbar}{\ol{G}}
\nc{\Jbar}{\ol{J}}
\nc{\Kbar}{\ol{K}}
\nc{\cKbar}{\ol{\cK}}
\nc{\Lbar}{\ol{L}}
\nc{\cLbar}{\ol{\cL}}
\nc{\Mbar}{\ol{M}}
\nc{\Nbar}{\ol{N}}
\nc{\Pbar}{\ol{P}}
\nc{\Qbar}{\ol{Q}}
\nc{\Rbar}{\ol{R}}
\nc{\Sbar}{\ol{S}}
\nc{\Tbar}{\ol{T}}
\nc{\Ubar}{\ol{U}}
\nc{\Vbar}{\ol{V}}
\nc{\cVbar}{\ol{\cV}}
\nc{\Wbar}{\ol{W}}
\nc{\cWbar}{\ol{\cW}}
\nc{\Xbar}{{\overline X}}
\nc{\Ybar}{{\overline Y}}
\nc{\Zbar}{{\overline Z}}
\nc{\cZbar}{{\overline \cZ}}

\nc{\epsbar}{\ol{\epsilon}}
\nc{\albar}{\ol{\al}}
\nc{\Albar}{\ol{\Al}}
\nc{\betabar}{\ol{\beta}}
\nc{\Betabar}{\ol{\Beta}}
\nc{\deltabar}{\ol{\delta}}
\nc{\etabar}{\ol{\eta}}
\nc{\lambar}{\ol{\lambda}}
\nc{\kapbar}{\ol{\kappa}}
\nc{\zetabar}{\ol{\zeta}}
\nc{\Zetabar}{\ol{\Zeta}}
\nc{\taubar}{\ol{\tau}}
\nc{\Taubar}{\ol{\Tau}}
\nc{\psibar}{\ol{\psi}}
\nc{\Psibar}{\ol{\Psi}}
\nc{\tpsibar}{\ol{\tpsi}}
\nc{\tPsibar}{\ol{\tPsi}}
\nc{\phibar}{\ol{\phi}}
\nc{\Phibar}{\ol{\Phi}}
\nc{\chibar}{\ol{\chi}}
\nc{\sigbar}{\ol{\sig}}
\nc{\Sigbar}{\ol{\Sig}}
\nc{\mubar}{\ol{\mu}}
\nc{\nubar}{\ol{\nu}}
\nc{\rhobar}{\ol{\rho}}
\nc{\ombar}{\ol{\om}}
\nc{\Ombar}{\ol{\Om}}
\nc{\Deltabar}{\ol{\Delta}}
\nc{\Thetabar}{\ol{\Theta}}
\nc{\xibar}{\ol{\xi}}
\nc{\Xibar}{\ol{\Xi}}

\nc{\Dthbar}{\ol{\rm D3}}

\nc{\fdot}{\dot{f}}
\nc{\gdot}{\dot{g}}
\nc{\pdot}{\dot{p}}
\nc{\qdot}{\dot{q}}
\nc{\rdot}{\dot{r}}
\nc{\sdot}{\dot{s}}
\nc{\tdot}{\dot{t}}
\nc{\udot}{\dot{u}}
\nc{\vdot}{\dot{v}}
\nc{\wdot}{\dot{w}}
\nc{\xdot}{\dot{x}}
\nc{\xddot}{\ddot{x}}
\nc{\ydot}{\dot{y}}
\nc{\zdot}{\dot{z}}
\nc{\yddot}{\ddot{y}}

\nc{\Adot}{\dot{A}}
\nc{\Bdot}{\dot{B}}
\nc{\Cdot}{\dot{C}}
\nc{\Pdot}{\dot{P}}
\nc{\Qdot}{\dot{Q}}

\nc{\Udot}{\dot{U}}
\nc{\Vdot}{\dot{V}}
\nc{\Wdot}{\dot{W}}

\nc{\taudot}{\dot{\tau}}
\nc{\phidot}{\dot{\phi}}
\nc{\psidot}{\dot{\psi}}
\nc{\chidot}{\dot{\chi}}
\nc{\sinp}{s_{\phi}}
\nc{\cosp}{c_{\phi}}
\nc{\tanp}{t_{\phi}}
\nc{\spone}{s_{\phi_1}}
\nc{\cpone}{c_{\phi_1}}
\nc{\tpone}{t_{\phi_1}}
\nc{\sptwo}{s_{\phi_2}}
\nc{\cptwo}{c_{\phi_2}}
\nc{\tptwo}{t_{\phi_2}}
\nc{\spth}{s_{\phi_3}}
\nc{\cpth}{c_{\phi_3}}
\nc{\tpth}{t_{\phi_3}}
\nc{\calp}{c_{\al}}
\nc{\salp}{s_{\al}}

\nc{\csch}{{\rm csch}}
\nc{\sech}{{\rm sech}}

\nc{\cothzlami}{\coth(z-\lam_i)}
\nc{\coshzlami}{\cosh(z-\lam_i)}
\nc{\sinhzlami}{\sinh(z-\lam_i)}

\nc{\cothzlamj}{\coth(z-\lam_j)}
\nc{\coshzlamj}{\cosh(z-\lam_j)}
\nc{\sinhzlamj}{\sinh(z-\lam_j)}

\nc{\cothlamij}{\coth(\lam_i-\lam_j)}
\nc{\coshlamij}{\cosh(\lam_i-\lam_j)}
\nc{\sinhlamij}{\sinh(\lam_i-\lam_j)}

\nc{\bah}{{\mathbf {\hat{A}}}}
\nc{\bX}{{\mathbf X}}
\nc{\ba}{{\bf a}}
\nc{\bb}{{\bf b}}
\nc{\bc}{{\bf c}}
\nc{\bd}{{\bf d}}
\nc{\bg}{{\bf g}}
\nc{\bk}{{\bf k}}
\nc{\bl}{{\bf l}}
\nc{\bm}{{\bf m}}
\nc{\bn}{{\bf n}}
\nc{\bo}{{\bf o}}
\nc{\bp}{{\bf p}}
\nc{\bq}{{\bf q}}
\nc{\br}{{\bf r}}
\nc{\bs}{{\bf s}}
\nc{\bt}{{\bf t}}
\nc{\bu}{{\bf u}}
\nc{\bv}{{\bf v}}
\nc{\bw}{{\bf w}}
\nc{\bx}{{\bf x}}
\nc{\by}{{\bf y}}
\nc{\bz}{{\bf z}}

\nc{\bH}{{\bf H}}
\nc{\bP}{{\bf P}}
\nc{\bQ}{{\bf Q}}

\nc{\bom}{{\bf \om}}
\nc{\bombar}{{\mathbf \ombar}}
\nc{\bPhi}{{\bf \Phi}}

\nc{\rma}{{\rm a}}
\nc{\rmb}{{\rm b}}
\nc{\rmc}{{\rm c}}
\nc{\rmd}{{\rm d}}
\nc{\rmg}{{\rm g}}
\nc{\rk}{{\rm k}}
\nc{\rml}{{\rm l}}
\nc{\rmm}{{\rm m}}
\nc{\rmn}{{\rm n}}
\nc{\rmo}{{\rm o}}
\nc{\rmp}{{\rm p}}
\nc{\rmq}{{\rm q}}
\nc{\rmr}{{\rm r}}
\nc{\rms}{{\rm s}}
\nc{\rmt}{{\rm t}}
\nc{\rmu}{{\rm u}}
\nc{\rmv}{{\rm v}}
\nc{\rmw}{{\rm w}}
\nc{\rmx}{{\rm x}}
\nc{\rmy}{{\rm y}}
\nc{\rmz}{{\rm z}}

\nc{\dal}{\dot{\al}}
\nc{\thadot}{\dot{\tha}}
\nc{\thab}{\bar{\theta}}
\nc{\thal}{\theta^{\al}}
\nc{\thdal}{\bar{\theta}^{\dal}}

\nc{\thsigthm}{\tha \sigma^m \thab}
\nc{\thsigthn}{\tha \sigma^n \thab}

\nc{\Dal}{D_{\al}}
\nc{\Ddal}{\bar{D}_{\dal}}
\nc{\CDal}{{\cal D}_{\al}}
\nc{\CDdal}{\bar{\cal D}_{\dal}}

\nc{\eq}[1]{{(\ref{#1})}}
\nc{\eqtwo}[2]{{(\ref{#1},\ref{#2})}}
\nc{\eqthree}[3]{(\ref{#1},\ref{#2},\ref{#3})}
\nc{\eqfour}[4]{(\ref{#1},\ref{#2},\ref{#3},\ref{#4})}
\nc{\eqfive}[5]{(\ref{#1},\ref{#2},\ref{#3},\ref{#4,\ref{#5}})}
\nc{\non}{\nonumber}
\nc{\Fzero}{F_{(0)}}
\nc{\Ftwo}{F_{(2)}}
\nc{\Ffour}{F_{(4)}}
\nc{\Fone}{F_{(1)}}
\nc{\Fthree}{F_{(3)}}
\nc{\Ffive}{F_{(5)}}
\nc{\Fn}{F_{(n)}}
\nc{\Fp}{F_{(p)}}

\nc{\tFzero}{\tF_{(0)}}
\nc{\tFtwo}{\tF_{(2)}}
\nc{\tFfour}{\tF_{(4)}}
\nc{\tFone}{\tF_{(1)}}
\nc{\tFthree}{\tF_{(3)}}
\nc{\tFfive}{\tF_{(5)}}
\nc{\tFn}{\tF_{(n)}}
\nc{\tFp}{\tF_{(p)}}

\nc{\Czero}{C_{(0)}}
\nc{\Ctwo}{C_{(2)}}
\nc{\Cfour}{C_{(4)}}
\nc{\Cone}{C_{(1)}}
\nc{\Cthree}{C_{(3)}}
\nc{\Cfive}{C_{(5)}}
\nc{\Cn}{C_{(n)}}

\nc{\equ}{{\rm eq}}

\nc{\vol}{{\rm vol}}
\nc{\Ainf}{A_{\infty}}
\nc{\End}{{\rm End}}
\nc{\Ext}{{\rm Ext}}
\nc{\IIB}{{\rm IIB}}
\nc{\Ad}{{\rm Ad}}
\nc{\IIA}{{\rm IIA}}
\nc{\AdS}{{\rm AdS}}
\nc{\CFT}{{\rm CFT}}
\nc{\diag}{{\rm diag}}
\nc{\Log}{{\rm Log}}
\nc{\Dslash}{\ensuremath \raisebox{0.025cm}{\slash}\hspace{-0.32cm} D}
\nc{\cDslash}{\ensuremath \raisebox{0.025cm}{\slash}\hspace{-0.32cm} \cD}
\nc{\omslash}{\om\!\!\!/}
\nc{\no}{\!:\!\!}
\nc{\ointdz}{\oint\frac{dz}{2\pi i}}
\nc{\ointdzone}{\oint\frac{dz_1}{2\pi i}}
\nc{\ointdztwo}{\oint\frac{dz_2}{2\pi i}}
\nc{\ointdzb}{\oint\frac{d\zbar}{2\pi i}}
\nc{\ointdzbone}{\oint\frac{d\zbar_1}{2\pi i}}
\nc{\ointdzbtwo}{\oint\frac{d\zbar_2}{2\pi i}}
\nc{\dz}{\frac{dz}{2\pi i}}
\nc{\dzb}{\frac{d\zbar}{2\pi i}}
\nc{\bpm}{\begin{pmatrix}}
\nc{\epm}{\end{pmatrix}}
 \nc{\bitem}{\begin{itemize}}
 \nc{\eitem}{\end{itemize}}
 \nc{\exercise}{\vskip 2mm \noindent {\bf Exercise:}}
 \nc{\definition}{\vskip 2mm \noindent {\bf Definition:}}
 
\begin{document}
\begin{center}
\vskip 2 cm

{\Large \bf  The Abelian Heterotic Conifold }
\vskip 1.25 cm 
{Nick Halmagyi$^*$, Dan Isra\"el$^*$, Eirik Svanes$^{*\dagger}$}\\
\vskip 5mm

\vskip0.5cm
\textit{
$^\dagger$Sorbonne Universit\'es, Institut Lagrange de Paris, \\
98 bis Bd Arago, 75014, Paris, France,\\
\vskip0.2cm
$^*$Sorbonne Universit\'es, UPMC Paris 06,  \\ 
UMR 7589, LPTHE, 75005, Paris, France \\
\vskip0.2cm
and \\
\vskip0.2cm
$^*$CNRS, UMR 7589, LPTHE, 75005, Paris, France}\\
\vskip0.5cm
halmagyi@lpthe.jussieu.fr \\ 
israel@lpthe.jussieu.fr \\
esvanes@lpthe.jussieu.fr 
\end{center}

\begin{abstract}
We study heterotic supergravity on the conifold and its $\ZZ_2$ orbifold with Abelian gauge fields and three-form flux. At large distances, these solutions are locally Ricci-flat, have a magnetic flux through the two-sphere at infinity as well as non-zero five-brane charge. For a given flux, our family of solutions has three real parameters, the size of the pair of two spheres in the IR and the dilaton zero mode. We present an explicit analytic solution for the decoupled near horizon region where for a given flux, the size of the cycles is frozen and the only parameter is the dilaton zero mode. We also present an exactly solvable worldsheet CFT for this near horizon region. When one of the two cycles has vanishing size, the near horizon region no longer exists but we obtain a solution on the (unorbifolded) resolved conifold.
\end{abstract}
%\end{titlepage}
%\newpage
%\tableofcontents

%%%%%%%%%%%%%%%%%%%%%%%%%%%%%%%%%%%%
\section{Introduction}
%%%%%%%%%%%%%%%%%%%%%%%%%%%%%%%%%%%%

Supergravity solutions with non-trivial flux profiles are a key tool in constructing four-dimensional, 
low energy models from string theory with reduced supersymmetry. Without flux, the low-energy models contain massless scalar fields;  
flux on the internal manifold provides a mechanism to potentially give a large mass to these fields. Breaking supersymmetry spontaneously, 
generating a small cosmological constant and providing conditions for slow roll inflation are several additional challenging aspects of 
low-energy model building.

Heterotic flux backgrounds have been the subject of consistent intensive research for a number of years, but concrete examples have been scarce. 
The BPS equations for $\mathcal{N}=1$ supersymmetry, which are known since the work of Hull and Strominger~\cite{Strominger:1986uh, Hull:1986kz}, indicate that the manifold is non-K\"ahler, being instead conformally balanced. Furthermore, the Bianchi identity is non-linear in the NS-NS three-form $\mathcal{H}$, hence particularly difficult to solve. Most of the works about the subject have considered a single type of such compactifications, given by  principal two-torus bundles over a warped K3 base, known as Fu-Yau compactifications. 
These solutions were first obtained from type IIB orientifolds by S-duality~\cite{Dasgupta:1999ss}, and subsequently studied by several authors, see $e.g.$~\cite{Goldstein:2002pg,Fu:2006vj,Becker:2009df}. Interestingly there exists a worldsheet description of such $\mathcal{N}=2$ compactifications as a gauged linear sigma model with torsion~\cite{Adams:2006kb}. Understanding more general solutions, and more specifically $\mathcal{N}=1$ solutions with $SU(3)$ structure of phenomenological interest, is for the most part an open problem 
(see however~\cite{Blaszczyk:2011ib,Quigley:2011pv,Quigley:2012gq,Adams:2012sh} for examples of gauged linear sigma-models relevant to this problem).

A useful tool for analyzing supergravity compactifications with flux is to consider non-compact internal manifolds which provide a local approximation to a compact model. This is especially relevant for heterotic supergravity since obtaining compact flux backgrounds is particularly difficult. The canonical example of such a non-compact manifold is the local conifold singularity in a Calabi-Yau threefold~\cite{Candelas:1989js}. The conifold has been studied in type II superstring theories in great detail with numerous interesting results; early results on non-perturbative effects~\cite{Strominger:1995cz} included a resolution of the singularity in the worldsheet conformal field theory of the conifold; the relationship between open and closed topological strings was actualized on the conifold~\cite{Gopakumar:1998ki}; the conifold provides very concrete models of holography~\cite{Klebanov:1998hh, Klebanov:2000hb}; finally, the conifold provides canonical examples of non-trivially wrapped D-brane configurations~\cite{Maldacena:2000yy, Atiyah:2000zz}. 

Our current work is very much in the same vein as~\cite{Klebanov:2000hb, Maldacena:2000yy} except that we will be studying heterotic strings, and that the main class of our solutions will require a $\ZZ_2$ orbifold of the conifold. It builds on earlier articles by one of the authors~\cite{Carlevaro:2008qf,Carlevaro:2009jx}, where solutions of this type, first based on Eguchi-Hanson space (hence providing local models of Fu-Yau compactifications, see also~\cite{Fu:2008ga}), second on the conifold. An important step towards obtaining these solutions was to define a {\it large charge} limit, in which the contribution of the tangent bundle curvature to the Bianchi identity can be consistently neglected. Furthermore, it was proven that the ``near-bolt" region of these solutions can be decoupled from the asymptotically locally Ricci-flat region, leading to smooth asymptotically linear dilaton solutions that admit exactly solvable worldsheet conformal field theory descriptions.  We will provide in this work important generalizations of these solutions, based on non-K\"ahler metrics on $T^{1,1}$ cones, that share most of these features.

These decoupling limits lead to asymptotically linear dilaton space-times, and are hence expected to have a holographic description, in terms 
of a four-dimensional $\mathcal{N}=1$ ``little string theory"~\cite{Aharony1998c}. The low-energy dynamics of these non-gravitational theories --~corresponding to the small $r$ regions of the dual geometries~-- depends on the theory at hand. 
In the blow-down limit of the conifold, the $E_8 \times E_8$ solution near the dilaton singularity 
is expected be lifted in $\mathcal{M}$-theory to a smooth $AdS_5$ eleven-dimensional solution, dual to an $\mathcal{N}=1$ superconformal 
theory in four dimensions.  In the $Spin(32)/\mathbb{Z}_2$ theory the physics is very different; the resolved $\mathbb{Z}_2$ orbifold 
of the conifold is expected to be dual to an $\mathcal{N}=1$ confining theory with matter; its field theory 
limit is appropriately described in the type I dual frame, where the confining string was identified 
in~\cite{Carlevaro:2009jx} as the fundamental string; in a way these solutions are heterotic versions  of the 
Chamseddine-Volkov-Maldacena-Nu\~nez solution~\cite{Chamseddine:1997nm, Chamseddine:1997mc, Maldacena:2000yy}. The more general 
solutions presented in the present article share the same qualitative features and enlarge significantly the parameter space of 
such holographically dual pairs.

Supersymmetry breaking is a key element in string model building. A very concrete mechanism was put forward in~\cite{Kachru:2002gs}, known as the KPV mechanism. Key to this mechanism is that in the Klebanov-Strassler background~\cite{Klebanov:2000hb} one may have D3-brane charge dissolved in flux or coming from explicit sources. Due to the heterotic Bianchi identity
\be
\di \cH_{(3)} = \al' (\Tr F\w F - R_+ \w R_+) +\delta (sources)
\ee
we can explore the possibilty that backgrounds such as ours may have the same conserved charges but where this charge is formed from different brane and flux configurations. There may then be tunneling between such vacua in the form of brane-flux annihilation~\cite{Kachru:2002gs}. 
We hope to return to this in the near future.

Our paper is organized as follows: In section two we setup the BPS equations for our ansatz. In section three we present our full set of solutions on the orbifolded conifold. In section four we present solutions on the unorbifolded conifold. In section five we present the worldsheet model for the near horizon solutions. In section six we discuss the various charges which identify our backgrounds.

\vspace{5mm}
\noindent {\bf Relation to Previous Works}: There are very few articles pertaining to local heterotic supergravity solutions for local throat geometries in the literature. Following \cite{Carlevaro:2009jx} which we have already mentioned, a further supergravity analysis on the conifold appeared in \cite{Chen:2013nma}. The solutions in that work have a singular dilaton in the IR and in our notation have $(r^2 H_2)'=0$, in addition they are performing a sort of expansion in small $a=r^2 H_2$. It is conceivable (despite some effort, we have not suceeded in checking this) that their work may be related to the blow-down limit of the near horizon solutions we present in section \ref{sec:nearhorizonSingular} but no others. We are confident that our analysis maps out the full parameter space of solutions within the large charge ansatz of section \ref{sec:Ansatz}. 

There are also some related works by Teng Fei \cite{Fei:2015kua, Fei:2015yaa} (see also \cite{Fu:2008ga} for local models on Eguchi-Hanson space). Any intersection with our work must be through a limit of these solutions which are a first order deformation around a Ricci-flat 
non-compact Calabi-Yau background. In general our solutions have non-vanishing flux at the zero-th order but in the limit where the size of the IR cycle is taken large, this flux becomes dilute and our solution is approximately Calabi-Yau.

%%%%%%%%%%%%%%%%%%%%%%%%%%%%%%%%%%%%
\section{Heterotic Supergravity on the Conifold}
%%%%%%%%%%%%%%%%%%%%%%%%%%%%%%%%%%%%
We begin by recalling ten-dimensional heterotic supergravity. The bosonic action reads~\cite{Bergshoeff:1989de}:
\begin{equation}
S=\int\mathrm{d}x^{10}\sqrt{-g}e^{-2\Phi}\left[\mathcal{R}+4\vert\mathrm{d}\Phi\vert^2-\frac{1}{2}\vert\cH_{(3)}\vert^2+\alpha'\left(\Tr\,\vert F\vert^2-\Tr\,\mathcal\vert{R}\vert^2\right)\right]\:,
\end{equation}
where $\mathcal{R}$ is the Ricci scalar, $\Phi$ is the dilaton, $\cH_{(3)}$ is the three-form flux, though gauge invariant by the Green-Schwarz mechanism \cite{Green:1984sg}, it satisfies a non-trivial Bianchi Identity to be defined below. $F$ is the curvature two-form of an $E_8\times E_8$ or $SO(32)$ gauge bundle. Furthermore $R$ is the curvature two-form of a tangent bundle connection, required by quantum consistency of the theory. Which connection this is will be detailed below. Our notation for the norm of a $p$-form $\alpha$ is
\begin{equation}
\vert\alpha\vert^2=\frac{1}{p!}\alpha_{m_1..m_p}\alpha^{m_1..m_p}\:.
\end{equation}
For completeness, we also recall the ten-dimensional supersymmetry variations
\begin{subequations}
\begin{align}
\delta\psi_m&=\nabla^-_m\epsilon=\nabla_m\epsilon-\frac{1}{8}\cH_{mab}\gamma^{ab}\epsilon\\
\delta\lambda&= (\gamma^a\nabla_a\phi-\frac{1}{12}\cH_{abc}\gamma^{abc})\epsilon\\
\delta\chi&=-\frac{1}{2}F_{ab}\gamma^{ab}\epsilon\:,
\end{align}
\end{subequations}
where $\psi_m$ is the gravitino, $\lambda$ is the dilatino and $\chi$ is the gaugino.
 
%%%%%%%%%%%%%%%%%%%%%%%%%%%%%%%%%%%%
\subsection{The BPS Equations}
%%%%%%%%%%%%%%%%%%%%%%%%%%%%%%%%%%%%

The BPS equations for heterotic supergravity dimensionally reduced on a general torsional background to four dimensions, while preserving $\cN=1$ supersymmetry, are given below in the form presented 
in~\cite{Gauntlett:2003cy}. In heterotic supergravity one needs to choose a connection on the tangent bundle and in the subsequent expressions 
we assume that this is the Hull connection~\cite{Hull:1986kz}:
\be
\label{eq:omegaplus}
(\Omega^+)^a_{\ b} = \omega^a_{\ b} +\frac{1}{2} \mathcal{H}^a_{\ b}\,,
\ee
where $\omega^a_{\ b}$ is the torsionless spin-connection one-form. 

Using the globally defined spinor on the internal six-manifold $\mathcal{M}_6$, one can construct a real two-form $J$ and a complex three-form  $\Om$ which satisfy 
the $SU(3)$ structure conditions:
\be 
\label{su3struct}
-\frac{i}{8}\Om\w \Ombar = \frac{1}{3!} J\w J \w J\,,\qquad J \w \Om=0\,.
\ee
Then the BPS equations which determine the geometry take the form of calibration conditions:
\begin{subequations}
\begin{align}
0&=\di (e^{-2\Phi} J\w J)\, ,  \label{dJeq} \\
0&= \di (e^{-2\Phi} \Om )\, , \label{dOmeq}  \\
\cH_{(3)}&= *_6 \, e^{2\Phi}\di \blp e^{-2\Phi} J \brp  \, .   \label{Hdefeq} 
\end{align}
\end{subequations}
Equation \eqref{dJeq} is usually referred as the conformally balanced condition. The three-form $\Omega$ defines an almost complex structure, given by
\begin{equation}
\label{eq:complexstr}
{J_m}^n=\frac{{I_m}^n}{\sqrt{-\frac{1}{6}\tr I^2}}\,,
\end{equation}
where the tangent bundle endomorphism $I$ is given by
\begin{equation}
{I_m}^n=(\textrm{Re}\Omega)_{mpq}(\textrm{Re}\Omega)_{rst}\,\epsilon^{npqrst}\,.\label{eq:CSnorm}
\end{equation}
The three-form $\Omega$ is then of type $(3,0)$ with respect to this almost complex structure. The conformally holomorphic condition~\eqref{dOmeq} 
is enough to ensure that ${J_m}^n$ is integrable, so that $\mathcal{M}_6$ is a complex manifold. The last equation~\eqref{Hdefeq} is often referred to as a calibration condition.

Next, the vector bundle over $\mathcal{M}_6$ should satisfy the instanton equations, namely
\begin{subequations}
\begin{align}
0&= F\w J \w J \, , \label{FJJeq} \\
0&= F \w \Om \, . \label{FOmEq}
\end{align}
\end{subequations}
The last of these equations imply that $F$ is of type $(1,1)$, which in turn implies that the vector bundle is holomorphic. 
Equation \eqref{FJJeq} is referred to as the Yang-Mills condition. These equations are often referred to as the 
zero-slope Hermitian-Yang-Mills equations. Indeed, Li and Yau proved in the non-K\"ahler case that a holomorphic vector 
bundle on a compact complex manifold admits a unique Hermitian-Yang-Mills connection if and only if it is zero-slope~\cite{Li:1986id}. 
These terms are however a bit misleading in our case, since we are not dealing with a compact space, nor need the 
connection on the bundle be hermitian in the sense that it is the Chern connection of a hermitian metric on the 
bundle. It should also be noted that the equations~\eqref{FJJeq}-\eqref{FOmEq} can be rephrased in terms of a single anti-self-duality equation
\be
\label{selfdual}
*F = - J \w F\:,
\ee
which is sometimes a more convenient form. One can show that \eqref{selfdual} requires $F$ to be type $(1,1)$ 
and primitive\footnote{A primitive $n$-form $\al$ on a symplectic manifold with symplectic form $J$ 
satisfies $J\llcorner \al=0 $.}~\cite{weil1958introduction}. The Bianchi identity ties 
this data together and is of course modified by the Green-Schwarz mechanism:\footnote{
We take the gauge connection to be hermitian. For abelian bundles this means that $F$ is real.}
\be
\di \cH_{(3)} = \al'\Bslb\Tr \, F \w F - \Tr\, R_+\w R_+ \Bsrb \, . \label{Bianchi}
\ee

As usual it is of some interest to confirm that the BPS equations and Bianchi identity imply the equations of motion. This is known to be true at $\cO(\al')$ 
when the Hull connection is used. However, should one choose to utilize a connection different from the Hull connection\footnote{See~\cite{Fu:2008ga} 
for a utilization of the Chern connection.} one must take into consideration additional corrections to the BPS equations and also the equations 
of motion. Bergshoeff and de Roo have computed  in~\cite{Bergshoeff:1989de} the first non vanishing corrections the heterotic action and supersymmetry variations; 
they found that the first correction to the action is at $\cO(\al'^3)$ and the first correction to the supersymmetry variations is $\cO(\al'^2)$. 
However, they worked explicitly with the Hull connection. Other choices of connections are possible, corresponding to different 
regularization schemes in the space-time effective action \cite{Hull:1985dx}, or different field choices from the 
supergravity point of view~\cite{Becker:2009df, Hull:1986xn, Sen:1986mg}. Replacing the Hull connection 
with a general connection $\nabla$ requires a correction to the gravitino and dilatino variations at $\cO(\al')$. These corrections are given by
\begin{subequations}
\begin{align}
\delta \Psi_m &= \delta_0 \Psi_m + \al' \delta_1 \Psi_m\,,\qquad 
\delta \lam = \delta_0 \lam + \al' \delta_1 \lam \\
\delta_1 \Psi_m&\sim  e^{2\Phi} \nabla^{-l} \Blp e^{-2\Phi} (R-R^+)_{ablm}\gam^{ab} \Brp \eps \\
\delta_1 \lam&\sim  e^{2\Phi} \nabla^{-L} \Blp e^{-2\Phi} (R-R^+)_{ablm}\gam^{abm} \Brp \eps \, ,
\end{align}
\end{subequations}
which agree with the corrections at leading order away from the Hull connection~\cite{delaOssa:2014msa}. Here $R$ is the curvature 
two-form of $\nabla$. The only subsequent change to the action is that the Hull connection is replaced by $\nabla$. 
This correction to the supersymmetry variations subsequently give a correction at $\cO(\al')$ to the Hull-Strominger 
system~\eq{dJeq}-\eq{Hdefeq} (which would of course vanish for the Hull connection). Combining this with the results 
of \cite{Ivanov:2009rh} one finds that the corrected BPS equations and 
Bianchi identity imply the equations of motion at $\cO(\al')$. If one was to use the 
$\cO(\al')$ BPS equations such as in \cite{Ivanov:2009rh}, one would find the need to enforce the condition
\be
*R^a_{\ b} = - J \w R^{a}_{\ b} \label{RInstanton}
\ee
to order $\cO(\al'^0)$, where the curvature two-form is evaluated for $\nabla$. One may refer to \eq{RInstanton} as the {\it instanton} 
equation since it can be shown to be equivalent to Hermitian Yang-Mills for the curvature two-form (with connection $\nabla$). 
In summary, for any connection which satisfies \eq{RInstanton} on-shell (such as the Hull connection) 
the $\cO(\al')$ corrections to the Hull-Strominger system necessarily vanish. It is perhaps of interest to note 
that the Chern connection does not necessarily satisfy \eq{RInstanton} and thus may not in 
general provide a solution $\cO(\al')$ corrected Hull-Strominger system.

When searching for exact solutions of the system, the Bianchi identity~\eq{Bianchi} provides complicated non-linearities.\footnote{In any case, one 
should not separate the formal $\alpha'$ expansion of the BPS equations and Bianchi identity from the $\alpha'$ expansion of the solution 
itself, see~\cite{Melnikov:2014ywa} for a recent discussion. In the present context the inverse of the magnetic charge will be the correct dimensionless parameter 
organizing this expansion.} Our strategy is to use a large charge limit such that the $\Tr\, R\w R$ term is suppressed by the $\Tr\, F\w F$ term. A further 
consequence is that both sides of \eq{RInstanton} are subleading in our large charge expansion, therefore Einstein's 
equation is satisfied to the order at which we work. This is in the same spirit as the {\it gauge solution} 
of Callan, Harvey and Strominger~\cite{Callan:1991at}. One may note that as a result~\eq{RInstanton} is subleading to our analysis.

%%%%%%%%%%%%%%%%%%%%%%%%%%%%%%%%%%%%
\subsection{The Ansatz}\label{sec:Ansatz}
%%%%%%%%%%%%%%%%%%%%%%%%%%%%%%%%%%%%
The metric ansatz for our non-compact heterotic solution with $SU(2)\times SU(2)\times U(1)$ isometry and four-dimensional 
$\mathcal{N}=1$ supersymmetry is
\begin{equation}
\di s_{10}^2 = \di s_{1,3}^2 + \frac{3H}{2}  \frac{\di r^2}{f^2}  + 
r^2\left[ \frac{H_1+H_2}{4}  (\sig_1^2+\sig_2^2)+\frac{H_1-H_2}{4} (\hsig_1^2+\hsig_2^2)+\frac{f^2H}{6} \eta^2  \right]\, .
\label{metans}
\end{equation}
Our conventions about the $SU(2)$ left-invariant one-forms on the conifold are given in appendix~\ref{app:conifold}. The NS-NS three-form ansatz is 
\be 
\cH_3=  \al' \Bslb h_1\, \eta \w (\Om_1-\Om_2)+h_2\,  \eta \w (\Om_1+\Om_2) \Bsrb \label{H3eval}\, ,\\
\ee
and finally we choose an Abelian gauge bundle which is locally of the form
\begin{equation}
F=  -\frac{1}{4} \Bslb \blp \Om_1 - \Om_2 - \di \bslb g_1\, \eta \bsrb \brp\,\bp  - \di \bslb g_2\, \eta\bsrb\, \bq\, \Bsrb \cdot \bH\,. 
\label{Fdef}
\end{equation}
In these expressions $(g_1,g_2,H,H_1,H_2,h_1,h_2)$ and the dilaton $\Phi$ are dimensionless functions of the radial coordinate $r$; we will later fix reparameterization of $r$ 
by eliminating $H$ in terms of $\{H_1,H_2\}$. 

The embedding of the Abelian gauge bundle into the Cartan subalgebra of $\frak{so}(32)$ or $\mathfrak{e}_8\times \mathfrak{e}_8$ is 
given in~(\ref{Fdef}) by $\bp$ and $\bq$, both vectors in $\RR^{16}$ whose quantization will be discussed in detail below. 
Furthermore and somewhat crucially, we assume that 
\be
\bp\cdot \bq=0.
\ee 

The complex frames on the internal space that we will use to get the $SU(3)$ structure, see eq.~(\ref{su3struct}), are taken to be
\begin{align}
E_1&= \sqrt{\frac{3 H}{2}}  \left( \frac{\di r}{f} + i \frac{f}{3} \eta\right)\,,\non \\
E_2 &= -i\, r \sqrt{\frac{H_1 + H_2}{4}}  \Blp \sig_1 + i \sig_2  \Brp\,,\\
E_3 &= -i\, r \sqrt{\frac{H_1 - H_2}{4}}  \Blp \hsig_1 + i \hsig_2  \Brp \non
\end{align}
which we use to define the fundamental forms
\be
\Om=E_1\w E_2 \w E_3\,,\qquad J=\frac{1}{2i} \sum_{i=1}^3(E_i\w \Ebar_i) \,.
\ee

%%%%%%%%%%%%%%%%%%%%%%%%%%%%%%%%%%%%
\subsection{Reducing the BPS Equations on the ansatz}
%%%%%%%%%%%%%%%%%%%%%%%%%%%%%%%%%%%%

We proceed by massaging these equations into a manageable form before choosing the radial gauge $H$. Expanding~\eq{dJeq} 
on our ansatz we get
\be
\log \Bslb \frac{e^{2\Phi}}{H_1^2-H_2^2}\Bsrb' =\frac{4}{r}\Bslb 1-\frac{H H_1}{ H_1^2-H_2^2}  \Bsrb\,. \label{dJEq2}
\ee
Then from~\eq{dOmeq} we get just one equation
\be
0=2+\frac{6 }{f^2} - \frac{8H H_1}{H_1^2-H_2^2} -r \log\bslb f^2 H/(H_1^2-H_2^2)\bsrb' \label{dOmEq2}
\ee
and from ~\eq{FJJeq} we obtain an analytic solution for $g_2$ 
\be
g_2= \frac{a^4}{r^4} \frac{e^{2(\Phi-\Phi_0)}}{H_1^2-H_2^2}
\ee
where $a$ is a real constant, as well as an equation for $g_1$
\be
rg_1' = \frac{4H(H_2-g_1H_1)}{H_1^2-H_2^2} \label{g1Eq1} \,.
\ee
At this point we have three equations~(\ref{dJEq2},\ref{dOmEq2},\ref{g1Eq1}) for six functions $\{\Phi, f,g_1,H,H_1,H_2\}$. The remaining two equations will come from the Bianchi identity and of course we must still choose the form of $H$.

After some computation, we find that equation~(\ref{Hdefeq}) for the three-form flux $\mathcal{H}$ defined in~(\ref{H3eval})  
reduces to:
\begin{equation}
\frac{1}{\al'}\, \di \cH_3 =  h'_1 \di r \w \eta \w (\Om_1-\Om_2) + h'_2 \di r\w \eta \w (\Om_1+\Om_2) -2 h_2 \Om_1\w\Om_2 \, ,
\end{equation}
with $h_{1,2}$ given by:
\be
\al' h_1 = -\frac{r f^2}{12} \blp r^2 H_2 \brp'\,, \qquad \al' h_2= -\frac{r f^2}{12} \blp -2rH+( r^2 H_1 )' \brp\,.
\ee
From the ansatz~\eq{Fdef} for the field strength of the gauge field we find that
\begin{equation}
\Tr\,  F\w F =f_1 \Om_1\w \Om_2 -\frac{1}{2}f'_1\, \di r\w \eta\w (\Om_1+ \Om_2)+f_2 \di r\w \eta\w (\Om_1- \Om_2) \Bsrb \,, \\
\end{equation}
with
\be
f_1 =   \frac{1}{4}\bslb \bp^2(-1+ g_1^2)+\bq^2 g_ 2^2\bsrb\,,\qquad
f_2=  -\frac{\bp^2}{4}  g'_1\, ,
\ee
and where we have used
\be
\bp\cdot \bq=0\,,\quad \Tr \, H^i H^j = 2\delta^{ij}\,, \quad \Tr (\bp\cdot \bH )^2 = 2 p^2\,,\quad \Tr (\bq\cdot \bH )^2 = 2 q^2\,.
\ee

A crucial point at this stage is that, as in~\cite{Carlevaro:2009jx}, we will consistently neglect the $\Tr\, R\wedge R$ contribution from 
the tangent bundle to the Bianchi identity~(\ref{Bianchi}). This may be viewed as an oversimplification but, as we will show, in a large 
charge limit $q=||\bf{q}||\gg 1$ this contribution is subleading in the $1/q$ expansion of the solution.

We get then in this regime two equations from the Bianchi identity~\eq{Bianchi}, relating the gauge field strength and NS-NS three-form 
ans\"atze:
\begin{subequations}
\begin{align}
f_1&=-2h_2 \label{Bianchi1} \\
f_2&= h'_1\label{Bianchi2}\,.
\end{align}
\end{subequations}
We can integrate immediately eq.~\eq{Bianchi2} to get
\be
g_1 =-\frac{4}{p^2} h_1+g_c\,. \label{g1Sol}
\ee 
where $g_c$ is an integration constant that will play an important role.

The dictionary between these supergravity parameters and those that will appear in the worldsheet 
theory for the near-bolt solution, in section~\ref{sec:worldsheet}, is given by:
\begin{equation}
\cosh \rho = (r/a)^4\,,\qquad k_1= (1+g_c) \,p^2 \,,\qquad k_2=(1-g_c) \, p^2\, . \label{k1k2gc}
\end{equation}

%%%%%%%%%%%%%%%%%%%%%%%%%%%%%%%%%%%%
\subsection{Summary of Equations}\label{sec:SummaryEqs}
%%%%%%%%%%%%%%%%%%%%%%%%%%%%%%%%%%%%

After this analysis we  summarize the BPS equations here. There are four unsolved equations for five functions $\{\Phi,f,H,H_1,H_2\}$:
\begin{subequations}
\begin{align}
\log \Bslb \frac{e^{2\Phi}}{H_1^2-H_2^2}\Bsrb' &=\frac{4}{r}\Bslb 1-\frac{H H_1}{ H_1^2-H_2^2}  \Bsrb\label{BPSEq1} \\
 \log\Bslb\frac{ f^2 H}{r^2(H_1^2-H_2^2)}\bsrb'&=\frac{6 }{rf^2} - \frac{8H H_1}{r(H_1^2-H_2^2)} \label{BPSEq2} \\
rg_1'&= \frac{4H(H_2-g_1H_1)}{H_1^2-H_2^2} \label{BPSEq3}\\
-2r f^2 \blp -2rH+( r^2 H_1 )' \brp&=   3\bslb p^2(-1+ g_1^2)+ q^2 g_ 2^2\bsrb \label{BPSEq4}
\end{align}
\end{subequations}
and two functions in the ansatz (\ref{metans}, \ref{Fdef}) have been solved for analytically giving two 
integration constants $\{a,g_c\}$:
\begin{subequations}
\begin{align}
%%%
g_1 &=\frac{r f^2}{3p^2} \blp r^2 H_2 \brp'+g_c \\
g_2&= \frac{a^4}{r^4} \frac{e^{2(\Phi-\Phi_0)}}{H_1^2-H_2^2} \,.
\end{align}
\end{subequations}
The functions appearing in the three form ansatz~(\ref{H3eval})  are given by
\begin{subequations}
\begin{align}
\al' h_1&= -\frac{r f^2}{12} \blp r^2 H_2 \brp'\,, \label{h1eqn1}\\
\al' h_2&= -\frac{r f^2}{12} \Big( -2rH+( r^2 H_1 )' \Big)\,.  
\end{align}
\end{subequations}

%%%%%%%%%%%%%%%%%%%%%%%%%%%%%%%%%%%%
\subsubsection{Rescaling the radial coordinate and Fixing the Radial Gauge}\label{sec:RescaledEquations}
%%%%%%%%%%%%%%%%%%%%%%%%%%%%%%%%%%%%

To analyze the solution space when $a\neq 0$, it will turn out to be convenient to use the dimensionless radial coordinate
\be
\rho=\frac{r}{a}\,.
\ee
In addition we will primarily use the gauge
\be
H=\frac{H_1^2-H_2^2}{H_1} \label{HRep}
\ee
Rescaling the remaining wrap factors as
\be
\tH_i=a^2 H_i\, ,\qquad i=1,2\, ,
\ee
we can then solve~\eq{BPSEq1} for the dilaton, 
and the system reduces to the coupled set of three nonlinear equations\footnote{here we use $'=\frac{\del}{\del\rho}$}:
\begin{subequations}
\begin{align}
 \log \left[\frac{ f^2 }{r^2 \tH_1}\right]'&=\frac{6 }{\rho f^2} - \frac{8}{\rho}  \label{BPSeqfix1}\, ,\\
(\rho^4g_1)' &= \frac{4\rho^3 \tH_2}{\tH_1} \label{BPSeqfix2}\, ,\\
-2\rho f^2 \left(  \frac{2\rho \tH_2^2}{\tH_1}+ \rho^2  \tH_1' \right) &=  3  \bslb p^2(-1+ g_1^2)+ q^2 g_ 2^2\bsrb \, , \label{BPSeqfix3}
\end{align}
\end{subequations}
in term of which the other functions in the solution ansatz~(\ref{metans},\ref{H3eval},\ref{Fdef}) are given by:
%%%
\begin{subequations}
\begin{align}
e^{2(\Phi-\Phi_0)} =& a^{-2}(\tH_1^2-\tH_2^2)\\
g_1 =&\frac{\rho f^2}{3p^2} \blp \rho^2 \tH_2 \brp'+g_c \,, \label{g1Eqn1}\\
g_2=& \frac{1}{\rho^4} \,,\\
\al' h_1=& -\frac{\rho f^2}{12} \blp \rho^2 \tH_2 \brp'\,, \\
\al' h_2=& -\frac{\rho f^2}{12} \Blp \frac{2\rho \tH_2^2}{\tH_1}+ \rho^2 \tH'_1 \Brp\, .
\end{align}
\end{subequations}
We will now consider in the following sections two classes of solutions to these equations with different topologies.  

%%%%%
%%%%%%%%%%%%%%%%%%%%%%%%%%%%%%%%%%%%
\section{Solutions from $T^{1,1}/\ZZ_2$}\label{sec:T11Z2}
%%%%%%%%%%%%%%%%%%%%%%%%%%%%%%%%%%%%

In this section we present solutions based on cones over $T^{1,1}/\ZZ_2$ endowed with non-K\"ahler metrics. The 
most straightforward solution is the flux-free solution with a Ricci-flat metric given in section~\ref{sec:RicciFlat}.  
Then in section~\ref{sec:nearhorizon} we present an analytic solution which we interpret as the 'near-bolt' region of the more general asymptotically locally Ricci-flat solution we find 
numerically in section~\ref{sec:FullBraneSolution} .

%%%%%%%%%%%%%%%%%%%%%%%%%%%%%%%%%%%%
\subsection{Ricci-flat metric}
\label{sec:RicciFlat}
%%%%%%%%%%%%%%%%%%%%%%%%%%%%%%%%%%%%

The simplest solution to the equations in section~\ref{sec:SummaryEqs} is that for the Ricci flat metric compatible with the K\"ahler structure on the anti-canonical bundle over $F_0=\PP^1\times \PP^1$.  The horizontal space at constant $r$ is $T^{1,1}/\ZZ_2$, where the orbifold is needed to avoid a conical singularity at the bolt, which has a finite four-cycle $\PP^1\times \PP^1$. This solution has 
of course $\cH_{(3)}=0$ and was found in some time ago in~\cite{Page1987}. In this case one cannot neglect the $\Tr\, R\wedge R$ 
term in the Bianchi identity~(\ref{Bianchi}) but one can choose the standard embedding of the spin connection into the 
gauge connection (then, eq.~\eq{BPSEq3} is identically zero). Working in the radial gauge~\eq{HRep} the solution is given by:
\begin{align}
H_1 &= a_1\sqrt{2 + \frac{1}{r^4}}\,, \non \\
H_2 &= \frac{a_1 }{ r^2} \,, \\
f^2&=\frac{1}{2r^8} \bslb (r^4-1)(2r^4+1)-a_2\sqrt{2r^4+1}\bsrb\,.\non
\end{align}
This solution has two constants $(a_1,a_2)$ which correspond to the blow up parameters of the pair of two-spheres. The bolt sits at
\be
r=a_3\,,\qquad a_2= (a_3-1)\sqrt{1+2a_3}\,.
\ee
and one easily checks that the periodicity of $\psi$ must be $\psi \sim \psi+ 2\pi$ for regularity, confirming that this is a 
resolution of a $\ZZ_2$ orbifold of the conifold. When $a_3=0$ and $a_2=-1$ one of the $S^2$'s degenerates to zero size at $r=0$, thus an entire $S^3/\ZZ_2$ shrinks. With these values, the unorbifolded geometry with $0\leqslant \psi < 4\pi$ is also regular and coincides with the resolved conifold.

A more conventional choice of radial co-ordinate for this Ricci-flat geometry is $H=1$ 
which leads to the following perhaps more familiar form of the Ricci-flat metric
\begin{align}
H_1+H_2&=1+ \frac{\al_1}{r^2}\,,\non \\
H_1-H_2 &= 1+\frac{\al_2}{r^2}\,, \\
f^2&= \frac{2r^4+3(\al_1+\al_2)r^2 + 6\al_1\al_2 }{2(r^2+\al_1)(r^2+\al_2)} \,,\non
\end{align}
with the bolt at $r=0$. The resolved conifold is recovered by setting $\al_1=0$ or $\al_2=0$.

%%%%%%%%%%%%%%%%%%%%%%%%%%%%%%%%%%%%
\subsection{Near-Bolt Solution}\label{sec:nearhorizon}
%%%%%%%%%%%%%%%%%%%%%%%%%%%%%%%%%%%%

We now present closed form `near-horizon' solutions for both finite and vanishing blow-up parameter. These solutions are not asymptotically 
locally Ricci-flat but our numerical analysis shows how they can be glued to a `far-brane' region which is locally Ricci-flat.
 
As in~\cite{Carlevaro:2009jx} the former solution can be viewed either as an approximation of the full solution in the regime 
$a^2 \ll \alpha' p^2$, $i.e.$ in a regime where the length scale associated with the backreaction of the three-form flux is much 
larger than the blow-up parameter $a$, or as obtained from the full solution through a double-scaling limit. Likewise, the singular 
solution, being strongly coupled in the IR, can be decoupled from the asymptotically locally Ricci-flat region by sending the asymptotic 
string coupling to zero. 

Both solutions admit an exact worldsheet conformal field theory description that will be discussed shortly in section~\ref{sec:worldsheet}.

%%%%%%%%%%%%%%%%%%%%%%%%%%%%%%%%%%%%
\subsubsection{The Regular Solution}\label{sec:nearhorizonRegular}
%%%%%%%%%%%%%%%%%%%%%%%%%%%%%%%%%%%%

Using the radial coordinate given by~\eq{HRep} we have found an analytic solution to the equations of section~\ref{sec:SummaryEqs} 
given by:
\begin{align}
& \left. {  \begin{array}{l} 
 \displaystyle H_1=\frac{p^2}{a^2 \rho^2}\\ 
\displaystyle H_2 =\frac{g_c p^2}{a^2 \rho^2}
\end{array} }
\right\} \ \implies \ 
H= \frac{p^2(1-g_c^2)}{a^2 \rho^2}\,, \non \\
&f^2=\frac{3}{4} \left( 1- \frac{ q^2 }{(1-g_c^2) p^{2}} \frac{1}{\rho^8}\right)\, ,\quad  g_1=g_c\,,\quad g_2=\frac{1}{\rho^4}\,, \\
& e^{2(\Phi-\Phi_0)}=\frac{p^4(1-g_c^2)}{a^4 \rho^4}\,.\non 
\end{align}
When the blow up parameter is non-vanishing ($a\neq0$) the full ten-dimensional solution is then given by:
\begin{subequations}
\begin{align}
\di s_{10}^2&=\di s_{1,3}^2 +\frac{2 \al' {\bf p}^2(1-g^2_c)}{R^2}  \left\{   \frac{\di R^2}{1-\frac{1}{R^8}} %\non \\
+ \frac{R^2}{8}\left( \frac{\sig_1^2+ \sig_2^2}{1-g_c} + \frac{\hsig_1^2 + \hsig_2^2}{1+g_c} 
+\frac{1}{2} \left(1-\frac{1}{R^8}\right) \eta^2 \right) \right\} \,,  \\
F&= -\frac{1}{4} \Bslb \blp (1+g_c) \Om_1 - (1-g_c) \Om_2  \brp\,\bp  - \frac{p}{q}\sqrt{1-g_c^2}\, 
\di \left( \frac{\eta}{R^4}\right)\, \bq\, \Bsrb \cdot \bH\,,   \\
\cH_3&=\frac{\al' p^2}{8} \left(1-\frac{1}{R^8}\right) (1-g_c^2) \, \eta \w (\Om_1+\Om_2)\,, \\
%%%
e^{\Phi}&= \frac{e^{\tPhi_0}}{R^2}\, ,  \label{eq:dilnh} 
\end{align}
\label{eq:nhsolution}
\end{subequations}
where we have absorbed the blow-up parameter by rescaling the radial co-ordinate and dilaton zero mode:
\be
\label{eq:rescaldil}
R^8=\frac{p^{2}(1-g_c^2)}{q^2}\, \rho^8\,,\qquad e^{2\tPhi_0}= e^{2\Phi_0}\frac{q^2a^8}{p^{2}(1-g_c^2)}\, ,
\ee 
thus demonstrating that the parameters $a$ and $q$ have been completely absorbed.\footnote{We have obtained 
that solution is invariant under rescalings of $\bf q$, however the embedding of the Abelian gauge 
group into the Cartan subalgebra that this vector specifies is of course still meaningful.} 
There remain just three parameters $\{p,\tPhi_0,g_c\}$. The range of the radial co-ordinate is $R\geqslant 1$; at $R=1$ there is a 
bolt and the internal manifold approximates
\be
\mathcal{M}_6\sim \RR^2 \times S^2 \times S^2\,.
\ee
Note that the dilaton is finite at the bolt. By analyzing the periodicity of $\psi$ at the bolt we find $0\leqslant \psi < 2\pi$ and conclude that the horizontal space away from the bolt is $T^{1,1}/\ZZ_2$, hence the total space is diffeomorphic to $\cO(-2)\ra F_{0}$.

As was discussed in~\cite{Carlevaro:2009jx}, this near-bolt solution can be decoupled from the asymptotically locally Ricci-flat solution, that we will present in~\ref{sec:FullBraneSolution}, using a \emph{double scaling limit}, defined as
\begin{equation}
g_s \to 0 \ , \quad \mu:=\frac{g_s \alpha'}{a^2} \ \text{fixed}\, ,
\end{equation}
where $g_s$ is the asymptotic string coupling.\footnote{In equation~(\ref{eq:dilnh}) the double scaling parameter 
$\mu$ is essentially the same as $e^{\tPhi_0}$, up to order one factors, giving the effective string coupling in this asymptotically 
linear dilaton background.}

%%%%%%%%%%%%%%%%%%%%%%%%%%%%%%%%%%%%
\subsubsection{The Singular Solution}\label{sec:nearhorizonSingular}
%%%%%%%%%%%%%%%%%%%%%%%%%%%%%%%%%%%%
When the blow up parameter vanishes ($i.e.$ when $a=0$) we obtain the {\it singular} solution, from which $\bf q$ disappears:
\begin{subequations}
\label{eq:singsol}
\begin{align}
\di s_{10}^2&=\di s_{1,3}^2 + \frac{2\al'p^2 (1-g^2_c)}{r^2}\left\{ \di r^2 + \frac{r^2}{8} \left[ \frac{1}{1-g_c} 
(\sig_1^2+ \sig_2^2)+ \frac{1}{1+g_c}(\hsig_1^2 + \hsig_2^2) +\frac{1}{2}\eta^2\right] \right\}  \\
F&= -\frac{1}{4}  \blp (1+g_c) \Om_1 - (1-g_c) \Om_2  \brp\,\bp  \cdot \bH  \\
\cH_3&=\frac{\al' p^2}{8} (1-g_c^2) \, \eta \w (\Om_1+\Om_2) \\
%%%
e^{\Phi-\tPhi_0}&= \frac{\al' p^2\sqrt{1-g_c^2}}{ r^2} \,.
\end{align}
\end{subequations}
and the metric is
\be
\mathcal{M}_6 \sim \RR \times T^{1,1}/\ZZ_n
\ee
where the metric on $T^{1,1}$ is non-Einstein.

While the metric is regular, this supergravity solution has a divergent dilaton for $r\to 0$.  We also note that since there is no bolt, the periodicity of $\psi$ is not fixed by 
regularity at the origin and we may also consider the horizontal space to be $T^{1,1}$ 
without an orbifold, we will do so in section \ref{sec:T11} and discover another branch of solutions.

%%%%%%%%%%%%%%%%%%%%%%%%%%%%%%%%%%%%
\subsection{Asymptotically Locally Ricci-flat Solution}\label{sec:FullBraneSolution}
%%%%%%%%%%%%%%%%%%%%%%%%%%%%%%%%%%%%

We now construct numerically the full solution including the region far from the brane which 
is asymptotically locally Ricci-flat (ALRF). For a fixed $g_c$, we find a parameter space of regular solutions of real dimension two.

%%%%%%%%%%%%%%%%%%%%%%%%%%%%%%%%%%%%
\subsubsection{Near-bolt Region}
%%%%%%%%%%%%%%%%%%%%%%%%%%%%%%%%%%%%

The boundary conditions near the bolt at $r=\al$ corresponding to the solutions of interest are given by:
\begin{subequations}
\begin{align}
f&\simeq f_0(\rho-\al)^{1/2} + \cO\blp(\rho-\al)^{3/2}\brp \, , \\
\tH_1&\simeq h_{1,0}+ \cO(\rho-\al) \, , \\
\tH_2&\simeq h_{2,0}+ \cO(\rho-\al)\, ,
\end{align}
\end{subequations}
so that the expansion of the metric on the internal manifold $\mathcal{M}_6$ near the bolt reads:
\begin{multline}
\di s_{M_6}^2 = \frac{ \alpha}{4} \, \Bigg\{
 (h_{1,0}-h_{1,0}^{-1}h_{2,0}^2) \left( \frac{\di \rho^2}{\rho-\al} 
%%%
+4 (\rho-\al) \eta^2 \right)  \\
%%%
+ \alpha \Big( (h_{1,0}+h_{2,0}) (\sig_1^2+ \sig_2^2)
%%%
+ (h_{1,0}-h_{2,0}) (\hsig_1^2+ \hsig_2^2) \Big) \Bigg\} + \ldots
\end{multline}
and we find that the BPS equations impose the position of the bolt to be given in terms of the integration constants as:
\begin{equation}
\al=\frac{1}{(1-g_c)^{1/8}} \frac{q^{1/4}}{p^{1/4}}\, ,
\end{equation}
as was found for the near horizon solution in subsection~\ref{sec:nearhorizon}. The IR expansion of the metric functions is 
then given in terms of two free parameters $\{ h_{1,0},h_{2,0} \} $:
\begin{subequations}
\begin{align}
f&= \frac{\sqrt{6}}{\al^{1/2}} (\rho-\al)^{1/2}  + \frac{\sqrt{3}\bslb 2p^2( h_{1,0}-g_c h_{2,0}) + \al^2 (7 h_{1,0}^2+2h_{2,0}^2)\bsrb }{2^{3/2} \al^{7/2} h_{1,0}^2} (\rho-\al)^{3/2} + \cO\blp (\rho-\al)^{5/2}\brp  \\
%%%
\tH_1&= h_{1,0} - \frac{2(\al^2  h_{2,0}^2+ p^2(h_{1,0} - g_c h_{2,0}))}{\al^3h_{1,0}} (\rho-\al)+\cO\blp (\rho-\al)^2\brp  \\
%%%%
\tH_2&= h_{2,0}- \frac{2(g_c p^2 h_{1,0}+(\al^2 h_{1,0}-p^2)h_{2,0})}{\al^3h_{1,0}} (\rho-\al) +\cO\blp (\rho-\al)^2\brp  
\end{align}
\end{subequations}
such that the function $g_1$ in the three-form $\mathcal{H}$ and the dilaton read:
\begin{subequations}
\begin{align}
g_1&=g_c + \frac{4(h_{2,0}-g_c h_{1,0})}{\al h_{1,0}} (\rho-\al) + \cO\blp (\rho-\al)^2\brp\,, \\
e^{2(\Phi-\Phi_0)} &= h_{1,0}^2-h_{2,0}^2 -\frac{4\blp  p^2 (h_{1,0}^2-2g_c h_{1,0} h_{2,0}+h_{2,0}^2) \brp }{\al^3 h_{1,0} }(\rho-\al) + \cO\blp  (\rho -\al)^2 \brp\,.
\end{align}
\end{subequations}

Using this parametrization, the analytic smooth near-bolt solution that was discussed previously is obtained 
by tuning the parameters to the values:
\begin{equation}
h_{1,0}= \frac{p^2}{\al^2}\,,\qquad h_{2,0}= \frac{g_c p^2 }{\al^2}\,,\qquad 0\leqslant g_c<1 \, , \label{nearhorvalues}
\end{equation}
which lie at the lower bound on the regular solution space. These near-bolt values of $(h_{1,0},h_{2,0})$ 
correspond to a discrete shift in the UV asymptotics much like the near horizon values in the family of 
solutions reviewed in appendix~\ref{sugraconifold}.

%%%%%%%%%%%%%%%%%%%%%%%%%%%%%%%%%%%%
\subsubsection{Asymptotic Region}\label{sec:farbrane}
%%%%%%%%%%%%%%%%%%%%%%%%%%%%%%%%%%%%
The complete large $r$ expansion of the functions $f$, $\tilde{H}_1$ and $\tilde{H}_2$ 
involves both polynomials and powers of logarithms:
\begin{subequations}
\begin{align}
f& \stackrel{r\gg 1}{\simeq} \sum_{i,j\geq 0}f_{i,j} \rho^{-2i} (\log \rho)^j \\
\tH_1& \stackrel{r\gg 1}{\simeq}\sum_{i,j\geq 0} h_{1,i,j} \rho^{-2i} (\log \rho)^j \\
\tH_2& \stackrel{r\gg 1}{\simeq} \sum_{i,j\geq 0}h_{2,i,j} \rho^{-2i} (\log \rho)^j\,.
\end{align}
\end{subequations}
The explicit expressions for these expansion parameters for $i\geqslant 2$ are rather involved. When $h_{1,0,0}\neq0$ and for 
$i< 2$ we find
\begin{subequations}
\begin{align}
f&= 1-\frac{3 p^2}{16 \rho^2 h_{1,0,0}}  + \cO(\rho^{-4})\\
\tH_1&= h_{1,0,0} + \frac{3 p^2}{4\rho^2}  + \cO(\rho^{-4})\\
\tH_2&= \frac{h_{2,1,0} -3g_c  p^2 \log \rho}{\rho^2}  + \cO(\rho^{-4})
\end{align}
\end{subequations}
which gives the gauge field function in eq.~(\ref{Fdef}) as:
\be
g_1=\frac{3g_c  p^2+2h_{2,1,0}-6g_c p^2 \log \rho}{h_{1,0,0}\, \rho^2} + \cO(\rho^{-4})
\ee
Note that with $g_c\neq0$ this decays slower than $\rho^{-4}$ and is therefore not a normalizable harmonic form.

To summarize, the large $r$ expansion of the ALRF solution involves four free parameters, which we found could be taken to be
\be
\{h_{1,0,0},h_{2,1,0},f_{3,0},h_{2,3,0}\}\,.
\ee
The near horizon solution has $h_{1,0,0}=h_{2,1,0}=f_{3,0}=h_{2,3,0}=0$ and different UV asymptotics, in particular $f=\sqrt{\frac{3}{4}}+ \cO(\rho^{-2})$.

%%%%%%%%%%%%%%%%%%%%%%%%%%%%%%%%%%%%
\subsubsection{Interpolating Numerical Solution} \label{sec:Numerical}
%%%%%%%%%%%%%%%%%%%%%%%%%%%%%%%%%%%%
We have found numerical asymptotically locally Ricci-flat solutions by shooting from the bolt in the IR. We display the key features of this numerical solution by focussing on the plots for $f(\rho)$, this displays clearly the feature that the near horizon asymptotics $f\simeq \sqrt{3/4}$ are realized for some period before the plot jumps smoothly to the ALRF value of $f\simeq 1$.  

\begin{figure}[bth!]
\centering
\begin{minipage}{.5\textwidth}
  \centering
  \includegraphics[width=.9\linewidth]{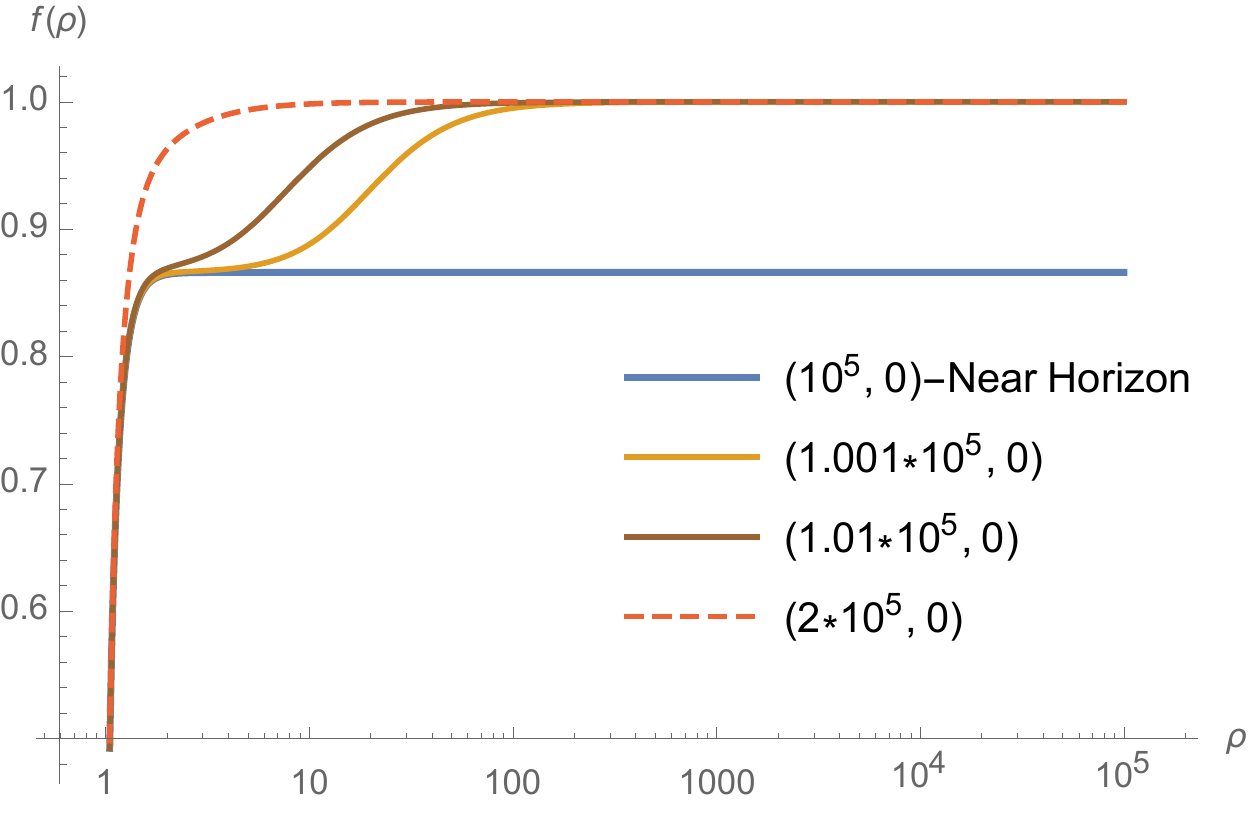}
\end{minipage}%
\begin{minipage}{.5\textwidth}
  \centering
  \includegraphics[width=.9\linewidth]{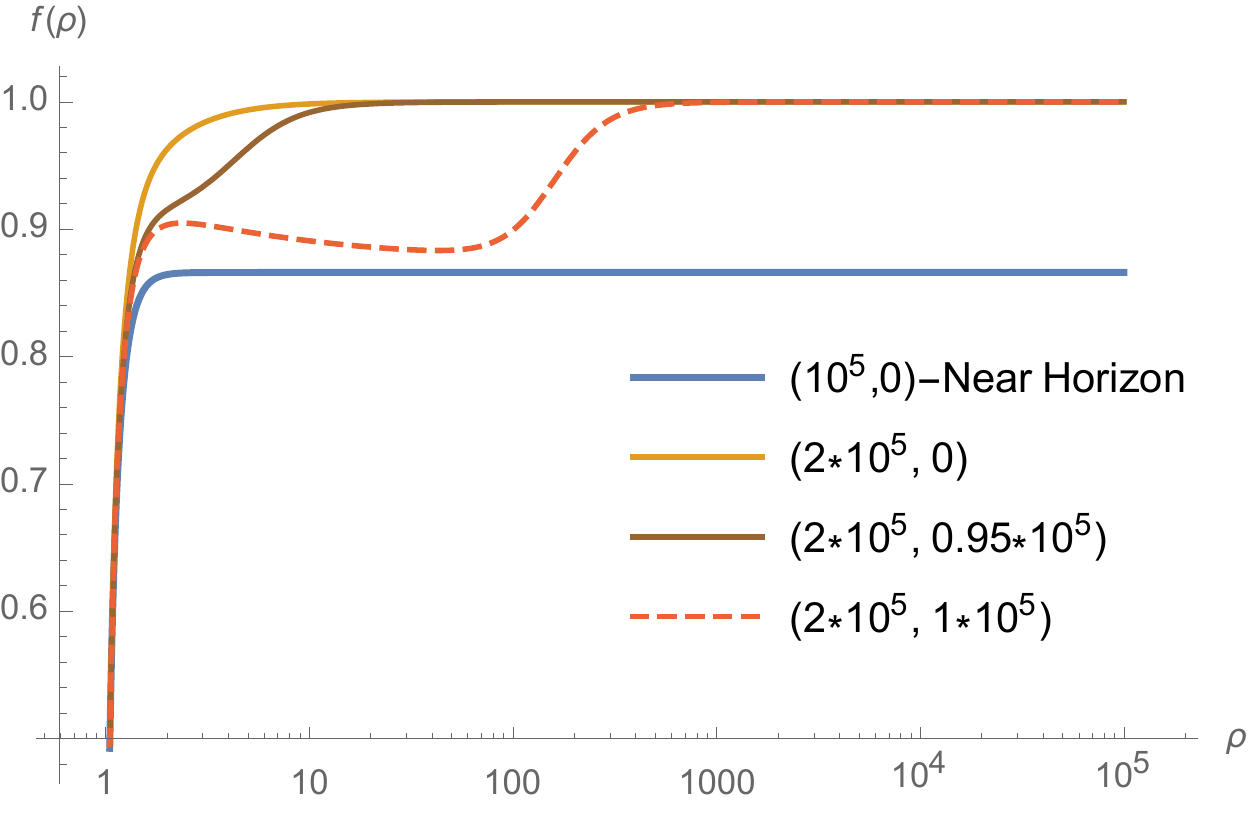}
\end{minipage}
  \caption{The metric function $f(\rho)$ with $(k_1,k_2)=(10^5,10^5)\Ra g_c=0$, vanishing $h_{2,0}$ (left) and nonvanishing $h_{2,0}$ (right) and four values of $h_{1,0}$.\label{gc0}}
\end{figure}

\begin{figure}[bth!]\label{gc13}
\centering
\begin{minipage}{.5\textwidth}
  \centering
  \includegraphics[width=.9\linewidth]{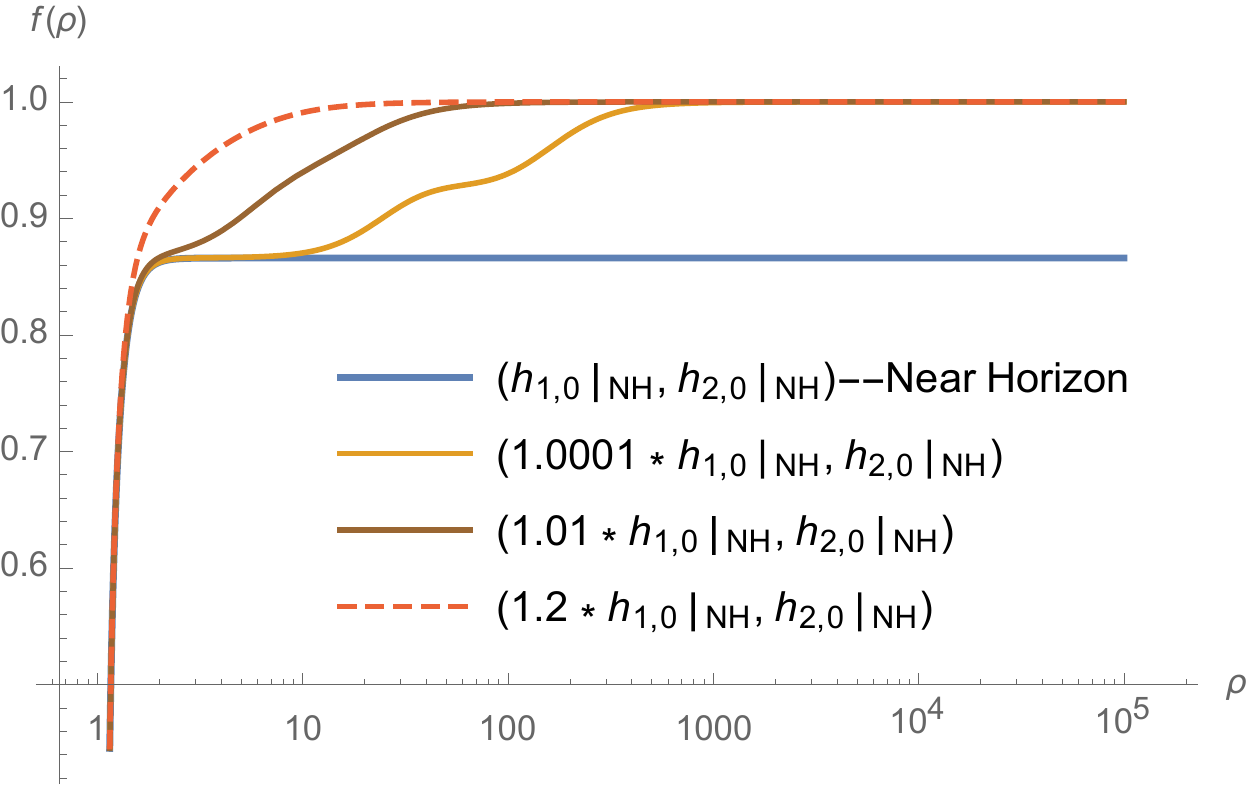}
\end{minipage}%
\begin{minipage}{.5\textwidth}
  \centering
  \includegraphics[width=.9\linewidth]{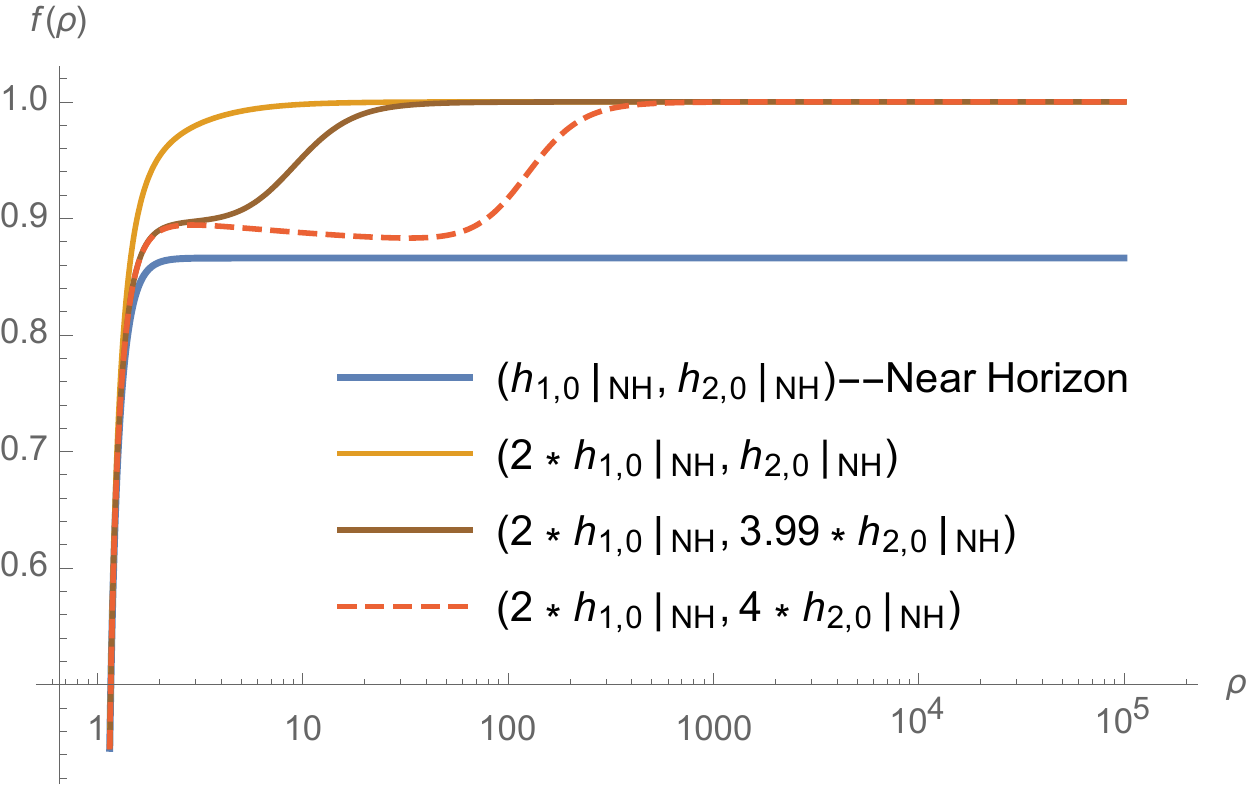}
\end{minipage}
  \caption{The metric function $f(\rho)$ with $(k_1,k_2)=(2\times 10^5,10^5)\Ra g_c=\frac{1}{3}$ with varying $h_{1,0}$ (left) and varying $h_{2,0}$ (right).}
\end{figure}

\begin{figure}[bth!]
\centering
\begin{minipage}{.5\textwidth}
  \centering
  \includegraphics[width=.9\linewidth]{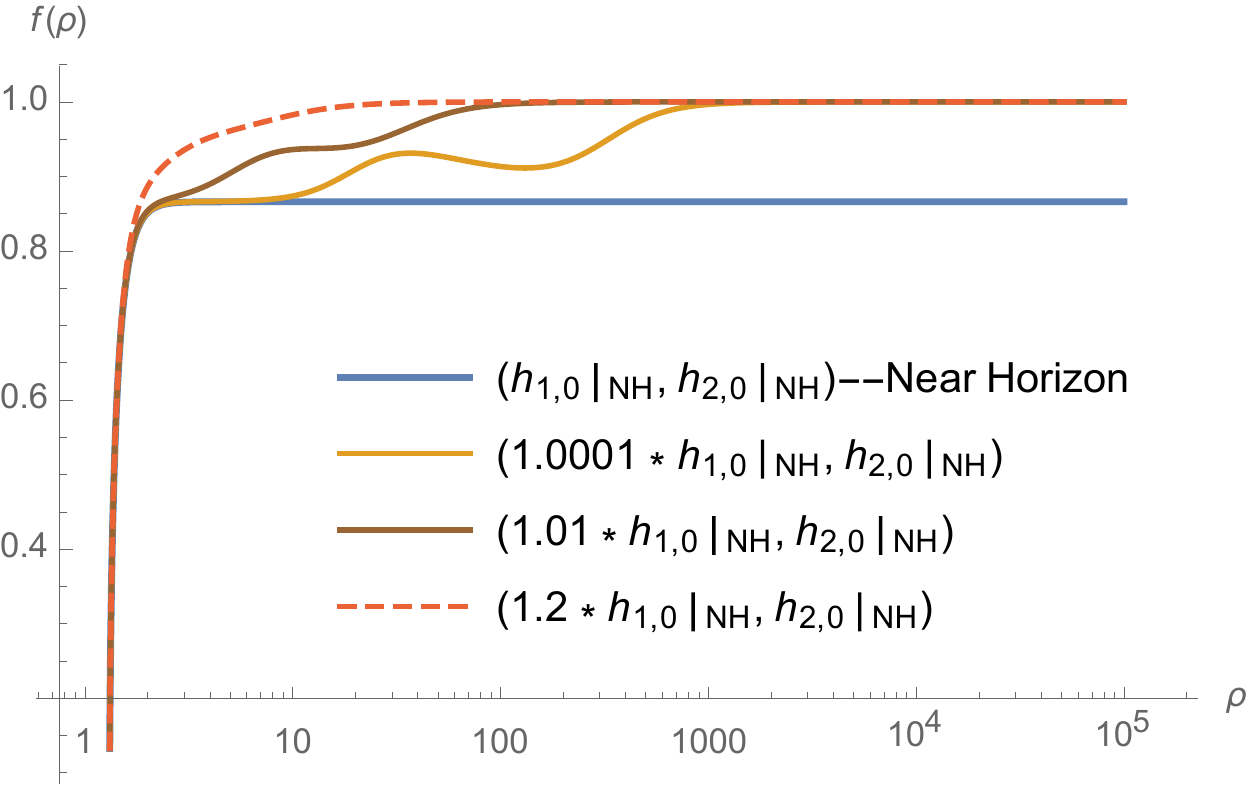}
\end{minipage}%
\begin{minipage}{.5\textwidth}
  \centering
  \includegraphics[width=.9\linewidth]{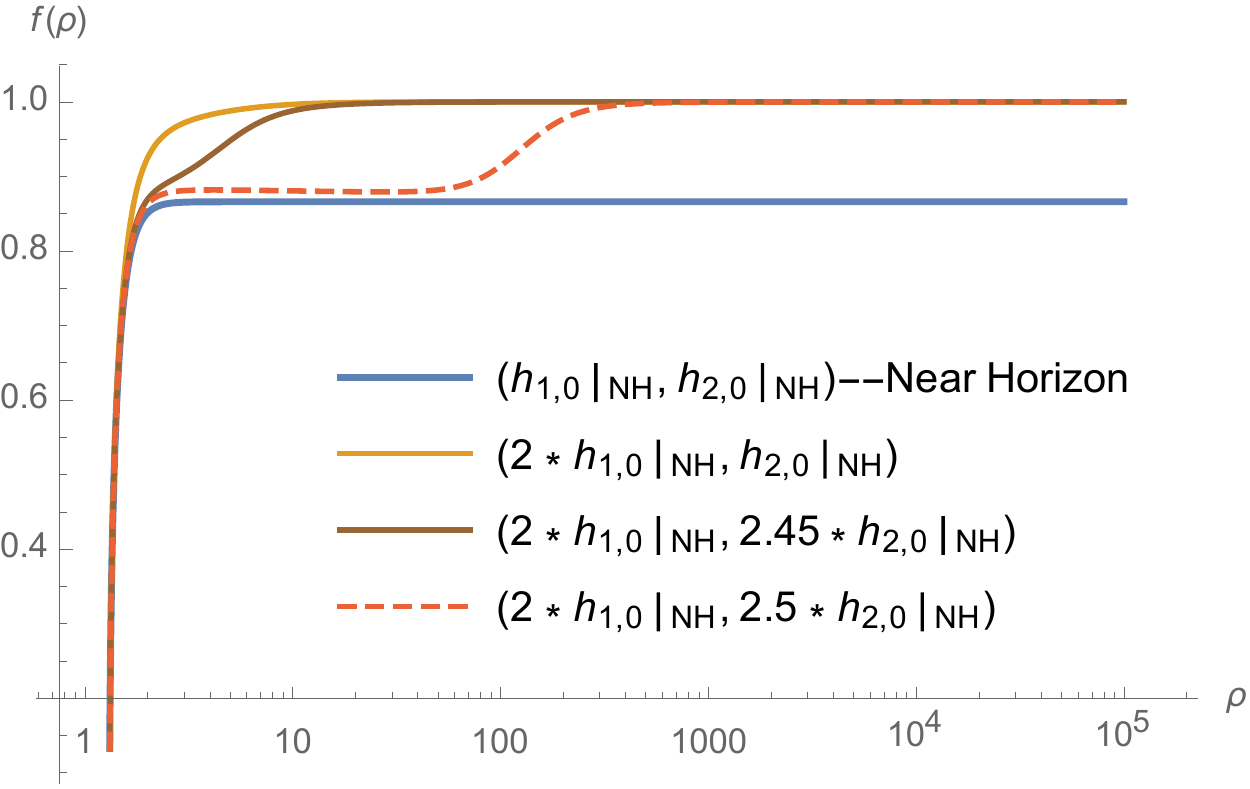}
\end{minipage}
  \caption{The metric function $f(\rho)$ with $(k_1,k_2)=(5\times 10^5,10^5)\Ra g_c=\frac{2}{3}$ 
with varying $h_{1,0}$ (left) and varying $h_{2,0}$ (right).\label{gc23}}
\end{figure}

In figures~\ref{gc0}-\ref{gc23} we have plotted $f(\rho)$ for $g_c=(0,\frac{1}{3},\frac{2}{3})$; one clearly observes a new feature emerge as $g_c$ in increased whereby a new dip forms between the two plateaus. As $g_c\ra 1^-$ this dip moves to the left ultimately pinching off the near horizon region. This limiting point will be the subject of section~\ref{sec:T11}.

Another key feature  is that for any given UV parameters $(k_1,k_2)$ the solution space it is two-dimensional, parameterized by $(h_{1,0},h_{2,0})$. The lower bound on $(h_{1,0},h_{2,0})$  is given by the near horizon values. There is an upper bound on $h_{2,0}$ which with some effort one can determine numerically, when $0\leqslant g_c<1$ this bound is less than $h_{1,0}$. Indeed when $h_{2,0}=h_{1,0}$ one of the finite IR two-spheres goes to zero size and one should recover the resolved conifold (which also allows for the periodicity $0\leqslant \psi < 4\pi$). However as will be elaborated on in section \ref{sec:T11} this limit requires in addition $g_c=1$. The decoupling limit of sections \eq{sec:nearhorizon} only exist for $h_{2,0}=0$. This can be seen in the plots below where $f(\rho)$ only approximates the near horizon solution if $h_{2,0}=0$.

The value of $h_{1,0}$ is unbounded and as it is taken large, the flux contribution to the equations is dilute and the metric approximates the Ricci-flat solution of section \ref{sec:RicciFlat}. So not only do the flux solutions in general interpolate in the radial direction between the near horizon solution and the Ricci-flat solution, in the {\it parameter space} of solutions, they interpolate between the exact near horizon and Ricci-flat solutions.

\vspace{5mm}
\noindent ${\bf g_c=0:}$ \\
Vanishing of $g_c$ implies equality between $k_1$ and $k_2$, we will take  $k_1=k_2=10^5$. The near horizon solution then has
\be
h_{1,0}|_{NH}= 10^5\,,\qquad h_{2,0}|_{NH}=0\,.
\ee
By allowing for $h_{2,0}$ to be nonzero, we generate a nontrivial profile for the metric function $H_2$ but as we see below, this rapidly falls to zero. This non-vanishing $H_2$ breaks spontaneously the $\ZZ_2$ symmetry which exchanges the pair of two-spheres. From the metric ansatz \eq{metans} we expect that in general $h_{2,0}\leqslant h_{1,0}$ which agrees with what we find numerically.

\begin{figure}[bth!]
\centering
  \includegraphics[width=.6 \linewidth]{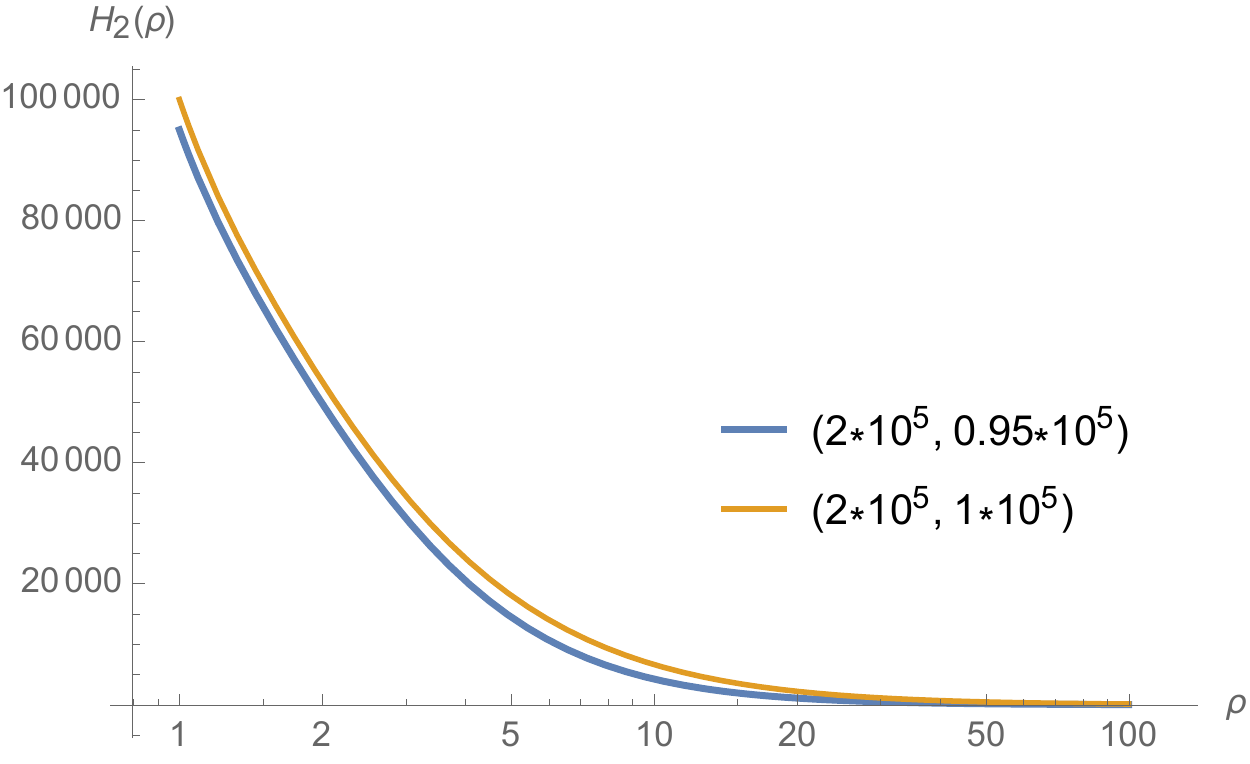}
  \caption{The metric function $H_2(\rho)$ with $(k_1,k_2)=(10^5,10^5)\Ra g_c=0$ but with non-vanishing $h_{2,0}$ and thus a spontaneously broken $\ZZ_2$ symmetry.}
\end{figure}

\newpage

\noindent ${\bf g_c=\frac{1}{3}:}$ \\
We obtain $g_c=\frac{1}{3}$ from $(k_1,k_2)=(2 \times 10^5,10^5)$. The near horizon values of $H_1$ and $H_2$ are
\be
h_{1,0}|_{NH}=122474\,,\qquad h_{2,0}|_{NH}= 40825\,.
\ee
We include one set of plots where we have fixed $h_{2,0}$ to the near horizon value and varied $h_{1,0}$ and another set of plots where we vary $h_{2,0}$.

\vspace{5mm}
\noindent ${\bf g_c=\frac{2}{3}:}$ \\
We obtain $g_c=\frac{2}{3}$ from $(k_1,k_2)=(5 \times 10^5,10^5)$. We now clearly see the development of 
non-monotonicity in $f(r)$ even when $h_{2,0}$ is set to the near-horizon value.

%%%%%%%%%%%%%%%%%%%%%%%%%%%%%%%%%%%%
\subsection{Curvature correction to the Bianchi identity}\label{sec:suppression}
%%%%%%%%%%%%%%%%%%%%%%%%%%%%%%%%%%%%

We have solved the Bianchi identity~\eq{Bianchi} by assuming that $\Tr\, F\w F$, the contribution from the vector bundle, dominates the 
$\Tr\, R_+ \w R_+$ term computed using the connection with torsion~(\ref{eq:omegaplus}) on the tangent bundle of the internal space. We now justify that this approximation is consistent in  the large charge limit $p^2,\, q^2 \gg 1$, by evaluating $\Tr\,  R_+\w R_+$ on-shell on 
the solution and demonstrating that it is subleading. For the near horizon solution of section~\ref{sec:nearhorizon}, we find that we 
explicitly get:
\begin{subequations}
\begin{align}
\Tr\, F\w F =& \frac{q^2}{\rho^8}\frac{\di \rho}{\rho}\w \eta \w (\Om_1+\Om_2)- 
\frac{q^2 + (1-g_c^2)^2 p^2 \rho^8}{4 \rho^8} \, \sig_1 \w \sig_2 \w\hsig_1\w \hsig_2  \\
%%%%%
\Tr\, R_+ \w R_+ =&- \frac{16(3+g_c^2)q^4}{ (1+g_c)(1-g_c^2)^2 p^4 \rho^{16}} \,\frac{\di \rho}{\rho}\w \eta \w (\Om_1+\Om_2) \non \\ 
&+\frac{16 g_c q^2 \bslb  (3+g_c^2)q^2 +  (1-g_c^2)^2 p^2 \rho^8 \bsrb}{(1+g_c)(1-g_c^2)^2 p^4\rho^{16}} \,
\frac{\di \rho}{\rho}\w \eta \w (\Om_1-\Om_2) \non \\
& + \frac{2\bslb (3+g_c)q^4-2g_c^2 (1-g_c^2)p^2 q^2 \rho^8 - (1-g_c^2)^3 p^4 \rho^{16}\bsrb}{(1-g_c^2)^2 p^4 \rho^{16}}\, 
\sig_1 \w \sig_2 \w\hsig_1\w \hsig_2
\end{align}
\end{subequations}
and we can see that in a large charge limit $p^2,\, q^2 \gg 1$ then $\Tr\, R_+ \w R_+$ is suppressed by $\Tr\, F\w F $. Beyond this large 
charge limit, the existence of the solution is ensured by the underlying exact worldsheet model that is discussed 
in section~\ref{sec:worldsheet}.

\begin{figure}[bth!]
\centering
  \includegraphics[width=.6 \linewidth]{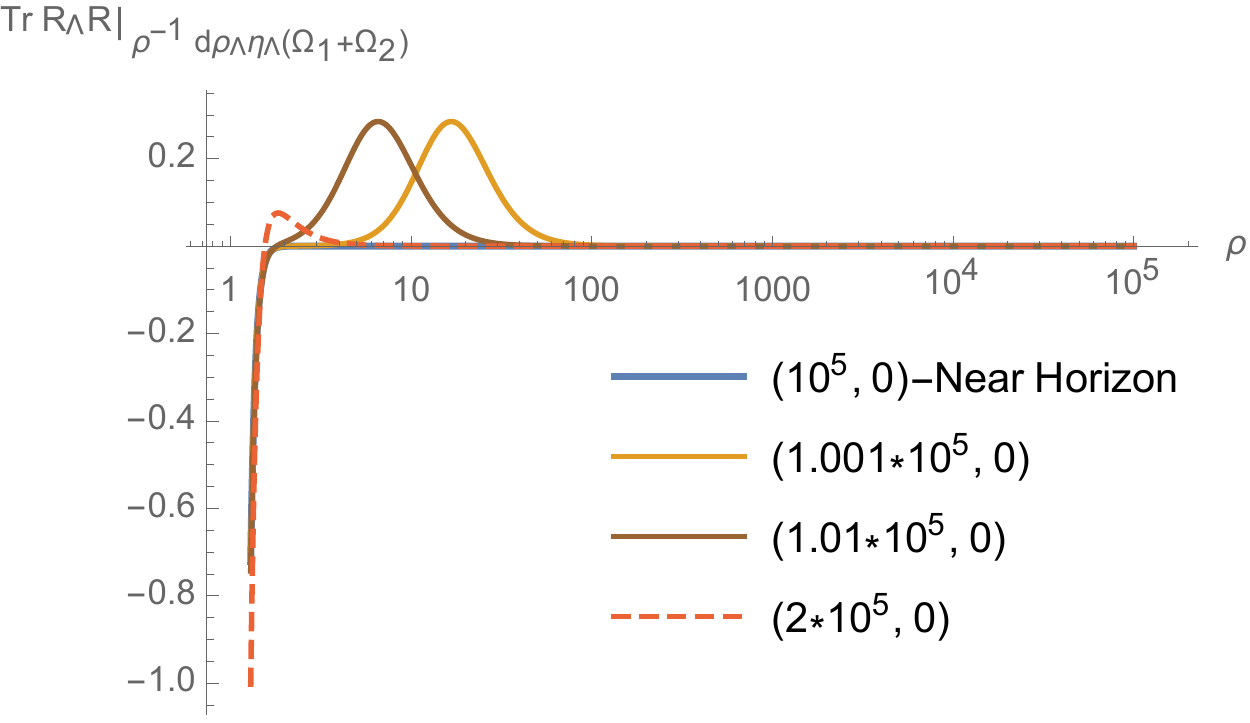}
  \caption{The coefficient of $\frac{\di \rho}{\rho} \w \eta \w (\Om_1+\Om_2)$ in $\Tr \, R_+\w R_+$ for $k_1=k_2$}
\end{figure}

The ALRF solutions are more difficult to analyze analytically. We have obtained analytic expressions in terms of the functions\footnote{by using the BPS equations we can eliminate the derivatives of these functions} $\{f,g_1,g_2,H_1,H_2\}$, which are rather more cumbersome than those in the $g_c\ra 0$ limit of \cite{Carlevaro:2009jx}. It is however quite instructive to plot the three inequivalent components 
of $\Tr\, R_+ \w R_+$ for some of the various parameters considered in~\ref{sec:Numerical}. We have included these plots in appendix \ref{app:NumericalBianchi} and they clearly demonstrate that $\Tr\, R_+ \w R_+$ remains $\cO(p^0)$ throughout and is thus suppressed 
by $\Tr\, F\w F$. By way of example we include here the results for the $\ZZ_2$ invariant case of $k_1=k_2$\footnote{The term 
proportional to $\di \rho\w \eta\w (\Om_1-\Om_2)$ vanishes when this $\ZZ_2$ symmetry is preserved}, 
which clearly show that $\Tr \, R_+\w R_+$ remains $\cO(p^0)$.

\begin{figure}[bth!]
\centering
  \includegraphics[width=.6 \linewidth]{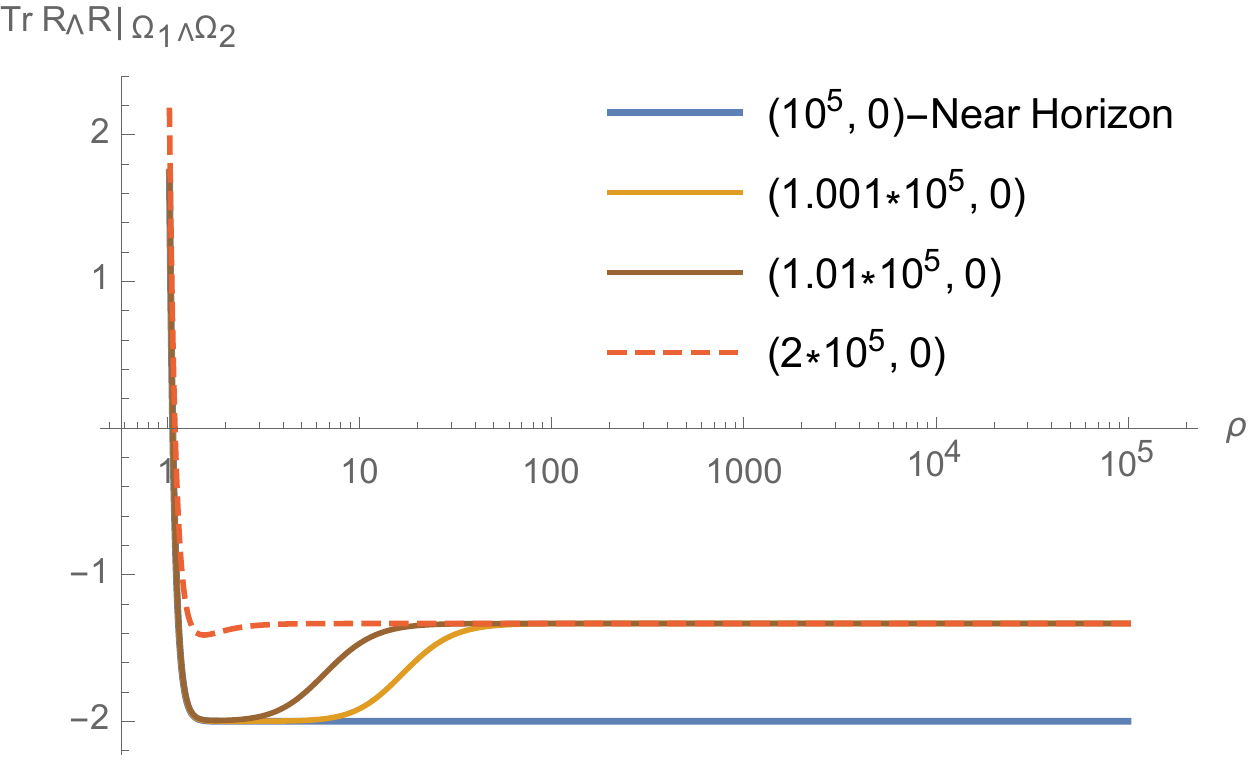}
  \caption{The coefficient of $ \Om_1\w \Om_2$ in $\Tr\, R_+\w R_+$ for $k_1=k_2$}
\end{figure}

%%%%%%%%%%%%%%%%%%%%%%%%%%%%%%%%%%%%
\section{A Heterotic Resolved Conifold}\label{sec:T11}
%%%%%%%%%%%%%%%%%%%%%%%%%%%%%%%%%%%%

In this section we present a class of numerical solutions which differ topologically from the previous ones.  In these 
geometries only one $S^2$ is of finite size at $r=0$ while a  full three-sphere shrinks; this is nothing but the familiar 
resolved conifold, however endowed with a non-K\"ahler metric due to the three-form flux.  

We find that these imposing these IR asymptotics is only possible at the boundary of the parameter space, namely by considering 
the limit:
\be
g_c\ra 1^-\,,\qquad a\ra 0^+ \, ,
\ee
which appears to be a singular limit of the near-bolt solution of section~\ref{sec:nearhorizon}.  This suggests that there is no decoupling limit, which would isolate the near-singularity region from the asymptotically locally Ricci-flat asymptotic region, for these solutions without a $\ZZ_2$ orbifold of the $T^{1,1}$ base.\footnote{or with the $\ZZ_2$ orbifold 
but with one two cycle shrinking to zero size at the tip.}

Indeed $g_c=1$ would give $k_2=0$ which is a singular limit of the worldsheet construction in section~\ref{sec:worldsheet}. In the previous section we found that tuning $(h_{1,0},h_{2,0})$ to their near-bolt values discretely changed the UV boundary conditions, here we find that tuning $g_c\ra 1$ discretely changes the IR boundary conditions.

The IR expansion of the equations~\eq{BPSeqfix1}-\eq{BPSeqfix3} compatible with these boundary conditions is of the form:
\begin{subequations}
\begin{align}
H_1&= \frac{h_{2,0}}{r^2} -   \frac{(p^2 -2 h_{2,0})h_{2,1}}{p^2}\, r^2 +\ldots \,,  \\
H_2&= \frac{h_{2,0}}{r^2} - h_{2,1}   \, r^2+\ldots  \,,  \\
f^2&= \frac{3}{4} + \frac{(2h_{2,0}-p^2 ) h_{2,1} }{4p^2 h_{2,0}}\,   r^4+\ldots  \, \\
e^{2(\Phi-\Phi_0)}&= \frac{4 h_{2,0}^2 h_{2,1}}{p^2} -  \frac{(7p^2 + 2h_{2,0})h_{2,1}^2}{6p^2} \, r^4 +\ldots \\
g_1&= 1- \frac{h_{2,1} r^4}{p^2} +\ldots \\
g_2&= 0\,.
\end{align}
\end{subequations}
where there appears to be two tunable parameters $(h_{2,0},h_{2,1})$. We can however eliminate one these two parameters by using the scaling symmetry of the BPS equations:
\bea
(r,H_1,H_2,e^{-\Phi_0})&\ra&   (\lam^{-1/2}r,\lam H_1,\lam H_2,\lam e^{-\Phi_0})\,.
\eea

We use this to set $h_{2,0}=p^2$ such that this branch of solution formally connects with the singular solution of section \ref{sec:nearhorizon}. The lowest order metric close to $r=0$ can be written as
\be
\di s_6^2 =  2 h_{2,1} p^2 \Bslb \di \hr^{\, 2}+\frac{ \hr^{\, 2}}{4} (\hsig_1^{\, 2}+ \hsig_2^{\, 2}+ \eta^2) + 
R^2_{S^2}(\sig_1^2+ \sig_2^2)    + \ldots \Bsrb
\ee
with $\hr=r^2$. We see that up to an overall constant, a round three sphere shrinks at $r=0$ and a two-sphere of radius $R_{S^2}^2=\frac{1}{4h_{2,1}}$ remains.

\begin{figure}[bth!]
\centering
\begin{subfigure}{.5\textwidth}
  \centering
  \includegraphics[width=\linewidth]{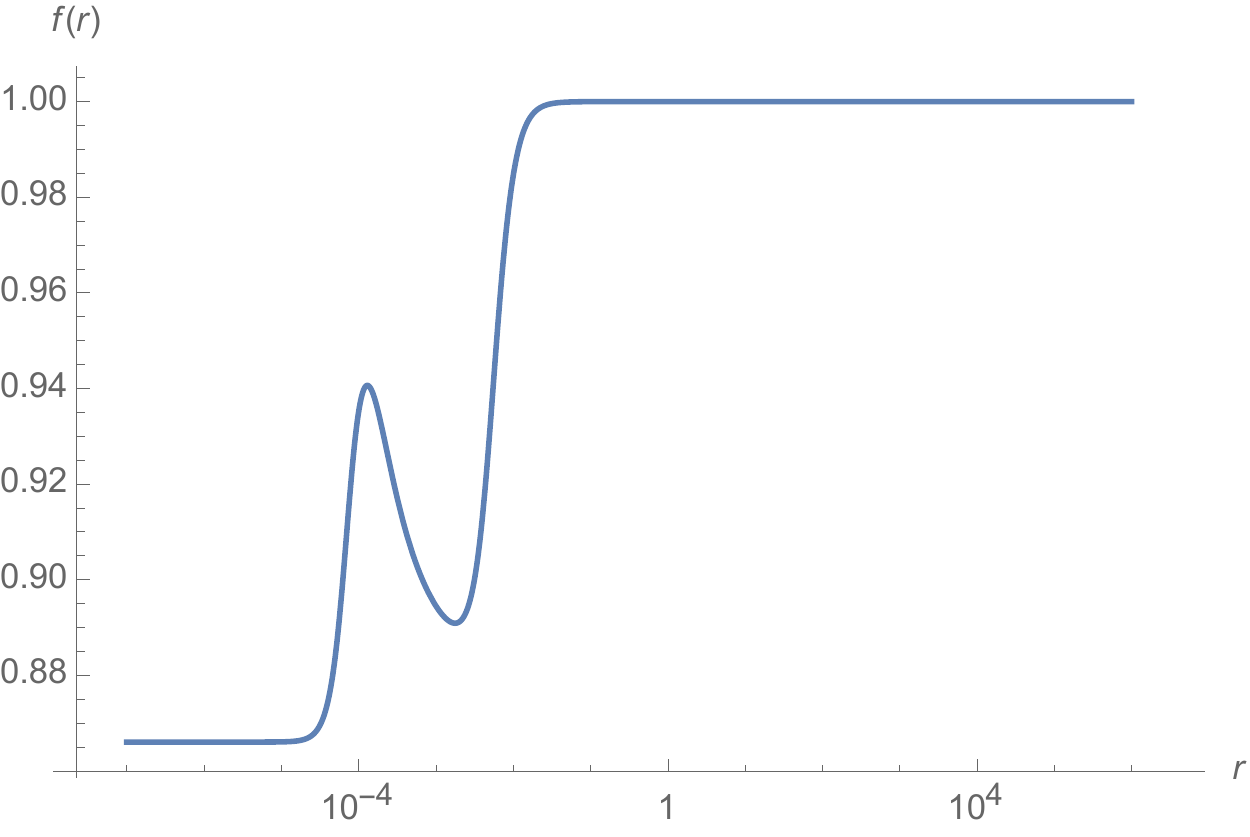}
  \caption{The metric function $f(r)$}
\end{subfigure}%
\begin{subfigure}{.5\textwidth}
  \centering
  \includegraphics[width=\linewidth]{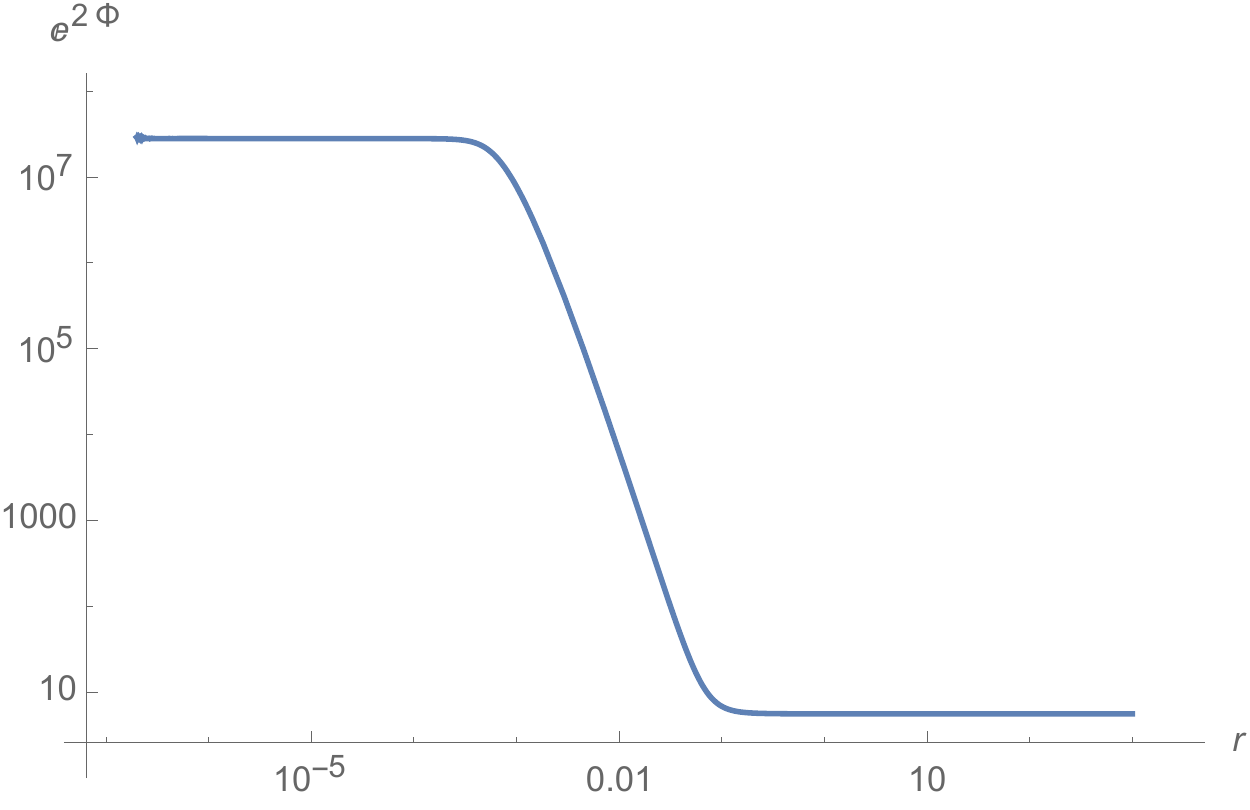}
  \caption{The dilaton}
\end{subfigure}
\caption{Numerical results for the metric function $f(r)$ and the dilaton $e^{2\Phi}$. We have set $p^2 =10^6$. The overall scale 
of $e^{2\Phi}$ can be absorbed into $\Phi_0$; only the ratio $e^{2\Phi|_{IR}}/ e^{2\Phi|_{UV}}\sim p^2$ is physical}
\label{fig:fDilPlots}
\end{figure}

Our numerical results support the existence of a one parameter family of asymptotically Ricci-flat solutions and are presented in 
figure~\ref{fig:fDilPlots}. We have found that $e^{2\Phi|_{UV}}/ e^{2\Phi|_{IR}}$ 
is bounded independently of the parameters of the solution and thus there is no possiblity of getting a divergent string coupling in the IR. As a result we 
have not found a near-horizon decoupling analogous to our solution in section~\ref{sec:nearhorizon}. We have evaluated the $\Tr\,  R\w R$ contribution 
to the Bianchi identity and found it to 
be $\cO(p^0)$, we have included these plots in appendix~\ref{app:NumericalBianchi}.

%%%%%%%%%%%%%%%%%%%%%%%%%%%%%%%%%%%%
\section{Worldsheet Model}\label{sec:worldsheet}
%%%%%%%%%%%%%%%%%%%%%%%%%%%%%%%%%%%%

In this section we will show that the near-bolt solution for the non-K\"ahler metric on 
$\cO(-2)\ra \mathbb{P}^1 \times \mathbb{P}^1$ with an Abelian vector bundle, given by eqns.~(\ref{eq:nhsolution}), corresponds 
to the target space of a solvable two-dimensional $(2,0)$ superconformal field theory. This results implies that this supergravity 
solution, that was found at leading order in a $1/p^2$ expansion, actually extends to an exact solution of heterotic string 
theory. This worldsheet formulation will also allow to clarify the issues of charge quantization that will be discussed in section~\ref{sec:charge} 
from the supergravity viewpoint.

%%%%%%%%%%%%%%%%%%%%%%%%%%%%%%%%%%%%%%%%%%%%%%%%%%%%%
\subsection{Near-bolt conifold from a gauged WZW model}
%%%%%%%%%%%%%%%%%%%%%%%%%%%%%%%%%%%%%%%%%%%%%%%%%%%%%
 
As a starting point we consider an $\cN=(1,0)$ Wess-Zumino-Witten (WZW) model in two dimensions for
\begin{equation}
G=  SL(2,\mathbb{R})_{k'}\times SU(2)_{k_1} \times SU(2)_{k_2} 
\end{equation}
and gauge asymmetrically an Abelian subgroup in order to get a  
gauged WZW model $G/H$ which will eventually provide a non-linear sigma-model with the 
conifold solution~(\ref{eq:nhsolution}) as target space. 

We will present a rough outline of this construction since many of the details have only minor changes to the analysis 
in~\cite{Carlevaro:2009jx} where $k_1=k_2$, $i.e.$ the $\mathbb{Z}_2$-symmetric solution with $g_c=0$,  was considered. 
Our technique for constructing anomaly free $(1,0)$ gauged WZW models is also closely related 
to models discussed in~\cite{Johnson:1994jw, Johnson:1994kh, Johnson:1994kv}.  Coset conformal field theories of this type are 
obtained through the following steps:
\begin{enumerate}
\item  Gauge the bosonic part of the WZW model under an Abelian subgroup in an asymmetric way and minimally couple their left-moving 
fermionic superpartners to the worldsheet gauge fields, in a supersymmetric way. In general this produces both a 'classical anomaly', 
$i.e.$ non-gauge invariant gauge couplings to the bosonic WZW model, and a  one-loop anomaly from the charged left-moving fermions. 
\item  Embed the Abelian subgroup into $U(1)^{16}$ (the Cartan subgroup of $SO(32)$  or $E_8\times E_8$) and minimally couple the 
corresponding 16 right-moving Weyl fermions. This minimal coupling results also in a one-loop anomaly.
\item Choosing the levels of the WZW models along with the embedding parameters of the gauge group from point 2, one should 
arrange for the total anomaly from these three contributions to cancel.
\end{enumerate}

We provide below some details of the construction in the present example. First, we parametrize 
the group-valued bosonic two-dimensional fields $(h(z,\bar z),g_1(z,\bar z) , g_2 (z,\bar z))$ 
of $SL(2,\mathbb{R})_{k'}\times SU(2)_{k_1}\times SU(2)_{k_2}$ in terms of Euler angles as follows:
\begin{subequations}
\label{eq:euler}
\begin{align}
h&= e^{i \frac{\tau_3}{2} (t+\varphi )} e^{\frac{\tau_1}{2} \varrho}e^{i \frac{\tau_3}{2} (t-\varphi)}\, ,  \\
g_\ell &= e^{i \frac{\tau_3}{2} \psi_\ell} e^{i\frac{\tau_1}{2} \theta_\ell}e^{i \frac{\tau_3}{2} \phi_\ell}\ , \quad 
\ell=1,2  
\end{align}
\end{subequations}
where $\tau_{1,2,3}$ are the Pauli matrices. The coordinate $\varphi$ has periodicity $\varphi \sim \varphi+2\pi$;
on the $N$-th cover of the group manifold $SL(2,\RR)$, the time-like coordinate $t$ 
has periodicity $t \sim t + 2\pi N$. 

The levels of the WZW models are chosen such that one gets at the end a $(2,0)$ SCFT with left central charge $c=9$, $i.e.$ compatible with 
four-dimensional Minkowski space-time (the right central charge being $\bar c = 6+ r_\textsc{v}$ where $r_\textsc{v}$ 
is the rank of the gauge bundle). This condition 
sets the level of the $SL(2,\mathbb{R})_{k'}$ model --~which does not need to be integer, unlike $k_1$ and $k_2$ which are 
levels of compact groups~-- to the value:\footnote{In our 
conventions the levels are those of the left-moving supersymmetric current algebras.}
\begin{equation}
\label{eq:levelcond}
c=9-\frac{6}{k_1} -\frac{6}{k_2} + \frac{6}{k'}=9 \ \implies \ k' = \frac{k_1 k_2}{k_1 + k_2}\, .
\end{equation}

As eventually a diagonal $\mathbb{Z}_2$ orbifold will be needed in order to avoid the conical singularity at the bolt, the levels $k_1$ and $k_2$ of both $SU(2)$ factors should be even integers as 
in~\cite{Carlevaro:2009jx} (this is imposed {\it in fine} by modular invariance of the torus partition function). We decompose these levels as follows: 
\begin{equation}
\label{eq:kdecomp}
k_1 =2 m s_1 \ , \quad k_2 = 2 m s_2 \ , \quad m,\, s_1, s_2 \in \mathbb{Z}_{>0} \ , \quad \text{gcd}\, (s_1, s_2) = 1\, ,
\end{equation}
such that $k' = 2m s_1 s_2/(s_1 + s_2)$. For instance if $k_1=k_2=2m$ we have $s_1=s_2=1$.

We then gauge asymmetrically an Abelian subgroup of $SL(2,\mathbb{R})\times SU(2)\times SU(2)$, defined by the group action:
\begin{subequations}
\begin{align}
U(1)_L\, :\ &
(g_1,g_2,h)  \longrightarrow 
(e^{\frac{i\tau_3}{2}\lambda(z,\bar z)} \, g_1,
e^{\frac{-i\tau_3}{2}\lambda(z,\bar z)}  \, g_2 , h ) \\
%%%
U(1)_N\, :\ & 
(g_1,g_2,h)  \longrightarrow 
(  e^{\frac{i\tau_3}{2}\alpha_1 \, \mu_1 (z,\bar z)} \, g_1,
 e^{\frac{i\tau_3}{2}\alpha_2 \, \mu_2(z,\bar z)}\, g_2 ,
e^{\frac{i\tau_3}{2}\beta \, \mu_1(z,\bar z)}\, h \, e^{\frac{i\tau_3}{2}\beta \, \mu_2(z,\bar z)})\, , 
\end{align}
\end{subequations}
where the integer parameters $\alpha_1$, $\alpha_2$ and $\beta$ parametrize the embedding of the Abelian subgroup $H$ into $G$. 

The gauging parametrized by $\lambda (z,\bar z)$ produces an anomaly --~both classical and quantum as explained before~-- which can be canceled at the quantum mechanical level by minimally coupling the right-moving fermions to the worldsheet gauge fields. We assemble the 
charges of the 16 currents in the Cartan subalgebra, $\{ \tilde{\lambda}^{2a-1}\tilde{\lambda}^{2a},\, a =1,\ldots,16 \}$, into 
a 16-component charge vector  ${\bf p}$. Then the gauging is anomaly-free provided that the following condition holds:\footnote{
Alternatively, one can replace the right-moving fermions $\{\bar{\chi}^a,a=1,\ldots,16 \}$ by a $SO(32)_1$ chiral WZW model; in this case there is no classical violation of gauge invariance.}
\begin{equation}
k_1 + k_2 = 2\, {\bf p}^{2}\, .\label{k1k2p}
\end{equation}
This gauging by itself produces a coset CFT corresponding to a non-Einstein metric on $T^{1,1}$ with H-flux (discarding the 
$SL(2,\mathbb{R})_{k'}$ factor).

The second gauging of the $U(1)_N$ factor, parametrized by $\{ \mu_\iota$, $\iota=1,2 \}$ is an example of {\it null gauging}, as the 
left and right anomalies can be made to cancel separately, see $e.g.$~\cite{Klimcik:1994wp}; as a consequence, one can choose independent gauge-fixing conditions for $\mu_1$ and $\mu_2$, effectively reducing the dimension of target space by two. 
The vanishing of the left classical anomaly implies that:
\begin{equation}
k_1 \alpha_1^2 + k_2 \alpha_2^2 = \beta^2 \frac{k_1 k_2}{k_1 + k_2}\ \Longrightarrow\ s_1 \alpha_1^{\, 2} + s_2 \alpha_2^{\, 2} = 
\frac{s_1 s_2}{s_1 + s_2} \beta^2\, .
\end{equation}
This condition is solved by chosing the embedding of the Abelian subgroup to be given by the parameters: 
\begin{equation}
\label{eq:embcond}
\alpha_1 = s_2 \ , \quad \alpha_2 = s_1  \ , \quad \beta = s_1 +s_2 \, .
\end{equation}
Then the right classical anomaly is canceled as before by minimally coupling the right-moving fermions provided that the 
corresponding charge vector ${\bf q}$ satisfies:\footnote{
Here $\frac{2m s_1 s_2}{s_1+s_2} +2$ is the bosonic level of the right-moving $\hat{\mathfrak{sl}}_2$ current algebra.}
\begin{equation}
\label{k1k2q}
\left(\frac{2 s_1 s_2}{s_1+s_2}\, m +2\right) (s_1+s_2)^2 = 2\, {\bf q}^{2}
\end{equation}
In order to avoid mixed gauge anomalies, one chooses the charge vectors to be orthogonal to each 
other in the Cartan subalgebra, $i.e.$ such that ${\bf p} \cdot {\bf q}=0$.

\begin{comment}
\begin{multline}
S = (k_1-2) S (g_1) + (k_2-2) S (g_2) + \left(\frac{k_1 k_2}{k_1 + k_2}+2 \right) S (h)\\
+ \frac{1}{8\pi} \int {\rm d}^2 z \, \Big\{
-2 \bar A \Big((k_1 - 2)(\partial \psi_1 + \cos \theta_1 \partial \phi_1) - 
(k_2 - 2) (\partial \psi_2 + \cos \theta_2 \partial \phi_2) \Big) 
+ (k_1 + k_2 -4) A \bar A\\
-2 B_1\, \beta (k' +2)\,  \Big( \bar \partial (t - \varphi) + \cosh \varrho \, \bar \partial (t+\varphi) \Big) \\
-2 \bar{B}_2 \Big(\alpha_1 (k_1 - 2) (\partial \psi_1 + \cos \theta_1 \partial \phi_1) + 
\alpha_2 (k_2 - 2) (\partial \psi_2 + \cos \theta_2 \partial \phi_2) 
+ \beta (k' +2)[\partial (t + \varphi) + \cosh \varrho \,  \partial (t-\varphi)]\Big)\\
+ [\alpha_1^2 (k_1-2) + \alpha_2^2 (k_2-2) ] B_2 \bar{B}_2 
-\beta^2(k' +2) [B_1 \bar B_1 + B_2 \bar B_2 + B_1 \bar B_2 \cosh \varrho]
\Big\}\, ,
\end{multline}
\end{comment}

Taking into account the relations~(\ref{eq:levelcond},\ref{eq:embcond}), defining the new angular variable $\psi :=\psi_1 + \psi_2$ and choosing the gauge-fixing conditions
\begin{equation}
\label{eq:gaugefix}
\psi_1 - \psi_2=0 \ , \quad t=0  \ , \quad \varphi = 0 \, ,
\end{equation}
The worldsheet action for the bosonic degees of freedom of the gauged WZW model is then of the form:
\begin{multline}
S = (2ms_1 -2) S (g_1) + (2m s_2-2) S (g_2) + \left(\frac{2ms_1 s_2}{s_1 + s_2}+2\right) 
\frac{1}{8\pi}\int {\rm d}^2 z \, \partial \varrho \bar \partial \varrho \\
+ \frac{1}{8\pi} \int {\rm d}^2 z \, \Big\{
-2 \bar A \Big((2ms_1 - 2) (\partial \psi/2 + \cos \theta_1 \partial \phi_1) - 
(2ms_2 - 2) (\partial \psi/2 + \cos \theta_2 \partial \phi_2) \Big) 
\\+ (2m(s_1 + s_2) -4) A \bar A
-2 \bar{B}_2 \Big(s_2(2ms_1 - 2) (\partial \psi/2 + \cos \theta_1 \partial \phi_1) + 
s_1 (2ms_2 - 2) (\partial \psi/2 + \cos \theta_2 \partial \phi_2) \Big)\\
+ 4 (s_1^{\, 2} +s_2^{\, 2}+ s_1 s_2) B_2 \bar{B}_2 
- (s_1+s_2)\left( 2m s_1 s_2 +2 (s_1+s_2)\right)\left( B_1 \bar B_1 + B_1 \bar B_2 \cosh \varrho \right)
\Big\}\, .
\end{multline}
where for instance $k_1 S(g_1)$ denotes the WZW action for $SU(2)_{k_1}$. 

In addition one needs to add minimally coupled fermions to get the full action of the model. The couplings of the 
left-moving Majorana-Weyl fermions $\{ \zeta_i, \, i=1,\ldots,6\}$ is dictated by worldsheet $(1,0)$ supersymmetry, while the 
couplings to the right-moving fermions $\{ \tilde{\lambda}^n, n=1,\ldots,32\}$ is only constrained by cancellation of 
gauge anomalies, see eqns.~(\ref{k1k2p},\ref{k1k2q}): 
\begin{multline}
S_F =\frac{1}{4\pi} \int {\rm d}^2 z \Big\{ \sum_{i=1}^6 \zeta_i \bar \partial \zeta_i 
-2 (\zeta^1 \zeta^2 - \zeta^3 \zeta^4) \bar A 
- 2 \Big( k_2\zeta^1 \zeta^2 - k_1\zeta^3 \zeta^4 - (k_1 + k_2) \zeta^5 \zeta^6\Big) \bar B_2\\
+ \sum_{n=1}^{32} \tilde{\lambda}^n \partial \tilde{\lambda}^n - A\,  \sum_{a=1}^{16} p_a\,  \tilde{\lambda}^{2a-1} \tilde{\lambda}^{2a}
- B_1  \, \sum_{a=1}^{16} q_a \, \tilde{\lambda}^{2a-1} \tilde{\lambda}^{2a}
\Big\}\, .
\end{multline}

\subsection{Large gauge transformations and charge quantization}

The gauge-fixing conditions~(\ref{eq:gaugefix}) allowed us to get rid of the gauge redundancy only for gauge transformations 
connected to the identity. There might be residual large gauge transformations that need to be carefully accounted for, as they 
would correspond to extra, undesirable orbifold actions. 

In terms of the Euler angles, see eqns.~(\ref{eq:euler}), the gauged Abelian 
subgroup corresponds to the following coordinate shifts:
\begin{equation}
\left\{ \begin{array}{ccc} 
\psi_1 & \to & \psi_1 + \lambda + s_2\, \mu_1\\
\psi_2 & \to & \psi_2 - \lambda + s_1\, \mu_1\\
t+ \varphi & \to & t+ \varphi +  (s_1+s_2) \mu_1 \\
t- \varphi & \to & t- \varphi + (s_1 + s_2)\mu_2
\end{array}
\right.
\end{equation}
Given that all these coordinates are periodic, gauge-fixing conditions generically leaves unfixed a discrete Abelian subrgroup. 

\subsubsection*{Singular conifold}

The gauge symmetry parametrized by $\lambda$, acting on $SU(2)_{k_1} \times SU(2)_{k_2}$, gives by itself (discarding 
the spectator $SL(2,\mathbb{R})$ factor) a non-Einstein $T^{1,1}$ with 
H-flux, which is used to build the singular conifold solution~(\ref{eq:singsol}). 

One can conveniently fix the gauge $\psi_1 - \psi_2=0$, the coordinate $\psi = \psi_1 + \psi_2$ being gauge invariant. 
This coordinate has the correct periodicity $\psi \sim \psi + 4\pi$ and  without remaining large gauge transformations that we should take care of.

\subsubsection*{Resolved orbifold of the conifold}

The gauge symmetry with parameter $\mu_1$ can be gauge-fixed by choosing a gauge with $t+\varphi=0$. The 
large gauge transformations that remain as a discrete gauge group are of the form $\mu_1 = \frac{2\pi}{s_1+s_2} (N a+ b)$, with $a,b \in \mathbb{Z}$, for 
the $N$-th cover of the $SL(2,\mathbb{R})$ group manifold. A $\mathbb{Z}_2$ subgroup acts non-trivially 
on the coordinate $\psi$ as $\psi \sim \psi + 2\pi$. This orbifold of $T^{1,1}$ is exactly what is needed to 
avoid a conical singularity at the bolt of the manifold $\cO(-2)\ra \mathbb{P}^1 \times \mathbb{P}^1$, that was 
discussed in the supergravity analysis. 

The gauge transformation $\psi_1-\psi_2  \to  \psi_1 -\psi_2 + 2\lambda + (s_2-s_1) \mu_1$ indicates that the gauge-fixing condition 
$\psi_1 - \psi_2=0$ that was chosen previously needs to be be modified to absorb the (partially gauge-fixed) $\mu_1$ action. Thus 
in the case of the coset CFT corresponding to the resolved geometry the remnant discrete subgroup, whose elements are of the form 
\begin{equation}
\lambda =  \frac{s_1-s_2}{s_1+s_2}  (N a+ b)\pi \ , \quad a,b \in \mathbb{Z}\, ,
\end{equation}
induces an orbifold action on the right-moving fermions minimally coupled to the gauge fields. As we eventually expect to get a heterotic non-linear sigma-model 
corresponding to the supergravity solution~(\ref{eq:nhsolution}) such orbifold is not expected.

Assuming that $s_1$ and $s_2$ (which are coprime positive integers) are both odd and otherwise generic, 
the unwanted orbifold action can be avoided at the expanse of quantizing the charges of the Abelian bundle as follows\footnote{One 
can  at most allow the action of a $\mathbb{Z}_2$ subgroup, as these right moving fermions are subject to a GSO projection 
onto even fermion number. In the $Spin(32)/\mathbb{Z}_2$ theory for instance,  
one can consider $\hat{\bf p}  \in (\mathbb{Z}+ \frac12)^{16}$ 
as well, corresponding to bundles without vector structure, see section~\ref{sec:charge}; the same holds for $\hat{\bf q}$.}
\begin{equation}\label{pphatintegral}
{\bf p} = (s_1 + s_2)\, \hat{\bf p} \ , \quad \hat{\bf p}  \in \mathbb{Z}^{16}\, .
\end{equation}

Finally, one can take the gauge $t-\varphi=0$ for the symmetry parametrized by $\mu_2$. The leftover discrete 
gauge symmetry transformations, of the form $\mu_2 = \frac{2\pi}{s_1+s_2} (N a- b)$ with $a,b \in \mathbb{Z}$ 
as before, will have generically a non-trivial action on the right-moving charged fermions (in terms of which the Cartan 
generators of the space-time gauge group are expressed). Likewise, one finds the quantization condition 
\begin{equation}\label{qqhatintegral}
{\bf q} = (s_1 + s_2)\, \hat{\bf q} \ , \quad \hat{\bf q}  \in \mathbb{Z}^{16}\, .
\end{equation}
The tadpole condition~(\ref{k1k2q}) is then given by 
\begin{equation}
\frac{m s_1 s_2}{s_1+s_2}  =  \hat{\bf q}^2-1\, . 
\end{equation}
In other words, given that $\hat{\bf q}^2$ is an  integer, we get the consistency condition 
\begin{equation}
\label{eq:k1k2quant}
\frac{s_1 s_2}{s_1 +s_2}m \in \mathbb{Z}_{>0} \ \Leftrightarrow \ \frac{k_1 k_2}{k_1 +k_2} \in 2\mathbb{Z}_{>0} \, .
\end{equation}

\subsection{Non-linear sigma-model from the gauged WZW model}

Obtaining the background fields at leading order in $\alpha'$ from this action requires to go through several 
steps:\footnote{It is also possible to obtain the exact expressions in $\alpha'$. For $k_1=k_2$ it has been argued in~\cite{IsraelHDR}, following 
the general method of~\cite{Tseytlin:1992ri} that these corrections vanish, hence that the non-linear 
sigma-model is not corrected at the perturbative level; we expect the same to hold here.} 
\begin{enumerate}
\item bosonizing all the fermions taking into account the gauge anomalies
\item integrating out classically the gauge field
\item going back to the fermionic variables
\end{enumerate}
Once the dust has settled, one can first read from this lengthy computation the metric of the near-horizon solution:
\begin{equation}
{\rm d}s^2 = \frac{\alpha' k_1}{4} (\sig_1^2+\sig_2^2) +\frac{\alpha' k_2}{4} (\hsig_1^2+\hsig_2^2) + 
\frac{\alpha'}{4}\frac{k_1 k_2}{k_1 + k_2} ({\rm d} \varrho^2 
+ \tanh^2 \varrho\, \eta^2 )\,.
\end{equation}
The  gauge field strength has first a non-normalizable part which reads, in terms of the 
vector of magnetic charges $\bp$:
\begin{equation}
F_\textsc{nn} = \frac{k_1 \Omega_1 - k_2 \Omega_2}{2(k_1 +k_2) } \, \bp \cdot \bH\, .
\end{equation}
The normalizable part which corresponds to the blow-up mode is locally of the form
\begin{equation}
F_\textsc{n} =  {\rm d} \left( \frac{\eta}{4\cosh \varrho} \right)  \, \bq \cdot \bH \, .
\end{equation}

%%%%%%%%%%%%%%%%%%%%%%%%%%%%%%%%%%%%
\section{Charges}
%%%%%%%%%%%%%%%%%%%%%%%%%%%%%%%%%%%%
\label{sec:charge}

In this final section we compute the charges of the solutions --~both the gauge and fivebrane charges~-- and discuss charge 
quantization.

%%%%%%%%%%%%%%%%%%%%%%%%%%%%%%%%%%%%
\subsection{Gauge Charges}
%%%%%%%%%%%%%%%%%%%%%%%%%%%%%%%%%%%%

All the solutions obtained from the ansatz~(\ref{metans},\ref{Fdef}) are characterized by an asymptotic magnetic monopole charge, 
given by:
\be
q_\textsc{uv}= \frac{1}{2\pi}\int_{S^2_\infty} F = \frac{1}{2}\, \bp \cdot \bH\, ,
\ee
with the two sphere at infinity is given by
\begin{equation}
S^2_\infty:\quad \psi=0\,,\quad \tha_1=\tha_2\,,\quad \phi_1=-\phi_2\,.
\end{equation}
The embedding of this magnetic charge in the Cartan of the gauge group given in terms of the 
charge vector $\bp$, and independent of the other charge vector $\bq$.

In the infrared of the regular solutions ($i.e.$  those diffeomorphic to $\cO(-2)\ra F_{0}$) there are two magnetic charges 
at the bolt $r=a$, through the pair of $\mathbb{P}^1$ cycles. They are given by:
\begin{subequations}
\begin{align}
q_{1,\textsc{ir}}&= \frac{1}{2\pi} \int _{S^2_1} F = 
 \frac{1}{2}\Big[ - (1+g_c) \bp - \frac{p}{q}\sqrt{1-g_c^{\, 2}}\, \bq \Big]\cdot \bH 
= \frac{1}{2}\Big[ - \frac{k_1}{p^2}\, \bp - \frac{\sqrt{k_1k_2}}{q p} \, \bq \Big]\cdot \bH\,, \label{q1} \\
q_{2,\textsc{ir}}&= \frac{1}{2\pi} \int _{S^2_2} F  
=\frac{1}{2}\Big[ (1-g_c) \bp - \frac{p}{q}\sqrt{1-g_c^{\, 2}}\, \bq \Big] \cdot \bH 
= \frac{1}{2}\Big[ \frac{k_2}{p^2}\, \bp - \frac{\sqrt{k_1k_2}}{q p}\, \bq \Big] \cdot \bH\,, \label{q2}
\end{align}
\end{subequations}
where the pair of two spheres are given by
\be
S^2_i:\quad R=1,\quad\tha_i=\tha_{i,0},\quad \phi_i=\phi_{i,0} 
\ee
and the integrals in \eq{q1} and \eq{q2} have been evaluated at the bolt $R=1$. In addition we have used the identifications from the worldsheet~\eq{k1k2gc}. 
In addition, using the decomposition~(\ref{eq:kdecomp}), the tapdole conditions~(\ref{k1k2p},\ref{k1k2q}) 
give (in the large charges limit $m \gg 1$ allowing to neglect the subleading term in~(\ref{k1k2q})):
\begin{subequations}
\begin{align}
q_{1,\textsc{ir}}&= - \left[ \frac{s_1}{s_1+s_2} \, \bp + \frac{1}{s_1+s_2} \bq \right] \cdot \bH \, ,\\
q_{2,\textsc{ir}}&= - \left[ \frac{s_2}{s_1+s_2} \, \bp - \frac{1}{s_1+s_2} \bq \right] \cdot \bH \, .
\end{align}
\end{subequations}
In order to get proper quantization conditions, we first rescale the magnetic charges as in section~\ref{sec:worldsheet}, 
eqns.~(\ref{pphatintegral},\ref{qqhatintegral}). We obtain then 
\begin{equation}
q_{1,\textsc{ir}} = -  \left(s_1 \, \hat{\bp} + \hat{\bq} \right) \cdot \bH \ , \quad 
q_{1,\textsc{ir}} = -  \left(s_2 \, \hat{\bp} - \hat{\bq} \right) \cdot \bH \, . 
\end{equation}
Let us consider for example the $Spin(32)/\mathbb{Z}_2$ theory. Imposing a Dirac quantization condition for the adjoint (two-index) representation of the 
gauge group, one finds two possible solutions
\begin{equation}
\forall \ell \in \{ 0,1,\ldots,16 \} \ , \quad 
\left\{ \begin{array}{l} s_1 \, \hat{p}_\ell + \hat{q}_{\ell} \in \mathbb{Z} \\ s_2 \, \hat{p}_\ell - \hat{q}_{\ell} \in \mathbb{Z}\end{array}\right. 
\quad \text{or} \quad 
\forall \ell \in \{ 0,1,\ldots,16 \} \ , \quad 
\left\{ \begin{array}{l} s_1 \, \hat{p}_\ell + \hat{q}_{\ell} \in \mathbb{Z}+\nicefrac{1}{2} \\ s_2 \, \hat{p}_\ell - \hat{q}_{\ell} \in \mathbb{Z}+ \nicefrac{1}{2} \end{array}\right.  \, .
\end{equation}
The former case corresponds to bundles with vector structure, and the latter to bundles without vector structure. Assuming that $s_1$ and $s_2$ are both odd, and given 
that $\mathbf{p}$ and $\mathbf{q}$ are orthogonal, we can simply choose 
$\hat{\bp},\, \hat{\bq}  \in \mathbb{Z}^{16}$ and $\hat{\bp},\, \hat{\bq}  \in (\mathbb{Z}+\nicefrac{1}{2})^{16}$ respectively.\footnote{Furthermore, 
one should impose that the first Chern class of the gauge bundle is even in order to allow for spinorial representations of the gauge group.} 

%%%%%%%%%%%%%%%%%%%%%%%%%%%%%%%%%%%%
\subsection{Five-brane charge}
%%%%%%%%%%%%%%%%%%%%%%%%%%%%%%%%%%%%

The five-brane charge in these backgrounds is somewhat subtle due to the non-closure of the three form $\cH_{(3)}$. There are two types of charge relevant to our 
solution\footnote{see \cite{Marolf:2000cb, Aharony:2009fc} for a review of these charges in type IIA string theory}, the Maxwell charge $\cQ_M$ and the Page charge $\cQ_P$:
\bea
\cQ_M&=& \frac{1}{2\pi^2 \al'}\int_{M_3} \cH_{(3)} \,,\qquad \cQ_{P}= \frac{1}{2\pi^2 \al'}\int_{M_3} \blp \cH_{(3)}- CS(A)\brp 
\eea
where $CS(A)$ is the Chern-Simons form for the Abelian connection $A$
\be
CS(A)= \Tr \, A\w F
\ee
and $M_3$ may be $S^3$ or $S^3/\ZZ_2$. Note that the Page charge is required to be quantized since \cite{Rohm:1985jv} 
\be
\frac{1}{2\pi^2 \al'} \int_{M_3} \text{d} B \in \ZZ
\ee
for any three manifold $M_3$.

%%%%%%%%%%%%%%%%%%%%%%%%%%%%%%%%%%%%
\subsubsection{Maxwell Charge}
%%%%%%%%%%%%%%%%%%%%%%%%%%%%%%%%%%%%
Recall that $b_3(T^{1,1})=1$, topologically there is a single three cycle and two canonical representatives of this class are
\bea
S_1^3:&& \tha_1=\tha_{1,0}\,,\ \phi_1=\phi_{1,0}\,, \qquad S_2^3:\quad  \tha_2=\tha_{2,0}\,,\ \phi_2=\phi_{2,0}\,,\non 
\eea
where $\{\tha_{i,0},\phi_{i,0}\}$ are constants. By computing the Maxwell charge with a Gaussian surface at large $r$, one hopes to enclose all the charge in the 
system but it is by now 
well-known that in local supergravity solutions such as ours the entire space has a non-zero charge density and thus the Maxwell charge is radially 
dependent~\cite{Klebanov:2000hb}. In addition we find that there is a nontrivial profile for the charge density on the $T^{1,1}$ itself at any 
non-zero distance from the bolt and thus the Maxwell charge depends on the representative three-cycle which is chosen.

From our expression for the three form $\cH_{(3)}$ in \eq{H3eval}  and taking into account the relative orientation of $S_1^3$ and $S^3_2$ we find that
\bea
\cQ_{M,1}=\frac{1}{2\pi^2 \al'}\int_{S^3_1/\ZZ_n} \cH_{(3)} &=& \frac{8}{n}(h_1+h_2)\,, \label{MaxCh1} \\
\cQ_{M,2}=\frac{1}{2\pi^2\al'}\int_{S^3_2/\ZZ_n} \cH_{(3)} &=&  \frac{8}{n}(h_1-h_2)\,.\label{MaxCh2}
\eea
where $n\in \{ 1,2\}$ account for the conifold and its $\mathbb{Z}_2$ orbifold respectively. 
From this we see that unless $h_2=0$, the Maxwell charge depends on the choice of three-cycle it is computed from. We can understand the difference as the 
integral of $\Tr F\w F$ over a 4-chain $M_4$ whose boundary is the different Lens spaces:
\be
\int_{S^3_1/\ZZ_n} \cH_{(3)} - \int_{S^3_2/\ZZ_n}\cH_{(3)}= \int_{M_4} \Tr F\w F
\ee
which is non-zero. We also note that both $(h_1,h_2)$ vanish at the bolt, so in particular 
\be
\int_{S^2_1\times S^2_2|_{bolt}} \Tr F\w F =0\,.
\ee

It is interesting to consider the radial dependence of the Maxwell charge. In particular when $g_c=0$ and the $\ZZ_2$ symmetry is preserved ($H_2=0$), the 
function $h_2$ is independent of the boundary condition $h_{2,0}$ and thus has the same profile for the near horizon solution as for the family of 
asymptotically Ricci flat solutions. This is perhaps counterintuitive since as explained in section \ref{sec:Numerical} the metric function $f(\rho)$ 
has quantitatively different UV behavior for these two classes of solutions. For the near horizon solution (with $n=2$) 
we have:\footnote{Recall that $g_c=\pm 1$ is not a regular limit of the near horizon solution.}
\be
h_1=0\,,\quad h_2|_{r=\infty}= \frac{p^2}{8}(1-g_c^2)=\frac{k_1k_2}{8 p^2} = \frac{k_1k_2}{4 (k_1+k_2)}\, .
\ee
Hence the Maxwell charge \eq{MaxCh1}-\eq{MaxCh2}, computed in the UV of the near-horizon solution, is always an (even) integer, see eq.~(\ref{eq:k1k2quant}) 
obtained from the 
worldsheet analysis.

When $0<g_c<1$ the flux profiles have differing UV asymptotics for the near horizon and asymptotically Ricci-flat solutions, in addition they are not monotonic. 
We display the flux profile for $g_c=\frac{1}{3}$ below:

\begin{figure}[bth!]
\centering
\begin{minipage}{.5\textwidth}
  \centering
  \includegraphics[width=.9\linewidth]{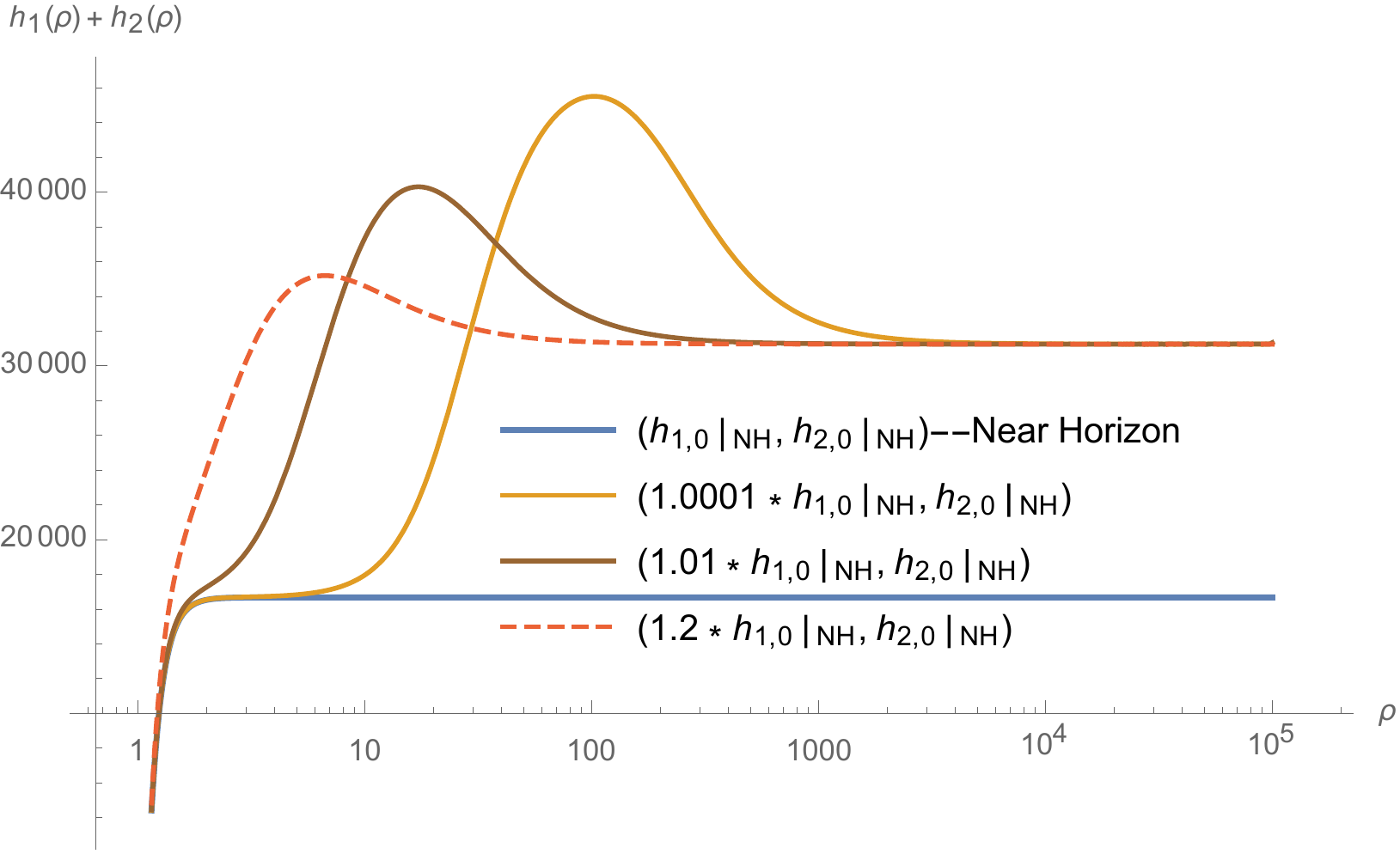}
\end{minipage}
\begin{minipage}{.5\textwidth}
  \centering
  \includegraphics[width=.9\linewidth]{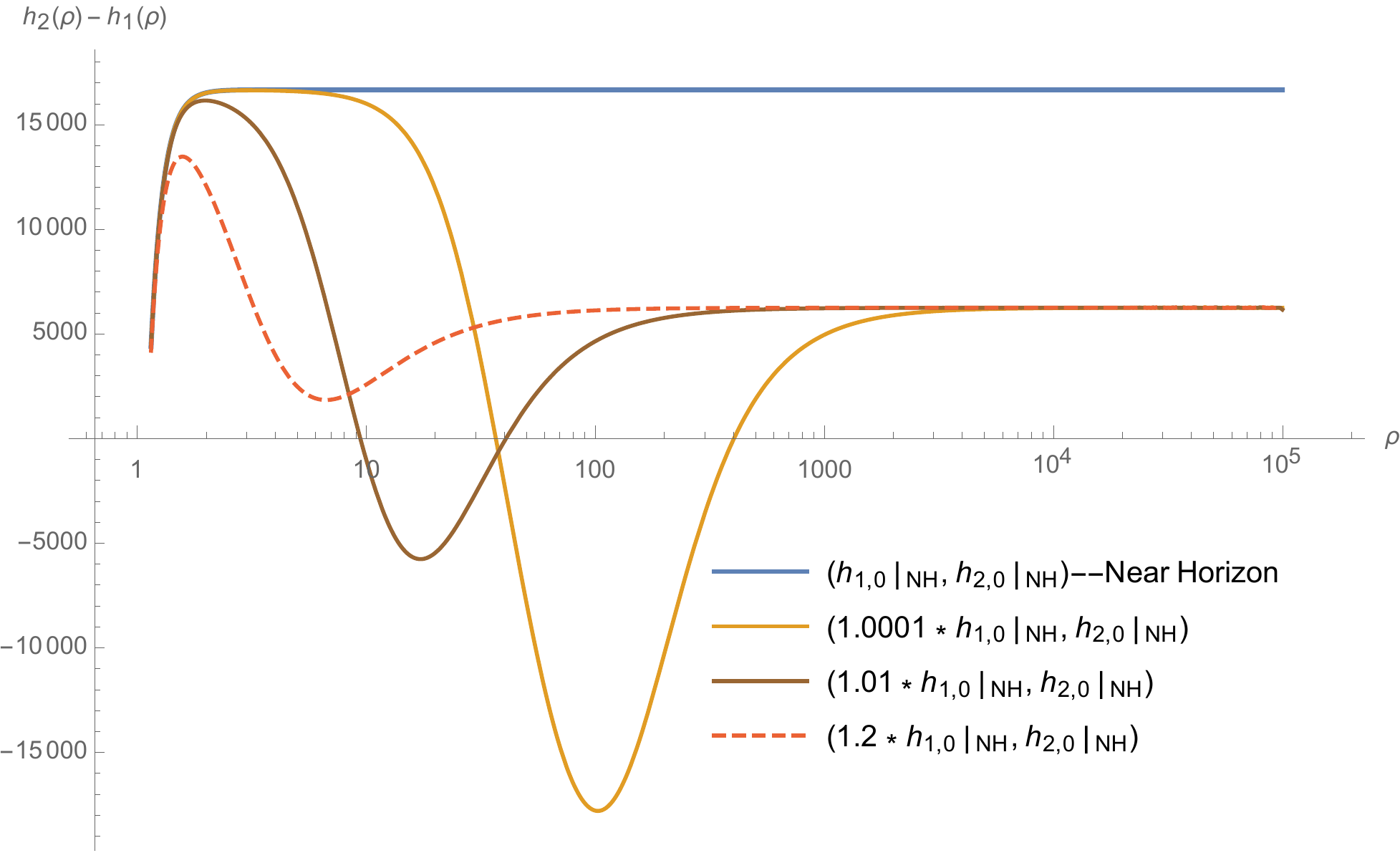}
\end{minipage}
  \caption{The flux functions $h_2(\rho)\pm h_1(\rho)$ with $ g_c=\frac{1}{3}$}
\end{figure}

%%%%%%%%%%%%%%%%%%%%%%%%%%%%%%%%%%%%
\subsubsection{Page Charge}
%%%%%%%%%%%%%%%%%%%%%%%%%%%%%%%%%%%%

The Page charge is the best signal of the presence of brane sources, although away from the blow down limit of our solutions, the 
fluxes are regular everywhere and thus we expect the Page charge to vanish. 

To compute the Page charge one must take care when choosing a gauge for the Maxwell field \eq{Fdef}, we find that two patches are sufficient to cover $T^{1,1}$:
\begin{subequations}
\begin{align}
A_{(1)}&= \frac{1}{4}\bslb \di \psi + \cos\tha_1 \di \phi_1- (-1+\cos\tha_2) \di \phi_2+ g_1\eta\bsrb \bp 
\cdot \bH + \frac{1}{4} g_2 \eta\, \bq\cdot \bH \,, \\
A_{(2)}&= \frac{1}{4}\bslb  -d\psi + (-1+\cos\tha_1) \di \phi_1- \cos\tha_2 \di \phi_2+  
g_1\eta \bsrb \bp \cdot \bH + \frac{1}{4} g_2 \eta\, \bq\cdot \bH \,.
\end{align}
\end{subequations}
The gauge field $A_{(1)}$ is well defined on all of $S^3_1$ and likewise $A_{(2)}$ is well defined on all of $S^3_2$ and away from $\tha_{i}=0,\pi$, 
they are related by a large gauge transformation. From this we compute
\begin{subequations}
\begin{align}
\Tr A_{(1)} \w F|_{\tha_2=0} =& -\frac{1}{8} \bslb p^2 (1+g_1)^2 + q^2 g_2^2 \bsrb (\di \psi+\cos \tha_1 \di \phi_1)\w \Om_1  \\
\Tr A_{(2)} \w F|_{\tha_1=0} =& -\frac{1}{8} \bslb p^2 (-1+ g_1)^2 + q^2 g_2^2 \bsrb (\di \psi+\cos \tha_2 \di \phi_2)\w \Om_2
\end{align}
\end{subequations}
and so 
\begin{subequations}
\begin{multline}
\cQ_{P,1}=\frac{1}{2\pi^2 \al'}\int_{S^3_1/\ZZ_n} \blp \cH_{(3)}- \Tr \, A_{(1)}\w F\brp 
\\= \frac{2}{n}\Bslb 4(h_1+h_2)+ \frac{1}{2} \bslb p^2 (1+g_1)^2 + q^2 g_2^2 \bsrb  \Bsrb   
= 2(1+g_c)\frac{p^2}{n}  
\end{multline}
\begin{multline}
\cQ_{P,2}=\frac{1}{2\pi^2 \al'}\int_{S^3_2/\ZZ_n} \blp \cH_{(3)}- \Tr\, A_{(2)}\w F\brp\\ 
=\frac{2}{n} \Bslb 4(h_1-h_2) - \frac{1}{2} \bslb p^2 (-1+g_1)^2 + q^2 g_2^2 \bsrb \Bsrb   = 2(-1+g_c)\frac{p^2}{n} 
\end{multline}
\end{subequations}
where we have used the Bianchi identity~\eq{Bianchi1} and~\eq{g1Sol}. Looking at the worldsheet analysis, we 
see from~\eq{k1k2p} and~\eq{k1k2gc} that
\bea
\cQ_{P,1}= \frac{2k_1}{n}\,,\qquad \cQ_{P,2}= -\frac{2k_2}{n}\,.
\eea
Thus integrality of the Page charge \cite{Rohm:1985jv} is consistent with the requirement that $k_1,k_2\in \frac{n}{2}\ZZ$ which we have also found from the worldsheet. 

Since $S_1^3$ and $S_2^3$ are in the same homology class, they must be the boundary of a four-chain $M_4$:
\be
\del (M_4) = S_1^3- S_2^3\,.
\ee
However we cannot use Stokes theorem to integrate the Page current 
\be 
*j_{P}=d\bslb \cH_{(3)} -\al' CS(A)\bsrb
\ee
on the four cycle since $CS(A)$ is not well defined on the four cycle. One can use Stokes theorem on each of the two patches and then 
subtract off the integral of the current on the overlap. This integral on the overlap reduces to 
\be
\int_{S^2\times S^1} \Tr \blp A_{(1)}-A_{(2)}\brp \w F = -\frac{4p^2}{n}
\ee
where the $S^2\times S^1$ is located at $\tha_1=\tha_2\,,\phi_1=-\phi_2$. In this manner we find that 
\be
\int_{M_4} * j_{P}=0\,.
\ee

This analysis seems sufficient to demonstrate that in the regular solutions there is no five brane Page charge. We will comment on the singular solution in the next section.

The source-free Klebanov-Strassler solution also has a vanishing Page charge but adding BPS D3-branes to the background is the crucial step 
in obtaining the supergravity dual of the Mesonic branch of the gauge theory. It would be interesting to understand the mechanism for adding source five-branes in this heterotic background.

%%%%%%%%%%%%%%%%%%%%%%%%%%%%%%%%%%%%
\subsubsection{Singular Solution}
%%%%%%%%%%%%%%%%%%%%%%%%%%%%%%%%%%%%

The intuition one has from the gauged CHS solution is that zero size instantons are given by explicit five-brane sources. Indeed in the local heterotic 
solutions of \cite{Carlevaro:2008qf} the blow down limit is precisely the $\ZZ_2$ orbifold of the zero size CHS solution and thus corresponds to explicit five-brane sources. 
When the blow up parameter of our solutions is taken to zero size (see section \ref{sec:nearhorizonSingular}) the dilaton and thus the Einstein frame metric is 
singular at $r=0$. In this blow-down limit the flux function $h_2(r)$ becomes a {\it constant} whereas for the regular solution the flux function vanishes at 
the bolt. We interpret this as implying that the contribution to the Maxwell charge in the blow-down limit comes from explicit brane sources i.e. the RHS of 
the Bianchi identity has an explicit $\delta$-function source.

%%%%%%%%%%%%%%%%%%%%%%%%%%%%%%%%%%%%
\subsection{Discrete Wilson Lines}\label{sec:WilsonLines}
%%%%%%%%%%%%%%%%%%%%%%%%%%%%%%%%%%%%
There is another gauge transformation which must be taken into account, the shift of the gauge potential by a flat connection. Since
\be
H_1(S^3/\ZZ_2,\ZZ)=H^2(S^3/\ZZ_2,\ZZ) =\ZZ_2
\ee
there exists discrete Wilson lines and as reviewed nicely in \cite{Gukov:2003cy}, their Chern-Simons invariants may be non-zero.  Reducible gauge connections are of the form 
\be
\label{eq:abWL}
U=\diag ( e^{\pi i \ell_1},\ldots ,e^{\pi i \ell_{16}} )\,.
\ee
and the Chern-Simons invariant for such a Wilson line is \cite{Witten:1985mj, Conrad:2000tk}
\be
\int_{S^3/\ZZ_2}CS(A)= \sum_{i=1}^{16} \frac{\ell_i^2}{4} \quad {\rm mod} \quad \ZZ\,.
\ee
More generally we quote the result from \cite{Nishi1998} (reviewed in \cite{Gukov:2003cy}) for an irreducible gauge connection on $S^3/\ZZ_2$ embedded into $SU(2)$, the answer being
\be
\int_{S^3/\ZZ_2}CS(A)= -\frac{m^2}{8} -\frac{\lam^2}{2} \quad {\rm mod}\quad \ZZ
\ee 
where $m\in \ZZ$ and $\lam\in\{0,\frac{1}{2}\}$.

 It is certainly possible to have the commutant of our gauge field from \eq{Fdef} in $SO(32)$ or $E_8\times E_8$ to be non-Abelian and contain an $SU(2)$. A Wilson line 
perpendicular to both $\bf{p}$ and $\bf{q}$ will thus give a constant shift of the Page charges proportional to the Chern-Simons invariant evaluated on the flat
 connection. Wilson lines with a component parallel to $\bf{p}$ or $\bf{q}$ seem to give a Page charge which is not independent of the radial coordinate, 
we interpret this as indicating that only Wilson lines perpendicular to both $\bf{p}$ and $\bf{q}$ can be activated. 

It may be possible to confirm this --~presumably in a 
weaker form~--in the near-horizon limit, using the worldsheet models presented in section~\ref{sec:worldsheet}. Indeed in that context an 
Abelian Wilson line as~(\ref{eq:abWL}) corresponds to an embedding of the $\mathbb{Z}_2$ orbifold of $T^{1,1}$ into the gauge lattice which should be compatible with modular invariance 
of the one-loop partition function.

\vskip 2cm
%%%%%%%%%%%%%%%%%%%%%%%%%%%%%%%%%%%%
\noindent {\bf Acknowledgements} We would like to thank Davide Cassani, Sheer El-Showk, Chris Hull and Ruben Minasian for useful conversations.

This work was conducted within the ILP LABEX (ANR-10-LABX-63) supported by French state funds managed by the 
ANR within the {\it Investissements d'Avenir} program (ANR-11-IDEX-0004-02) and by the project QHNS in the program ANR Blanc 
SIMI5 of Agence National de la Recherche. 
%%%%%%%%%%%%%%%%%%%%%%%%%%%%%%%%%%%%
\newpage
%%%%%%%%%%%%%%%%%%%%%%%%%%%%%%%%%%%%
\begin{appendix}
%%%%%%%%%%%%%%%%%%%%%%%%%%%%%%%%%%%%

%%%%%%%%%%%%%%%%%%%%%%%%%%%%%%%%%%%%
\section{The Conifold}\label{app:conifold}
%%%%%%%%%%%%%%%%%%%%%%%%%%%%%%%%%%%%

We include here some well known information on the conifold \cite{Candelas:1989js}, if only to establish notation. The 
singular conifold is a cone over $T^{1,1}$; the Ricci-flat metric, with $SU(2)\times SU(2)\times U(1)$ isometry group, is 
given by:
\be
\di s^2= \di r^2 + r^2 \Bslb \frac{1}{6}\blp \sig_1^2 +\sig_2^2 +  \hsig_1^2 +\hsig_2^2 \brp+ \frac{1}{9}\, \eta^2    \Bsrb \, ,
\ee
where
\be
\eta=\sig_3 + \hsig_3=\di \psi + \cos \tha_1 \di \phi_1 + \cos \tha_2 \di \phi_2 \, ,
\ee
and the left invariant one forms $\{\sig_i,\hsig_i\}$ are defined below in \ref{sec:InvForms}. We note that 
\be 
\di \eta= -(\Om_1+\Om_2) \, ,
\ee
where we also use the two-forms
\be
\Om_1=-\sig_1\w \sig_2=\sin \tha_1 \di \tha_1 \w \di \phi_1\,,\qquad \Om_2 = -\hsig_1\w \hsig_2=\sin \tha_2 \di \tha_2 \w \di \phi_2\,.
\ee

The five-manifold $T^{1,1}$ is diffeomorphic to $S^3\times S^2$. 
Two representatrepresentativesives of $H_3(T^{1,1},\ZZ)$ which we will find useful are given by
\bea
S_1^3:&& \tha_1=\tha_{1,0}\,,\ \phi_1=\phi_{1,0}\,, \qquad S_2^3:\quad  \tha_2=\tha_{2,0}\,,\ \phi_2=\phi_{2,0}\,, 
\eea
where $\tha_{i,0}$ and $\phi_{i,0}$ are constants. A representative 2-cycle in $H_{2}(T^{1,1},\ZZ)$ is given by
\be
S^2:\quad \psi=0\,,\quad \tha_1=\tha_2\,,\quad \phi_1=-\phi_2
\ee
but of course one is free to take any other representatives. 

The volume form on $S^3$ is 
\\
\be
\om_3=\half \eta\w (\Om_1-\Om_2)\,,\qquad \int_{S^3} \om_3 = 8 \pi^2
\ee
while the volume form on $S^2$ is 
\be
\om_2=\frac{1}{2}(\Om_1-\Om_2)\,,\qquad \int_{S^2} \om_2 = 2\pi\,.
\ee
The range of the angles on $T^{1,1}$ is
\be
0\leqslant \psi < 4\pi\,,\quad 0\leqslant \tha_1,\tha_2 \leqslant \pi\,,\quad 0\leqslant \phi_1,\phi_2  < 2\pi\,.
\ee

In this paper we study the $\ZZ_2$ orbifold which shrinks the Hopf fiber of the $S^3$ by a factor of two such that:\footnote{One might like to recall that this $S^3/\ZZ_2$ is diffeomorphic to $\RR\PP^3$ and is also called the Lens space $L(2,1)$}
\be
0\leqslant \psi < 2\pi\,.
\ee

%%%%%%%%%%%%%%%%%%%%%%%%%%%%%%%%%%%%
\subsection{Invariant Forms} \label{sec:InvForms}
%%%%%%%%%%%%%%%%%%%%%%%%%%%%%%%%%%%%
The left invariant one-forms on $T^{1,1}$ are given by
\begin{subequations}
\begin{align}
\sig_i &=-i\, \Tr \bslb \tau_i \, g^{-1} \di g\bsrb \,,\qquad g= e^{\frac{i}{2} \phi_1 \tau_3}e^{\frac{i}{2} \tha_1 \tau_1}e^{\frac{i}{4} \psi \tau_3} \\
\hsig_i &=-i\, \Tr \bslb \tau_i\,  \hg^{-1} d\hg\bsrb \,,\qquad \hg= e^{\frac{i}{2} \phi_2 \tau_3}e^{\frac{i}{2} \tha_2 \tau_1}e^{\frac{i}{4} \psi \tau_3} \, ,
\end{align}
\end{subequations}
where $\{ \tau_i , \, i=1,2,3\}$ are the usual Pauli matrices and $0\leqslant \psi < 4\pi$. 
Note that we have used a common third Euler angle $\psi$ for both sets of one forms. Explicitly we have (and similarly for $\hsig_i$)
\begin{subequations}
\begin{align}
\sig_1 &= \cos \frac{\psi}{2} \di \tha_1 + \sin\frac{\psi}{2} \sin \tha_1 \di \phi_1 \\
\sig_2 &= \sin \frac{\psi}{2} \di \tha_1 - \cos\frac{\psi}{2} \sin \tha_1 \di \phi_1  \\
\sig_3&= \frac{1}{2}\di \psi + \cos \tha_1 \di \phi_1  \\
\sig_1\pm i\sig_2 &= e^{\pm i\psi/2} (\di \tha_1 \mp i \sin \tha_1 \di \phi_1) \, ,
\end{align}
\end{subequations}
which satisfy
\be
\di \sig_1 = \sig_2 \w \sig_3 \, ,
\ee
and cyclic permutations thereof.

%%%%%%%%%%%%%%%%%%%%%%%%%%%%%%%%%%%%
\section{$SU(3)$-Torsion Classes}
%%%%%%%%%%%%%%%%%%%%%%%%%%%%%%%%%%%%
Six-dimensional supersymmetric heterotic compactifications to four-dimensional Minkowski space
\bea
M_{1,3}\times X
\eea
give rise to $SU(3)$-structure manifolds $(X,J,\Omega)$ with a holomorphic bundle satisfying the BPS-conditions \eqref{dJeq}-\eqref{Hdefeq}. The fundamental forms satisfy the relations
\bea
\Omega\wedge J&=&0 \\
-\frac{i}{8}\Om\w \Ombar &=& \frac{1}{3!} J\w J \w J\:.
\eea

\subsection{Intrinsic Torsion}
The intrinsic torsion $T$ of a generic $SU(3)$-structure takes values in\footnote{See e.g \cite{Gauntlett:2003cy} for an overview of $G$-structures and intrinsic torsion.}
\bea
T\in\Lambda^1(X)\otimes\mathbf{su}(3)^\perp&=&W_1\oplus W_2\oplus W_3\oplus W_4\oplus W_5 \\
(\mathbf{3}+\mathbf{\bar 3})\otimes(\mathbf{1}+\mathbf{3}+\mathbf{\bar 3})&=&(\mathbf{1}+\mathbf{1})\oplus(\mathbf{8}+\mathbf{8})\oplus(\mathbf{6}+\mathbf{\bar 6})\oplus(\mathbf{3}+\mathbf{\bar 3})\oplus(\mathbf{3}+\mathbf{\bar 3})\:,
\eea
where $\mathbf{su}(3)^\perp\oplus\mathbf{su}(3)=\mathbf{spin}(6)$. The torsion classes $\{W_i\}$ are given by
\bea
{\rm d} J&=&-\frac{3}{2}{\rm Im}(W_1^{(\mathbf{1})}\bar\Omega)+W_4^{(\mathbf{3}+\mathbf{\bar 3})}\wedge J+W_3^{(\mathbf{6}+\mathbf{\bar 6})}\\
{\rm d} \Omega&=& W_1^{(\mathbf{1})}J\wedge J+W_2^{(\mathbf{8})}\wedge J+W_5^{(\mathbf{\bar 3})}\wedge\Omega\:.
\eea
Note that the Calabi-Yau condition is given by ${\rm d} J={\rm d}\Omega=0$, i.e. all torsion classes vanish. In terms of supersymmetric heterotic compactifications, the torsion classes are given by
\bea
W_1&=&W_2=0\\
2W_4^{(\mathbf{3}+\mathbf{\bar 3})}&=&W_5^{(\mathbf{3}+\mathbf{\bar 3})}=2{\rm d}\Phi\:.
\eea
In particular, the torsion class $W_3^{(\mathbf{6}+\mathbf{\bar 6})}$ is given by
\bea
W_3^{(\mathbf{6}+\mathbf{\bar 6})}=e^{\Phi}{\rm d}\left(e^{-\Phi}J\right)\:.
\eea
We can compute $W_3$ explicitly in terms of the frame ansatz. We find
\bea
W_3&=&\frac{r}{4}{\rm d}r\wedge{\rm d}\theta_2\wedge{\rm d}\phi_2\, \sin(\theta_2)\left(-2 H + 2 H_1 - 2 H_2 +r H_1' - rH_2' - r\left(H_1 - H_2\right)\Phi'\right)\\
&+& \frac{r}{4}{\rm d}r\wedge{\rm d}\theta_1\wedge{\rm d}\phi_1\, \sin(\theta_1) \left(-2 H + 2 H_1 + 2 H_2 + r H_1' + r H_2' -r\left(H_1 + H_2\right) \Phi'\right)\:.
\eea

If we want our space to be conformally K\"ahler, we need $W_3$ to vanish. Taking the sum and difference between the terms in parenthesis, we find that we need to require
\bea
2H - r\,H_1' - H_1\left(2 - r\,\Phi'\right)&=&0\\
 r\, H_2' + H_2\left(2 - r\,\Phi'\right)&=&0\:.
\eea
By an appropriate choice of radial coordinate given by $H$, we see that we can satisfy the first equation. The second equation will in general impose an extra constraint on the solution, which seems hard to solve in general. In particular, for the Near Horizon solution this reduces to the constraint that $g_c=0$, which implies that $H_2=0$.

We can compute the torsion class for the radial coordinate choice given by \eqref{HRep}. We find

\bea
W_3=\frac{r \left( H_1^2 (2 H_2 + r H_2')-2 H_2^3 - r H_1 H_2 H_1'\right)}{4(H_1^3-H_1 H_2^2)}&\times&\Big((H_1+ H_2) {\rm d}r\wedge{\rm d}\theta_1\wedge{\rm d}\phi_1\sin(\theta_1)\notag\\ &-& (H_1-H_2) {\rm d}r\wedge{\rm d}\theta_2\wedge{\rm d}\phi_2\sin(\theta_2)\Big)\:.
\eea
It follows that in order to have a conformally K\"ahler space, we should require the additional equation 
\be
H_1^2 (2 H_2 + r H_2')-2 H_2^3 - r H_1 H_2 H_1'=0\:,
\ee
again potentially over-constraining the system. In particular, the Near Horizon solution gives
\be
W_3=\frac{p^2\,g_c}{2r}\Big((1 + g_c) {\rm d}r\wedge{\rm d}\theta_1\wedge{\rm d}\phi_1\sin(\theta_1) - (1 - g_c) {\rm d}r\wedge{\rm d}\theta_2\wedge{\rm d}\phi_2\sin(\theta_2)\Big)\:,
\ee
which vanishes iff $p^2\,g_c=0$. This corresponds to the Near Horizon solution of \cite{Carlevaro:2009jx}, where $H_2=0$.

%%%%%%%%%%%%%%%%%%%%%%%%%%%%%%%%%%%%
\section{Supergravity on the Conifold}\label{sugraconifold}
%%%%%%%%%%%%%%%%%%%%%%%%%%%%%%%%%%%%

In this appendix we review solutions of Type I supergravity on the deformed conifold and type IIA supergravity on the resolved conifold, both with non-trivial flux profiles. Our rationale for including a review of these solutions is to point out a comparison to our Hetorotic solutions on $T^{11}/\ZZ_2$ found in section \ref{sec:T11Z2}. The existence of a decoupled near horizon region with an closed form analytic solution is common to all these solutions as is the additional parameter corresponding to the stringy size of the appropriate cycle.

%%%%%%%%%%%%%%%%%%%%%%%%%%%%%%%%%%%%
\subsection{$\cN=1$ Supergravity on the Deformed Conifold}\label{sec:CVMN}
%%%%%%%%%%%%%%%%%%%%%%%%%%%%%%%%%%%%

%%%%%%%%%%%%%%%%%%%%%%%%%%%%%%%%%%%%
\subsubsection{The Regular Solution}

There is a family of regular solutions of ten dimensional $\cN=1$ supergravity on the deformed 
conifold~\cite{Chamseddine:1997nm, Chamseddine:1997mc, Maldacena:2000yy, Butti:2004pk, Casero:2006pt} 
which was nicely reviewed and elaborated on in~\cite{Maldacena:2009mw}, some of which we reproduce here. The solution is 
\begin{subequations}
\begin{align}
\di s^2_{{\rm str}}=&\di  s_{1,3}^2 + \frac{\al' M }{4} \di s_6^2 \non \\
\di s_6^2=& c' \big(\di r^2 + (\sig_3+ \hsig_3)^2\big)  + \frac{c}{\tanh r} \big( \sig_1^2+\sig_2^2+\hsig_1^2+ \hsig_2^2 \big) + 
\frac{2c}{\sinh r} \big(\sig_1\hsig_1+ \sig_2 \hsig_2 \big) \non \\
& +\Big[ \frac{r}{\tanh r}-1\Big] ( \sig_1^2+\sig_2^2-\hsig_1^2- \hsig_2^2)   \\
H_{(3)}=& \frac{\al' M }{4} \Big\{ (\sig_3+\hsig_3) \w \bslb \sig_1\w \sig_2 + \hsig_1\w \hsig_2 + 
\frac{r}{\sinh r} \sig_1\w \hsig_2 + \hsig_1 \w \sig_2\bsrb  \non \\
& + \frac{r\coth r-1}{\sinh r} dr \w (\sig_1\w \hsig_1+\sig_2 \w \hsig_2)\Big\}  \\
e^{2(\Phi-\Phi_0)} =& \frac{f^{1/2} c'}{\sinh^2 t} 
\end{align}
\end{subequations}
where the two functions $\{f,c\}$ satisfy
\begin{subequations}
\begin{align}
f'&= 4 \sinh^2r \,c  \\
c'&= \frac{1}{f} \bslb  c^2 \sinh^2 r -(r \cosh r-\sinh r)^2\bsrb\,.
\end{align}
\end{subequations}

As explained in~\cite{Maldacena:2009mw}, there is a two-parameter family of asymptotically Ricci-flat solutions where the two parameters are the dilaton zero mode $\Phi_0$ and another parameter $\gam\geqslant 1$ which can be thought of as the IR size of the finite three-cycle. Generically this family of solutions approaches the Ricci-flat conical metric of~\cite{Candelas:1989js} at large $r$ with constant dilaton and should be thought of analogous to our numerical solutions in section~\ref{sec:FullBraneSolution}. When $\gam=1$ the large $r$ asymptotics cross over to be non-Ricci flat and with a divergent dilaton and is analogous to our 
heterotic solution in section~\ref{sec:nearhorizon}.

Explicitly, the expansion of the functions $c(r)$ and $f(r)$ in the IR, $i.e.$ close to $r=0$, is given in terms 
of the parameter $\gamma$ as follows:
\begin{subequations}
\begin{align}
c(r) &= \gam^2 r + \frac{1-\gam^4}{15\gam^2} r^3 + \cO(r^5)\,, \label{cIR} \\
f (r) &= \gam^2 r^4 + \frac{2(6\gam^4-1)}{45\gam^4} r^6  + \cO(r^8) \label{fIR}\,.
\end{align}
\end{subequations}
When $\gam=1$ the expansion \eq{cIR} of $c(r)$ truncates and the exact solution is that found in four dimensional gauged supergravity by Chamsedine-Volkov~\cite{Chamseddine:1997nm, Chamseddine:1997mc} and uplifted to a ten dimensional solution by 
Maldacena-Nunez~\cite{Maldacena:2000yy}:
\begin{subequations}
\begin{align}
c(r) &= r \,,  \\
f (r) &= r^2 \sinh^2 r  -(r \cosh r-\sinh r)^2\,.
\end{align}
\end{subequations}
This particular solution is not asymptotically Ricci-flat, in particular the dilaton diverges at large $r$. For any representative of the family discussed above, this CV-MN solution corresponds to an approximation to the near horizon region. It is a non-trivial fact that this near horizon region decouples and exists as an independent exact solution. 

%%%%%%%%%%%%%%%%%%%%%%%%%%%%%%%%%%%%
\subsubsection{The Singular Solution}

We can obtain a seemingly singular limit of the CV-MN solution by taking the large $r$ limit, the resulting solution is itself an exact solution. We  find
\begin{subequations}
\begin{align}
c(r)&=r  \\
f(r) &= \frac{r}{2}  e^{2r}  
\end{align}
\end{subequations}
and
\begin{subequations}
\begin{align}
\di s_6^2&= \di r^2 + (\sig_3+ \hsig_3)^2  + 2r( \sig_1^2+\sig_2^2) + (\hsig_1^2+ \hsig_2^2)   \\
H_{(3)}&= \frac{\al' M }{4}  (\sig_3+\hsig_3) \w \bslb \sig_1\w \sig_2 + \hsig_1\w \hsig_2 \bsrb  \\
e^{2(\Phi-\Phi_0)} &= r^{1/2} e^{-r}\, .
\end{align}
\end{subequations}

This is the exact but singular solution of the BPS equations found in~\cite{Maldacena:2000yy} as a precursor to the regular solution, by analogy to our singular solution in~\ref{sec:nearhorizonSingular} we interpret this as the solution with where the size of the three sphere is taken to zero. It was noticed in~\cite{Maldacena:2000yy} that this does not satisfy the critera of~\cite{Gubser:2000nd} for a good singularity whereas our analogous heterotic solution in section~\ref{sec:nearhorizonSingular} has a description in terms of an exactly solvable worldsheet conformal field theory and is thus regular in string theory. It would of course be interesting to provide a worldsheet description of the CV-MN solution and discover that this singular limit is regular in string theory.

%%%%%%%%%%%%%%%%%%%%%%%%%%%%%%%%%%%%
\boldmath
\subsection{$G_2$ Holonomy Manifolds}\label{sec:G2Holonomy}
\unboldmath
%%%%%%%%%%%%%%%%%%%%%%%%%%%%%%%%%%%%

A two parameter family of $G_2$ holonomy metrics  with $SU(2)^2\times U(1)$ symmetry was found in~\cite{Brandhuber:2001kq}:
\begin{equation}
\di s_7^2  =E^2 \di r^2+ A^2 \bslb \sig_1^2 + \sig_2^2 \bsrb + B^2 \sig_3^2 + C^2 \bslb (\hsig_1-f \sig_1)^2 + (\hsig_2-f \sig_2)^2\bsrb + D^2 (\hsig_3-g \sig_3)^2  \label{BrandhuberMetric}
\end{equation}
with
\begin{align}
A^2 &= \frac{a'}{4a} (4 a^2 (b+ r_0^3)-b^3)/\Om  \non \\
B^2&= \frac{b b'}{a'} A^2\,, \qquad C^2 = \frac{aba'}{\Om} \non \\
D^2&=  \frac{a^2 b'}{\Om} \,,\qquad E^2 = \frac{(a')^2 b'}{\Om } \non \\
f&= \frac{b}{2a} \,,\qquad g=1-2f^2 \non \\
\Om^3 &= \frac{bb' (a')^2\blp 4 a^2 (b+ r_0^3)-b^3 \brp}{4}\,.
\end{align}
There is one second order differential eqation for the two functions $(a,b)$
\be
0=4 a' b' \bslb ab(b+r_0^3)a' +(b^3 -a^2 (a^2 (r_0^3+2b)))b'\bsrb + b\blp b^3 -4a^2(r_0^3 +b) \brp (a' b''-a''b')\,.\label{BrandhuberEq}
\ee
and in addition one of $(a,b)$ must still be fixed using reparameterization of the radial co-ordinate. 

A particular simple solution to~\eq{BrandhuberEq} is $a=b$, and gives a one parameter family with enhanced $SU(2)^3$ 
symmetry~\cite{BryantSalamon, Gibbons:1989er}, which we refer to as the GPP/BS solution. Conventionally one takes
\be
a=\frac{4}{3}(r^3-r_0^3)
\ee
giving
\be
\di s_7^2 = 12 \frac{\di r^2}{1-\frac{r_0^3}{r^3}} + r^2 \sig^a \sig^a + \frac{r^2}{3} \Blp 1- \frac{r_0^3}{r^3}\Brp (\sig^a + \hsig^a)^2\,.\label{GPPBSmetric}
\ee
The one real parameter is $r_0$. By rescaling $r\ra r_0 r$ we can normalize the size of the finite three-sphere to unity. In addition by setting $r_0=0$ we obtain a singular conical metric but it is understood that M-theory on this singularity is 
well-defined~\cite{Atiyah:2001qf, Acharya:2004qe}.

To see the more general two parameter family of \cite{Brandhuber:2001kq}, it is convenient to choose a slightly different radial coordinate, one then has an expansion at small $r$ given by
\begin{subequations}
\begin{align}
a&= r^3 \\
b&= (1-y) r^3 -\frac{1}{3r_0^3} (2y-5y^2 +4y^3-y^4) r^6 + \cO(r^9)\,.
\end{align}
\end{subequations}
with $0\geq y \geq1$\,. One interesting feature of this family is that with $y=0$, the expansion truncates at first order and 
one has the GPP/BS solution \eq{GPPBSmetric} in a different radial coordinate. Thus at $y=0$ 
one has an asymptotically conical metric while for $0<y\leqslant 1$ one finds very 
different asymptotics, namely that the M-theory circle stabilizes at finite size. 
 
%%%%%%%%%%%%%%%%%%%%%%%%%%%%%%%%%%%%
\subsection{IIA Supergravity with $F_2$ flux and Dilaton}
%%%%%%%%%%%%%%%%%%%%%%%%%%%%%%%%%%%%

One can reduce any $G_2$ holonomy manifold with a freely acting $U(1)$ isometry to 
IIA supergravity with $F_2$ flux and dilaton. In string frame~\eq{BrandhuberMetric} gives:
\begin{align}
\di s_{{\rm str}}^2= &\frac{\blp B^2 + (1-g)^2 D^2\brp^{1/2}}{2} \Bigg\{ \di s_{1,3}^2 +  
E^2 \di r^2+ A^2 \left(\sig_1^2 + \sig_2^2 \right)  \non \\&
+ C^2 \bslb (\hsig_1-f \sig_1)^2 + (\hsig_2-f \sig_2)^2\bsrb + \frac{B^2D^2}{B^2 + (1-g)^2 D^2} (\hsig_3- \sig_3)^2 \Bigg\}  \non \\
F_{(2)}=& \sig_2\w \sig_3 + \hsig_2\w \hsig_3 + \di \left[ \frac{B^2-(1-g^2)D^2}{B^2+ (1-g)^2D^2}\blp \sig_3+\hsig_3\brp \right]  \\
e^{\Phi-\Phi_0}=& \frac{1}{2^{3/2}} \bslb B^2 +(1-g)^2 D^2\bsrb^{3/4}\, .
\end{align}

The IIA solution arising from dimensional reduction of the GPP/BS metric with $r_0$ normalized to unity is
\begin{align}
\di s_{{\rm str}}^2 =& \frac{1}{2} \Bigg\{ \left(\frac{4  r^3 - 1}{3r}\right)^{1/2} \Bslb \di s_{1,3}^2 
+  \frac{12r^3}{r^3-1}  \di r^2+ r^2 \left(\sig_1^2 + \sig_2^2 \right)   \non \\
& + \frac{16(r^3-1)^2}{9r^2} \left[ (\hsig_1-\tfrac{1}{2} \sig_1)^2 + (\hsig_2-\tfrac{1}{2} \sig_2)^2\right] 
+ \frac{4r^2 (r^3-1)}{4r^3-1} (\hsig_3- \sig_3)^2 \Bigg\}  \non \\
F_{(2)}=& \sig_2\w \sig_3 + \hsig_2\w \hsig_3 + \di \Bslb\frac{3 }{4r^3-1}\blp \sig_3+\hsig_3\brp\Bsrb  \\
e^{\Phi-\Phi_0}=& \frac{1}{2^{3/2}} \bslb \frac{4 r^3-1}{3r}\bsrb^{3/4}\, .
\end{align}
This has a divergent dilaton at large $r$, a direct  consequence of the fact that this $G_2$ holonomy metric is asymptotically conical. We interpret this solution as the near horizon limit of backreacted wrapped D6-branes, in much the same way as the CV-MN solution of 
appendix~\ref{sec:CVMN} has a divergent dilaton and is the near horizon limit of backreacted wrapped NS5 branes.

The more general solutions with reduced symmetry have an asymptotically constant dilaton much like the more general solutions 
of appendix~\ref{sec:CVMN}. One may consider these two families of solution to be mirror duals~\cite{Halmagyi:2010st}. The reason we have reviewed this material is to highlight the features which are common to these type II flux solutions and the heterotic torsional solutions of section~\ref{sec:T11Z2}.

%%%%%%%%%%%%%%%%%%%%%%%%%%%%%%%%%%%%
\section{Numerical Evaluation of the Bianchi Identity}\label{app:NumericalBianchi}
%%%%%%%%%%%%%%%%%%%%%%%%%%%%%%%%%%%%
In section \ref{sec:suppression} we promised further evidence for the supression of the $\Tr R_+\w R_+$ in the Bianchi identity. In lieu of a slick argument for for the scaling of $\Tr R_+\w R_+$ with the charges $\{p^2,q^2\}$ we resort to numerically evaluating this term on shell. For the various choices of parameters presented in section \ref{sec:Numerical} our results are as follows:

\begin{figure}[bth!]
\centering
\begin{minipage}{.5\textwidth}
  \centering
  \includegraphics[width=.9\linewidth]{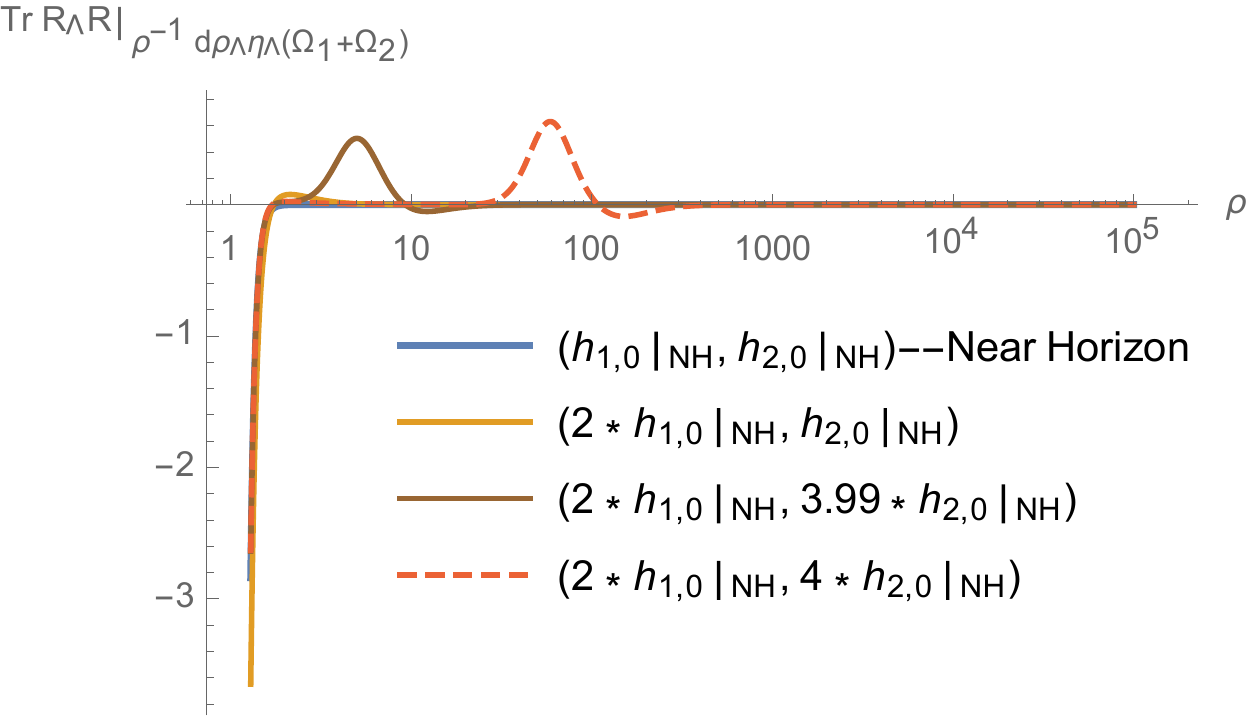}
\end{minipage}%
\begin{minipage}{.5\textwidth}
  \centering
  \includegraphics[width=.9\linewidth]{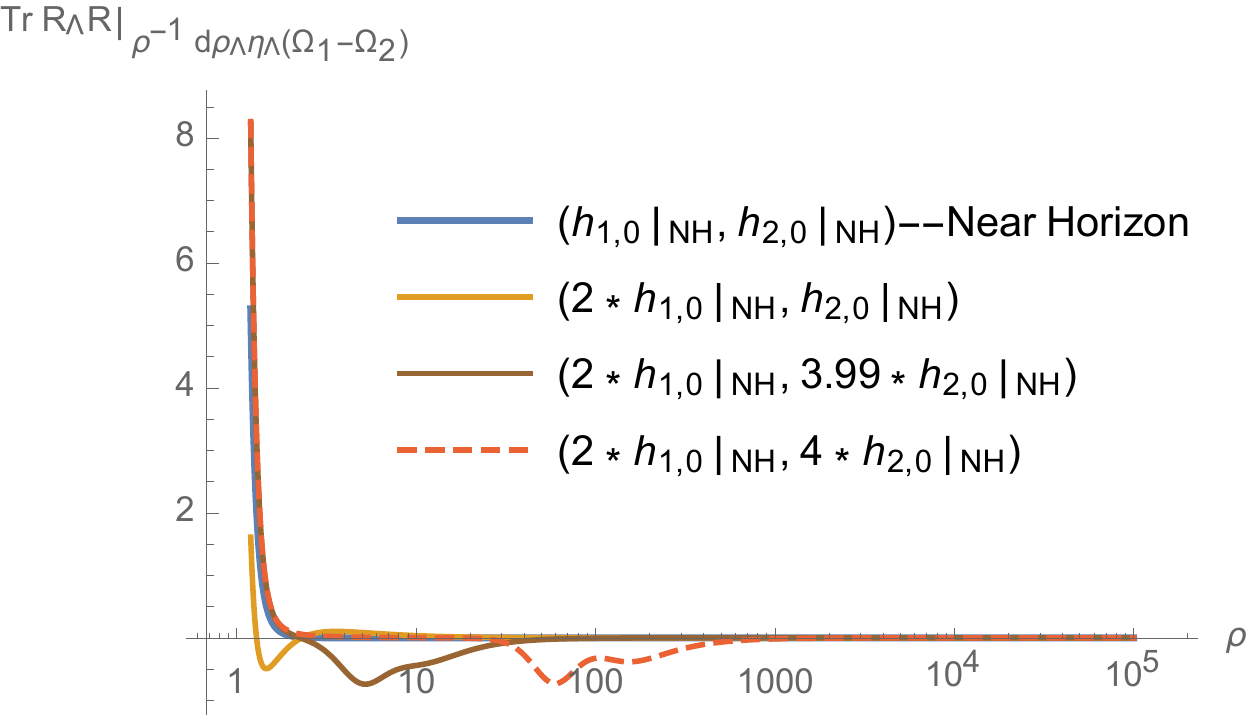}
\end{minipage}
\vskip5mm
\includegraphics[width=.4\linewidth]{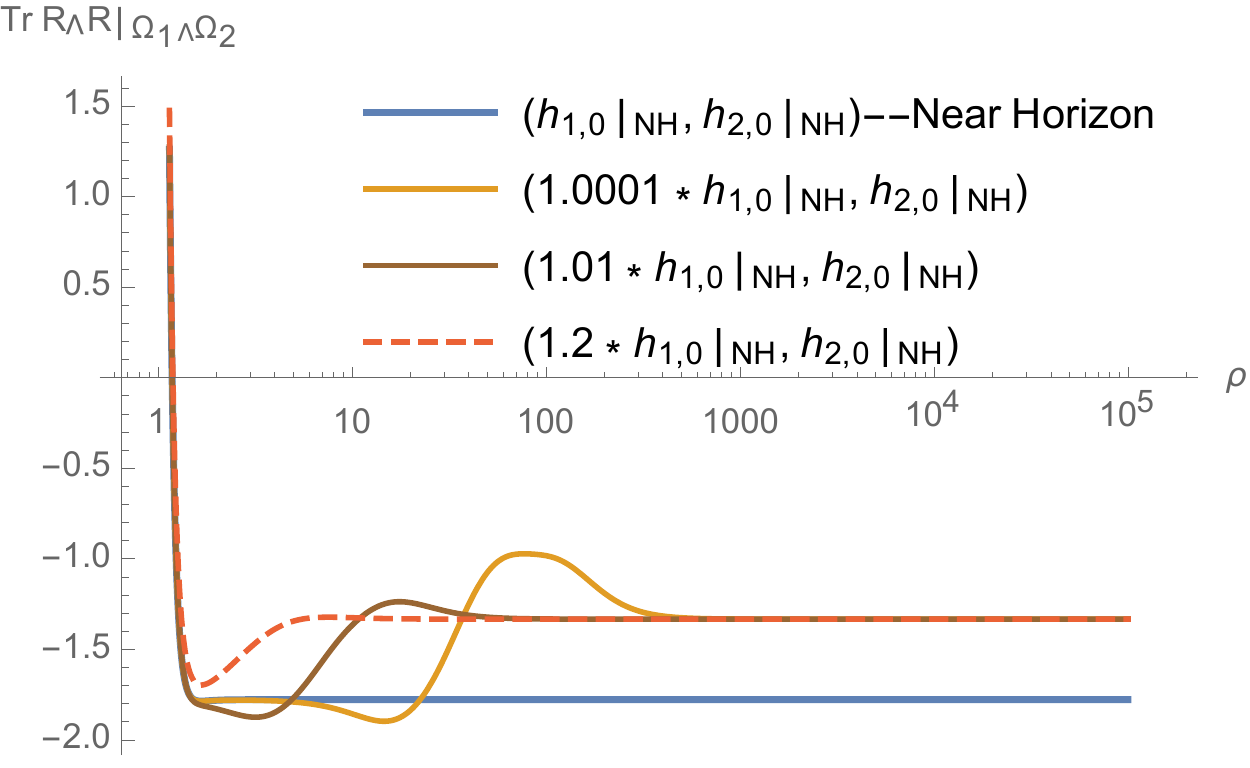}
\caption{\underline{\boldmath $g_c=\frac{1}{3}\, :\quad (k_1,k_2)=(2\times 10^5,10^5)$}}
\label{RwRth1th2_gc13}
\end{figure}
\vspace{5mm}

\begin{figure}[bth!]
\centering
\begin{minipage}{.5\textwidth}
  \centering
  \includegraphics[width=.9\linewidth]{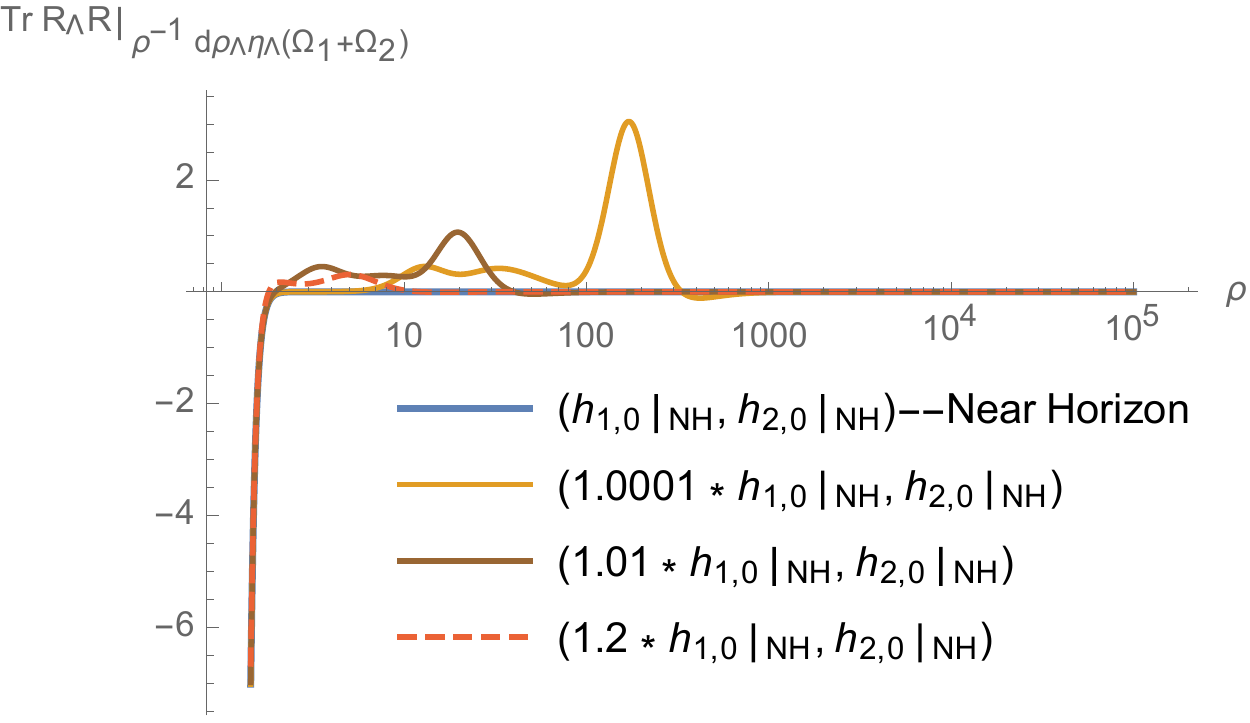}
\end{minipage}%
\begin{minipage}{.5\textwidth}
  \centering
  \includegraphics[width=.9\linewidth]{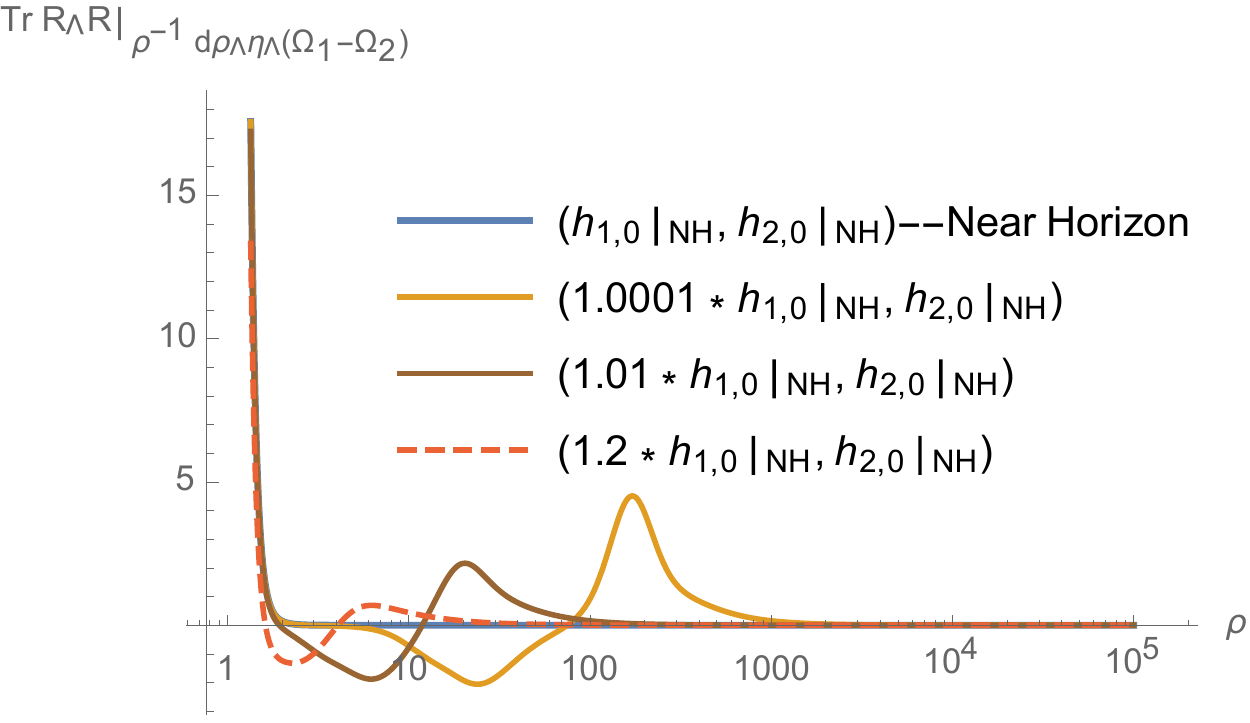}
\end{minipage}
\vskip5mm
\includegraphics[width=.4\linewidth]{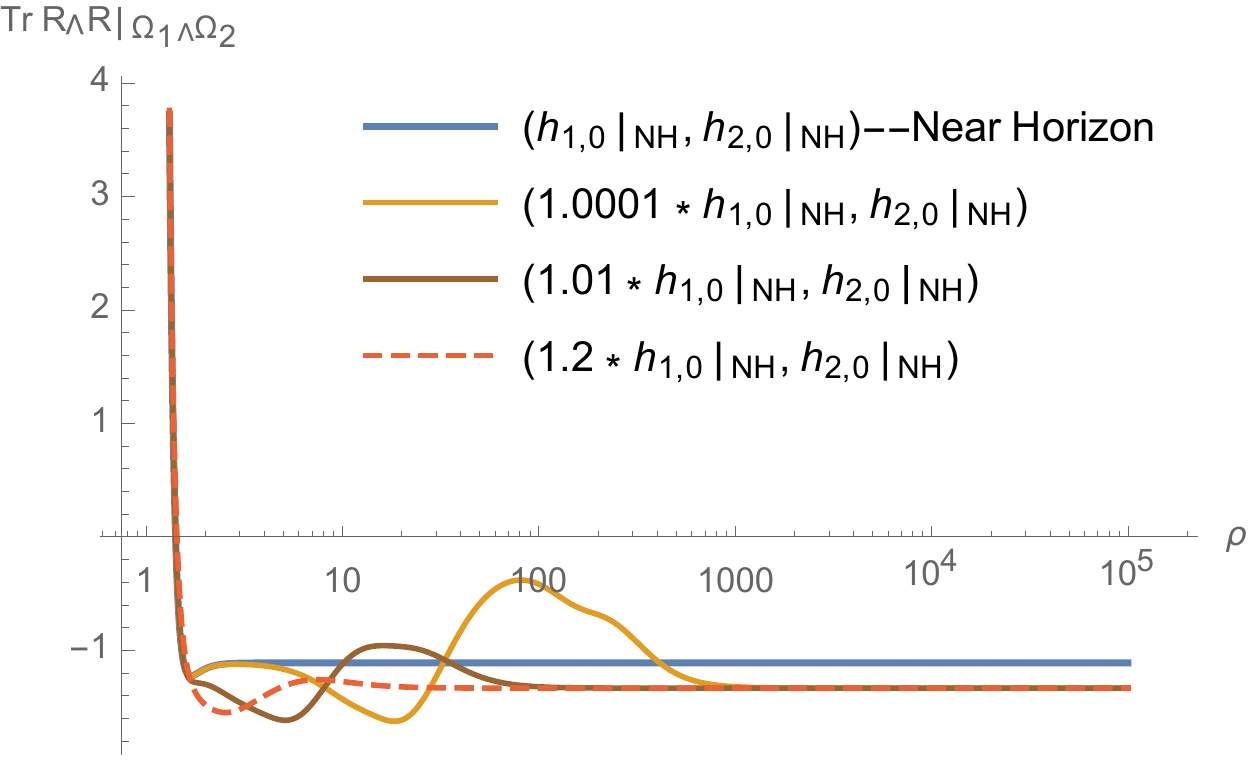}
\caption{\boldmath \underline{$\mathbf{g_c=\frac{2}{3}}:\quad (k_1,k_2)=(5\times 10^5,10^5)$}}
\end{figure}

\begin{figure}[bth!]
\centering
\begin{minipage}{.5\textwidth}
  \centering
  \includegraphics[width=.9\linewidth]{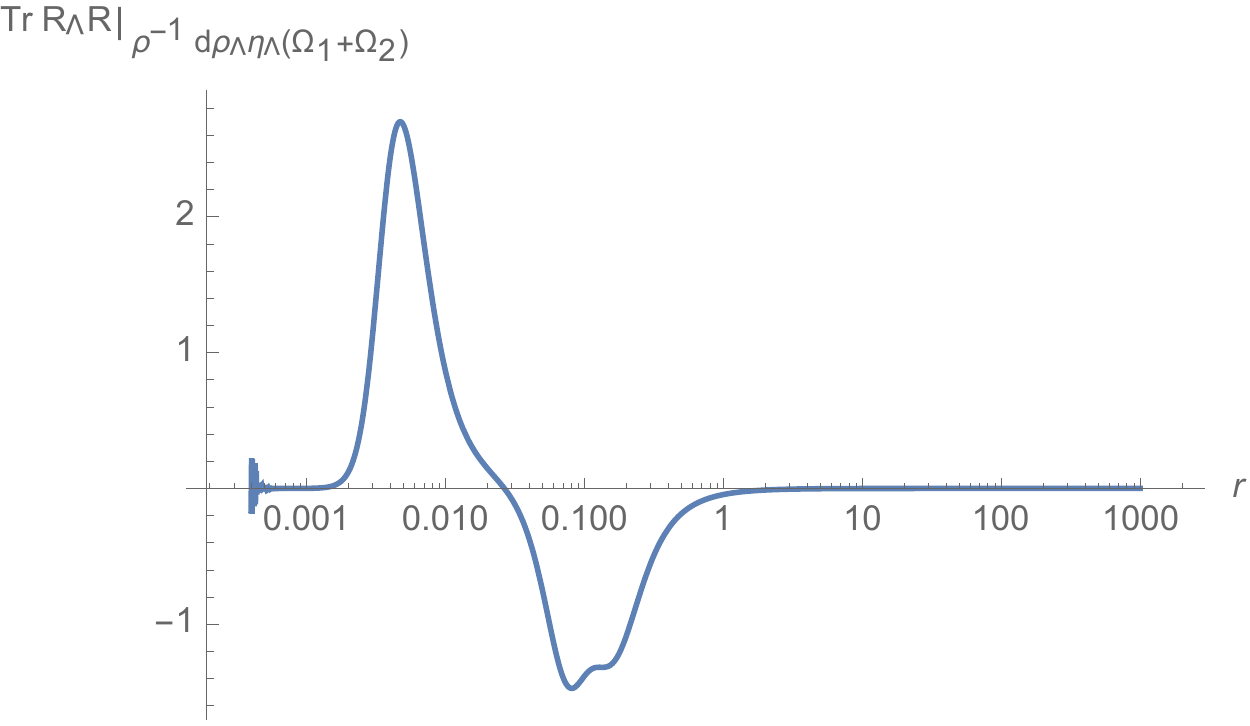}
\end{minipage}%
\begin{minipage}{.5\textwidth}
  \centering
  \includegraphics[width=.9\linewidth]{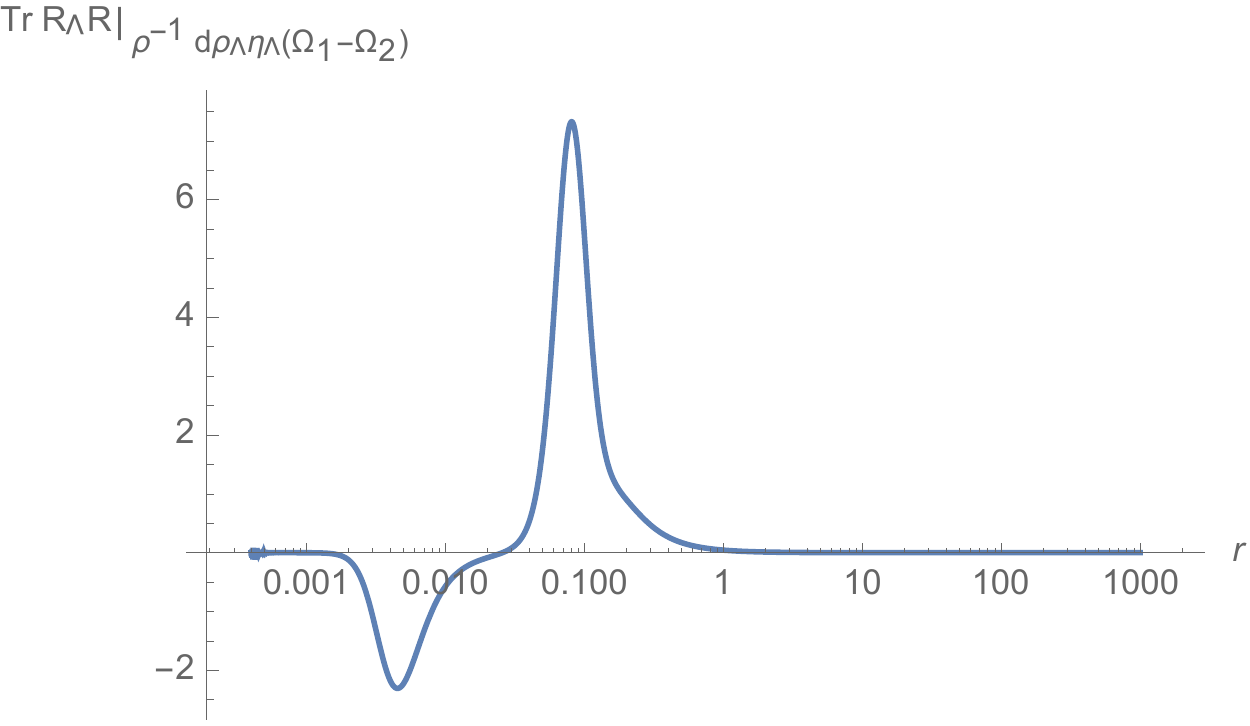}
\end{minipage}
\vskip5mm
\includegraphics[width=.4\linewidth]{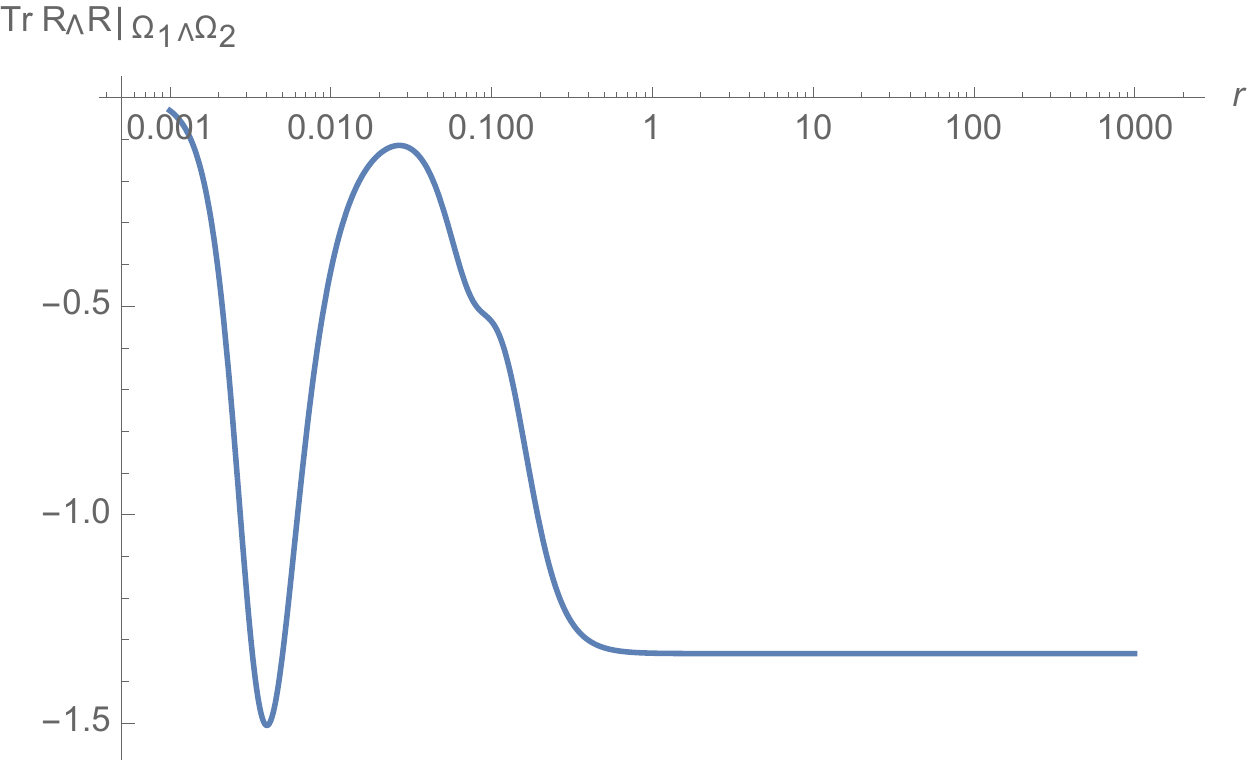}
\caption{\boldmath \underline{$\mathbf{g_c=1}$. The Resolved Conifold}}
\label{RwRth1th2_gc23}
\end{figure}

\newpage
%%%%%%%%%%%%%%%%%%%%%%%%%%%%%%%%%%%%
\end{appendix}
%%%%%%%%%%%%%%%%%%%%%%%%%%%%%%%%%%%%
\newpage
\newpage
%%%%%%%%%%%%%%%%%%%%%%%%%%%%%%%%%%%%
%%%%%%%%%%%%%%%%%%%%%%%%%%%%%%%%%%%%

\newpage
\bibliographystyle{JHEP} 

\providecommand{\href}[2]{#2}\begingroup\raggedright\endgroup

%%%%%%%%%%%%%%%%%%%%%%%%%%%%%%%%%%%%
%%%%%%%%%%%%%%%%%%%%%%%%%%%%%%%%%%%%

%%%%%%%%%%%%%%%%%%%%%%%%%%%%%%%%%%%%
\end{document}